\documentclass[11pt]{article}

\usepackage{bigints}
\usepackage{tabstackengine}
\stackMath
\usepackage{jheppub}
\usepackage{graphbox}
\usepackage[usenames,dvipsnames]{xcolor} 
\usepackage{tikz}	
\usepackage{dirtytalk}
\usepackage{dsfont}
\usepackage{float}
\usepackage{mathdots}
\usepackage{xfrac}

\usepackage{bbm, dsfont}
\usepackage[normalem]{ulem}
\usepackage{slashed}
\usepackage{mathtools}
\usepackage{cancel}
\usepackage{cleveref}
\usepackage{orcidlink}
\usepackage{enumerate}
\usepackage{fontawesome}
\usepackage{adjustbox}
\usepackage{tcolorbox}
\usepackage{amsmath}
\usepackage{amssymb}
\usepackage{amsthm}
\usepackage{shuffle}
\usepackage{enumitem}
\usepackage{color}
\usepackage{hyperref}
\usepackage{bbm}
\usepackage{multirow}
\usepackage{tikz}
    \usetikzlibrary{positioning}
    \usetikzlibrary{arrows}
    \usetikzlibrary{shapes}
    \usepackage{pgfplots}
    \usetikzlibrary{calc}
    \usetikzlibrary{decorations.markings}
     \usetikzlibrary{decorations.pathmorphing}
    \tikzset{snake it/.style={decorate, decoration=snake}}
\def\centerarc[#1](#2)(#3:#4:#5) 
    { \draw[#1] ($(#2)+({#5*cos(#3)},{#5*sin(#3)})$) arc (#3:#4:#5); }
    \usepackage{booktabs}
\def\beq{\begin{equation}}
\def\eeq{\end{equation}}
\def\bsp#1\esp{\begin{split}#1\end{split}}

\renewcommand{\th}{\th}

\renewcommand{\d}{\text{d}}

\tikzset {_hq8agl9h9/.code = {\pgfsetadditionalshadetransform{ \pgftransformshift{\pgfpoint{-19.5 bp } { -37.5 bp }  }  \pgftransformrotate{0 }  \pgftransformscale{2 }  }}}
\pgfdeclarehorizontalshading{_4sn6rfw6e}{150bp}{rgb(0bp)=(1,1,1);
rgb(50.5bp)=(1,1,1);
rgb(72.5bp)=(0.92,0.01,0.11);
rgb(100bp)=(1.3,0.01,0.11)}
\tikzset{every picture/.style={line width=0.75pt}}
\tikzset {_bhh1yolhd/.code = {\pgfsetadditionalshadetransform{ \pgftransformshift{\pgfpoint{11.5 bp } { -8 bp }  }  \pgftransformrotate{-90 }  \pgftransformscale{2 }  }}}
\pgfdeclarehorizontalshading{_90qpx1llj}{150bp}{rgb(0bp)=(0.82,0.01,0.11);
rgb(10.5bp)=(0.82,0.01,0.11);
rgb(52.5bp)=(1,1,1);
rgb(100bp)=(1.3,1,1)}
\tikzset{every picture/.style={line width=0.75pt}}
\tikzset {_h90ekjryl/.code = {\pgfsetadditionalshadetransform{ \pgftransformshift{\pgfpoint{11.5 bp } { -8 bp }  }  \pgftransformrotate{-90 }  \pgftransformscale{2 }  }}}
\pgfdeclarehorizontalshading{_k8eguni3i}{150bp}{rgb(0bp)=(0,0.11,1);
rgb(10.5bp)=(0,0.11,1);
rgb(52.5bp)=(1,1,1);
rgb(100bp)=(1.3,1,1)}
\tikzset{every picture/.style={line width=0.75pt}}
\tikzset {_xqcddvd9p/.code = {\pgfsetadditionalshadetransform{ \pgftransformshift{\pgfpoint{-4 bp } { -27.5 bp }  }  \pgftransformrotate{0 }  \pgftransformscale{2 }  }}}
\pgfdeclarehorizontalshading{_tl6yhambn}{150bp}{rgb(0bp)=(1,1,1);
rgb(50.5bp)=(1.3,1,1);
rgb(72.5bp)=(0,0.11,1);
rgb(100bp)=(0,0.11,1)}
\tikzset{every picture/.style={line width=0.75pt}}
\newcommand\wedgedot[1][1]{
\,
\tikzset{every picture/.style={line width=0.75pt}}      
\begin{tikzpicture}[x=0.75pt,y=0.75pt,yscale=-1,xscale=1]
\draw   (100,130.12) -- (105.06,120) -- (110.12,130.12) ;
\draw  [fill={rgb, 255:red, 0; green, 0; blue, 0 }  ,fill opacity=1 ] (104,127) .. controls (104,126.45) and (104.45,126) .. (105,126) .. controls (105.56,126) and (106.01,126.45) .. (106.01,127) .. controls (106.01,127.56) and (105.56,128.01) .. (105,128.01) .. controls (104.45,128.01) and (104,127.56) .. (104,127) -- cycle ;
\end{tikzpicture}
\,
}
\makeatletter

\def\env@sqcases{
  \let\@ifnextchar\new@ifnextchar
  \left\lbrack
  \def\arraystretch{1.2}
  \array{@{}l@{\quad}l@{}}
}
\makeatother
\definecolor{darkGreen}{cmyk}{1,.2,1,0.2}

\title{On canonical differential equations for Calabi-Yau multi-scale Feynman integrals}

\abstract{We generalise a method recently introduced in the literature, that derives canonical differential equations, to multi-scale Feynman integrals with an underlying Calabi-Yau geometry.
We start by recomputing a canonical form for the sunrise integral with all unequal masses.
Additionally, we compute for the first time a canonical form for the three-loop banana integral with two unequal masses and for a four-loop banana integral with two unequal masses. 
For the integrals we compute, we find an $\epsilon$-form whose connection has at most simple poles.   
We motivate our construction by studying the Picard-Fuchs operators acting on the integrals considered.
In the appendices, we give a constructive explanation for why our generalisation works.}

\author[a,\orcidlink{0000-0002-5266-9449}]{Sara Maggio,}
\emailAdd{smaggio@uni-bonn.de}
\affiliation[a]{Bethe Center for Theoretical Physics, Universität Bonn, D-53115, Germany
}

\author[b,\orcidlink{0009-0007-0651-0676}]{Yoann Sohnle}
\emailAdd{yoann.sohnle@physics.uu.se}
\affiliation[b]{
	Department of Physics and Astronomy, Uppsala University, Box 516, 75120 Uppsala, Sweden
}

\allowdisplaybreaks
\begin{document}
\preprint{\begin{tabular}{r}
UUITP–14/25 \\
\vspace{-25pt} BONN-TH-2025-16
\end{tabular}}

\maketitle
\normalem
\allowdisplaybreaks
\raggedbottom
\newpage

\section{Introduction}

Feynman integrals are central quantities in Quantum Field Theory, as they are essential building blocks for computing scattering amplitudes which are one of the bridges between theory and experiment~\cite{Travaglini:2022uwo}. 
Integrals of similar type have also appeared in the description of the scattering of compact objects in classical General Relativity in the so-called post-Minkowskian expansion~\cite{Goldberger:2004jt, Porto:2016pyg, Kalin:2020mvi,  Mogull:2020sak, Kalin:2020fhe,Kalin:2022hph, Jakobsen:2022psy, Dlapa:2024cje}.
It is therefore necessary to develop sophisticated techniques which help us compute this type of integrals, in particular, by understanding their  
mathematical structure and underlying geometrical information. 
It turns out that at the core of all these integrals are classes of special functions related to complex manifolds where the simplest is the Riemann sphere where
multiple polylogarithms (MPLs)~\cite{Kummer, Remiddi:1999ew, Goncharov:1995, Goncharov:1998kja, Vollinga:2004sn, Duhr:2011zq} appear. 
However, diagrams needed for cutting-edge computations give rise to special functions beyond the ones that live on the Riemann sphere.
These functions beyond MPLs are associated to periods of different geometries.
The first example being the massive sunrise diagram in even dimensions which can be associated to an elliptic curve (which is isomorphic to a torus)~\cite{Bloch:2013tra,Broadhurst:1987ei,Bauberger:1994by,Bauberger:1994hx,Laporta:2004rb}.
Recent progress has been made to go beyond the torus, on the one hand by studying Feynman integrals associated to higher genus Riemann surfaces~\cite{Georgoudis:2015hca,Duhr:2024uid} and their function space \cite{DHoker:2023vax,Baune:2024biq,DHoker:2025szl}.
On the other hand by studying Feynman integrals associated to higher dimensional Calabi-Yau manifolds~\cite{Primo:2017ipr,Klemm:2019dbm,Bourjaily:2018yfy} and their function space~\cite{Bonisch:2024nru,Duhr:2025ppd,Duhr:2025tdf}.
\\Feynman integrals typically diverge and need to be regulated. 
The most commonly used regularisation scheme is dimensional regularisation~\cite{tHooft:1972tcz,Bollini:1972ui}, where the integrals are computed in $D=D_0-2\epsilon$ dimensions. 
They can then be expanded around $\epsilon = 0$. 
An extremely powerful technique to compute the integrals is through the differential equations (DEs)~\cite{Kotikov:1990kg,Remiddi:1997ny,Gehrmann:1999as}, which are derived using integration-by-parts (IBPs) identities~\cite{Tkachov:1981wb,Chetyrkin:1981qh} among Feynman integrals.
This method is extremely convenient because we are not interested in the solution for generic values of $\epsilon$, but rather as a Laurent series in $\epsilon$.
Moreover, in the last decade, it has been observed that it is often possible to find bases of Feynman integrals whose differential equations take a specific form, often referred to as \emph{canonical form}~\cite{Henn:2013pwa}. The dependence on $\epsilon$ can then be factorised out of the differential equation system, which also makes the analytic properties of their solutions close to $\epsilon=0$ completely manifest: each order in the Laurent series can be expressed as Chen iterated integrals~\cite{Chen:1977oja} over the differential forms appearing in the connection.
However, a completely general approach to finding canonical bases, even in the polylogarithmic
case is not yet fully understood.
Moreover, starting at two-loops, finding the canonical basis will involve a gauge transformation dependent on period functions of geometries beyond the Riemann sphere.
\\In recent years, substantial effort has been
dedicated to extend the definition of canonical differential equations to Feynman integrals beyond polylogarithms, and much progress has been made, in particular
in the elliptic and one-parameter Calabi-Yau cases~\cite{Gorges:2023zgv,Pogel:2022yat,Adams:2016xah,Adams:2018bsn,Pogel:2022ken,Ahmed:2024tsg,Frellesvig:2021hkr,Dlapa:2022wdu,
Frellesvig:2023iwr,Chen:2025hzq,Pogel:2022vat}.
In particular, ref.~\cite{Pogel:2022vat} provides an ansatz to bring equal mass banana integrals, with underlying single-scale Calabi-Yau geometry, at any loop order, into $\epsilon$-factorised form.
Moreover, very recently in ref.~\cite{Duhr:2025lbz}, a generalisation of the method of ref.~\cite{Gorges:2023zgv} has been developed for generic integrals with an underlying Calabi-Yau geometry (and beyond), providing explicit calculations for cutting edge Calabi-Yau one-scale integrals.
The applicability of this algorithm has been explicitly demonstrated for many state-of-the-art problems containing integrals of elliptic type~\cite{Gorges:2023zgv, Becchetti:2025rrz, Becchetti:2025oyb,Duhr:2024bzt,Forner:2024ojj,Marzucca:2025eak} and for their generalisation to Calabi-Yau varieties~\cite{Duhr:2024bzt, Forner:2024ojj, Klemm:2024wtd, Driesse:2024feo, Driesse:2024xad, Duhr:2025lbz} and higher-genus surfaces~\cite{Duhr:2024uid}.
\\In this paper, we propose a generalisation of the method introduced in ref.~\cite{Pogel:2022vat}, to multi-scale Calabi-Yau integrals. In the preparation of this work, we have realised that our generalisation coincides with the method from refs.~\cite{Gorges:2023zgv,Duhr:2025lbz}, applied to multi-scale Calabi-Yau Feynman integrals. In fact, this is the first application of refs.~\cite{Gorges:2023zgv,Duhr:2025lbz} to this class of integrals.
To support our claim, we derive locally at the MUM-point, for the first time, a canonical differential equation for banana integrals with unequal masses.
\\This paper is organised as follows. 
In section~\ref{sec:2} we set our conventions and review the instruments that will be needed for the procedure we propose. 
Following this, in section~\ref{sec:4}, we present our procedure for multi-variate Feynman integrals with underlying Calabi-Yau geometry. 
In section~\ref{sec:sunrise}, we recompute a canonical form for the unequal mass sunrise integral using the method presented before. 
Finally, in section~\ref{Computing new integrals}, we apply our construction to compute a canonical form for integrals with underlying multi-variate Calabi-Yau geometry, namely, the three-loop banana with two unequal masses (two-parameters K3) and the four-loop banana with two unequal masses (two-parameters CY three-fold). 
Additionally, we include three appendices.
In the first two we prove the validity of our procedure for the elliptic and Calabi-Yau two-fold case.
In the third, we briefly review the modular bootstrap method of Ref. \cite{Giroux:2022wav}.
The main quantitative results will be compiled in the ancillary files {\bf{\texttt sunrise.nb}}, {\bf{\texttt banana\_3\_1.nb}}, {\bf{\texttt banana\_2\_2.nb}} and {\bf{\texttt banana\_4\_1.nb}}.

\section{Definitions and Review}
\label{sec:2}

\subsection{Feynman integrals and their differential equations}\label{sec:2.1}
Let us consider $\text{L}$-loop scalar Feynman integrals with $e$ external legs.
We work in the framework of dimensional regularisation where we pick (unless specified otherwise) the dimension to be $D=2-2\epsilon$, since our examples are two-point functions. 
In this framework $\epsilon$ tracks the UV and IR behaviour of the integral.
A given Feynman integral can be expressed by the formula:
\begin{equation}\label{int_def}
    I_{\nu_1,\nu_2,\dots,\nu_r}(\underbrace{\{p_i\cdot p_j,m_i^2\}}_{{\bf z}};\epsilon)=e^{\text{L}\gamma_E \epsilon}\int\left(\prod_{j=1}^\text{L}\dfrac{\text{d}^Dl_j}{i\pi^{D/2}}\right)\dfrac{1}{D_1^{\nu_1}D_2^{\nu_2}\dots D_r^{\nu_r}}\,,
\end{equation}
where $\gamma_E$ is the Euler-Mascheroni constant, the exponents of the propagators $\nu_j$ are integers and
the propagators $D_j$ are defined through:
\begin{equation}
    D_j=\left(\sum_{k=1}^\text{L} \alpha_{jk}l_k+\sum_{k=1}^{e-1}\beta_{jk}p_k\right)^2-m_j^2\,,
\end{equation}
where $\alpha_{jk} \beta_{jk} $ can be chosen to lie in $\{0,\pm 1\}$.
Any integral of the form (\ref{int_def}) can be reduced to a finite linear combination of master integrals (MIs) which has fixed $\nu_j$.
Every integral belongs to a \emph{sector}, defined as the set of integrals from the family that share exactly the same propagators, i.e., $I_{\nu_1,\dots,\nu_{r}}$ and $I_{\nu'_1,\dots,\nu'_{r}}$ belong to the same sector if $\theta(\nu_i)  =\theta(\nu'_i)$, for all $1\le i\le r$, where $\theta$ denotes the Heaviside step function, $\theta(x)=1$ if $x>0$ and $\theta(x)=0$ otherwise. 
Determining these MIs and their associated differential equation can be achieved through the Laporta algorithm, where different software implementations are available \cite{Lee:2012cn,Smirnov:2008iw,Smirnov:2019qkx,Klappert:2020nbg,Maierhofer:2017gsa,Peraro:2019svx,Guan:2024byi}.
This leads to a system of linear first-order differential equations for the vector of MIs $\bf I$ that can be shown to be \cite{Kotikov:1991hm,Kotikov:1991pm,Henn:2013pwa,Kotikov:1990kg,Gehrmann:1999as}:
\begin{equation}\label{deq_start}
\text{d}{\bf{I}}({\bf z};\epsilon)={\bf{A}}({\bf z};\epsilon)\cdot{\bf{I}}\left({\bf z};\epsilon\right),
\end{equation}
where ${\bf{A}}$ is a matrix of rational one-forms. 
To be able to solve for $\bf{I}$, we aim to find a gauge transformation ${\bf{U}}$:
\begin{align}\label{gauge_trafo}
    {\bf{M}}&={\bf{U}}\cdot {\bf{I}}\,,\notag\\
    \tilde{\bf{A}}&={\bf{U}}\cdot{\bf{A}}\cdot{\bf{U}}^{-1}+\text{d}{\bf{U}}\cdot{\bf{U}}^{-1}\,,
\end{align}
such that the differential equation is $\epsilon$-factorised:
\begin{equation}\label{eps_fact}
    \text{d}{\bf{M}}\left({\bf z};\epsilon\right)=\epsilon\tilde{\bf{A}}\left({\bf z}\right)\cdot{\bf{M}}\left({\bf z};\epsilon\right)\,.
\end{equation}
Once this is achieved, one can solve for ${\bf{M}}$:
\begin{align}
    {\bf{M}}\left({\bf z};\epsilon\right)&=\mathcal{P}\hspace{-0.8mm}\exp\left(\epsilon\int_{{\bf z}_0}^{{\bf z}}\tilde{\bf A}({\bf z}')\right){\bf{M}}({\bf z}_0;\epsilon)\,,
\end{align}
obtaining a Laurent series by expanding the path-ordered exponential, that can be easily truncated at the desired order.
Note that unless specified otherwise the entries of ${\bf \tilde{A}}$ are closed one-forms.
This will lead to iterated integrals over algebraic one-forms.

\subsection{Feynman integrals and their geometry}\label{sec:2.2}

Of special interest in this paper will be the so-called \textit{maximal cuts}, which correspond to integrals where all propagators are on-shell. 
As pointed out in ref.~\cite{Duhr:2025lbz}, in a single maximal cut there can be more geometries, and the correct framework to use is the \textit{mixed Hodge structure} (MHS). 
Let us briefly review this.

\paragraph{Mixed Hodge structure:}
First, since we want to evaluate the master integrals in 
\\${D=D_0-2\epsilon}$, we need to make a good choice for $D_0$. 
It turns out that analysing the integrand in Baikov representation often reveals which is the right integer dimension $D_0$ to choose.
It may be possible to identify more than one geometry from a given maximal cut. 
For example, the family of ice-cone integrals with equal masses hides two different Calabi-Yau varieties in the same maximal cut~\cite{Duhr:2022dxb}. 
\\Let us now focus on only one of the geometries in the maximal cut and let us assume that it describes an $l$-dimensional manifold. 
Each geometry on the maximal cut gives rise to a set of (independent) differential $l$-forms, which generate (a subspace of) the $l^{\textrm{th}}$ cohomology group~$H^l(X,\mathbb{Q})$.
From algebraic geometry~\cite{PMIHES_1971__40__5_0, PMIHES_1974__44__5_0} we know that the cohomology of an algebraic variety always carries a mixed Hodge structure (MHS). 
This roughly means that the~$l^{\textrm{th}}$ cohomology group $H^l(X,\mathbb{Q})$ of an algebraic variety is always equipped with two filtrations.
There is an increasing filtration called the \textit{weight filtration},
\beq
0=W_{-1}\subseteq W_0 \subseteq W_1\subseteq \ldots W_{2l} = H^l(X,\mathbb{Q})\,,
\eeq 
and a decreasing filtration on the complexification called the \textit{Hodge filtration}
\beq\label{eq:Hodge_filtration_DEQ}
0\subseteq\ldots\subseteq F^p \subseteq F^{p-1} \subseteq \ldots \subseteq F^1\subseteq F^0 = H^l(X,\mathbb{C})\,.
\eeq 
The graded pieces $W_k/W_{k-1}$ naturally carry a pure Hodge structure of weight $k$ induced by the Hodge filtration. The pure Hodge structures can be seen as the MHS carried by the cohomology of projective smooth varieties, in which case the MHS is concentrated in weight~${l: 0=W_{l-1} \subset W_l = H^l(X,\mathbb{Q})}$.
\\It was argued in ref.~\cite{Duhr:2025lbz}, that the MHS captures precisely how to find a good starting basis. In particular, the weight filtration captures the fact that some forms may have simple poles. This is because the weight-graded pieces tell us how the cohomology of $X$ contains pieces coming from `simpler' varieties. We refer the reader to ref.~\cite{Duhr:2025lbz} for a more in-depth review and explanation.

\paragraph{Some examples:} In all the examples we consider, we find that the maximal cut only features one geometry.
In cases where there are more than one geometry, we expect the differential equation to split into different sub-blocks, provided that we choose a good starting basis as predicted by the MHS.
Equivalently as putting the propagators on-shell, a maximal cut of a Feynman integral is obtained by evaluating the integrand in Baikov representation on a contour that encircles all the propagator poles. 
The integration contour varies according to the geometry of the Feynman integral considered. 
For instance, at one loop we encounter complex curves of genus zero, namely Riemann spheres.
In this case, maximal cuts, in the appropriate integer dimension, evaluate to algebraic functions and the differential one-forms obtained in the differential equation are the so-called $dlog$-forms, which after integration evaluate to multiple polylogarithms (MPLs)~\cite{Kummer, Remiddi:1999ew, Goncharov:1995, Goncharov:1998kja, Vollinga:2004sn, Duhr:2011zq}. 
\\Starting from two-loops, we encounter more involved geometries~\cite{Broadhurst:1987ei,Bauberger:1994by,Bauberger:1994hx,Laporta:2004rb,Kniehl:2005bc,Aglietti:2007as,Bloch:2016izu,Adams:2018kez}, the simplest example of which appears in the sunrise graph with equal non-zero masses. 
When computing the maximal cut of this integral in $D_0=2$, we get
\begin{equation}
 \mathrm{MaxCut}(I)\biggr|_{D=2}\sim\int\dfrac{\text{d}z}{\sqrt{P_4(z)}}\,,
\end{equation}
where $P_4(z)$ is a polynomial of degree four and therefore has four different roots. 
The polynomial $P_4(z)$ defines the zero-locus of a (complex) elliptic curve, thus two independent cycles can be chosen as integration contours around/across branch cuts and we get two independent functions, which correspond to the \textit{periods} of the elliptic curve~\cite{Bloch:2013tra,Broadhurst:1987ei,Bauberger:1994by,Bauberger:1994hx,Laporta:2004rb}. 
Since complex elliptic curves are isomorphic to complex tori, the standard geometric coordinates for a Feynman integral with $n$ moduli, are $\tau, z_1, ..., z_{n-1}$, where $\tau$ is the modular parameter defined as the ratio of the two periods and $z_1, ..., z_{n-1}$ are additional marked points (one of them can be fixed using translation symmetry). 
In that case, differential one-forms can be written in the following form:
\begin{subequations}
\begin{align}
    &\omega_k^{\mathrm{modular}}=2\pi i f_k(\tau)\text{d}\tau\,,\\
    &\omega_k^{\mathrm{Kronecker}}=(2\pi i)^{2-k}\left[g^{(k-1)}(z,\tau)\,\text{d}z+(k-1)\,g^{(k)}(z,\tau)\dfrac{\text{d}\tau}{2\pi i}\right]\,,\label{kronecker}
\end{align}
\end{subequations}
where $f_k(\tau)$ is a modular form of weight $k$ for some congruence subgroup and $g^{(k)}(z,\tau)$ are the coefficients of the expansion of the Kronecker function. 
A similar analysis applies when the degree of the polynomial under the square root in the maximal cut is three~\cite{Broedel:2017kkb,Broedel:2017siw}. 
For these elliptic cases, the $W_1$ part of the weight filtration coincides with the cohomology $H^1(\mathcal{E},\mathbb{Q})$, where $\mathcal{E}$ is the elliptic curve.
The complete space of master integrals can be identified with the $W_2$ part of the weight filtration. Working modulo linear combinations of the two master integrals in $W_1$, it leads us to consider the quotient $W_2/W_1$, which dimension depends on the number of additional master integrals in the maximal cut, capturing their existence.
\\Two ways to go beyond elliptic curves have appeared within the framework of Feynman integrals. 
One is given by curves associated to Riemann surfaces of higher genus~\cite{Marzucca:2023gto,Duhr:2024uid}.
The other corresponds to higher-dimensional manifolds, in particular Calabi-Yau manifolds~\cite{Brown:2010bw,Primo:2017ipr,Bourjaily:2018ycu,Klemm:2019dbm,Bonisch:2021yfw,Duhr:2022pch}.
The simplest example where one observes this geometry are the so-called $\text{L}$-loop banana integrals for $\text{L}\geq 2$~\cite{Bloch:2014qca} (the sunrise integral at $L=2$ has an underlying geometry of a one-dimensional Calabi-Yau manifold which in turn is a torus).
We will review their $\epsilon$-factorisation in section~\ref{sec:3.1}.
\\To summarise, we have seen that Feynman integrals are related to different geometries and that one way to study these geometries and their associated one-forms amounts to studying the maximal cut of the integral. 
Moreover, as we will see in section \ref{sec:2.3}, the maximal cut can be shown to solve a homogeneous higher-order Picard-Fuchs (PF) equation, that translates into a coupled block of the differential equation matrix. Hence, finding the transformation to a canonical basis for maximal cuts is an important step for finding the canonical basis of the full Feynman integral family.

\subsection{Picard-Fuchs operators}\label{sec:2.3}
It is possible to rewrite the system of first order differential equations (\ref{deq_start}) into a system of inhomogeneous higher-order differential equations, which will take the form:
\begin{equation}
    L^{(k)}(\mathbf{z},\epsilon)\, I_{\nu_1,\dots,\nu_r}(\mathbf{z};\epsilon)=R_{\nu_1,\dots,\nu_r}(\mathbf{z};\epsilon)\,,
    \label{pfoperator}
\end{equation}
where $k$ is the order of the differential operator and the inhomogeneity $R_{\nu_1,\dots,\nu_r}(\mathbf{z};\epsilon)$ contains master integrals from lower sectors.
Hence, the differential equation for the maximal cut is  determined by setting $R_{\nu_1,\dots,\nu_r}(\mathbf{z};\epsilon) $ to zero.
Differential equations of Feynman integrals are expected to have only regular singularities i.e. Eq.~(\ref{pfoperator}) is a \textit{Fuchsian} differential equation. The differential operator $L^{(k)}(\mathbf{z},\epsilon)$ is referred as the $\textit{Picard-Fuchs}$ operator and takes the form
\begin{equation}
    L^{(k)}(\mathbf{z},\epsilon)=\sum_{i_1,...,i_n\geq 0}a_{k,i_1,...,i_n}(\mathbf{z};\epsilon)\partial_{z_1}^{i_1}...\partial_{z_n}^{i_n}\,,
\end{equation}
where $k=\mathrm{Max}\{i_1+i_2+\dots+i_n\}$  and 
$a_{k,i_1,...,i_n}(\mathbf{z};\epsilon)$ are polynomials in $\mathbf{z}$ and $\epsilon$.  
The PF operator splits into two parts: $L^{(k)}(\mathbf{z},\epsilon)=L_{0}^{(k)}(\mathbf{z})+L_{\epsilon}^{(k)}(\mathbf{z},\epsilon)$,\footnote{From now we leave the $\mathbf{z}$ and $\epsilon$ dependence of the operators implicit.} where $L_{0}^{(k)}$ contains the $\mathcal{O}(\epsilon^0)$ part of the operator and therefore annihilates the maximal cuts at $\epsilon=0$.
\\Let us briefly review some strategies to solve the homogeneous equation $L_0^{(k)}f(\mathbf{z})=0$, where we follow ref.~\cite{Bonisch:2021yfw}. 
Starting from a one-parameter PF operator, we write:
\begin{equation}
    L_0^{(k)}=q_k(z)\,\partial_z^k+q_{k-1}(z)\,\partial_z^{k-1}+...+q_0(z)\,, \qquad q_k(z)\neq 0\,,
\end{equation}
where $q_i(z)$ are polynomials and we assume that the $q_i(z)$ do not have any common zero. The leading coefficient $q_k(z)=:\mathrm{Disc}(L_0^{(k)})$ is called the discriminant. It is convenient to also write the operator in an equivalent form
\begin{equation}
    L_0^{(k)}=\tilde{q}_k(z)\,\theta^k+\tilde{q}_{k-1}(z)\,\theta^{k-1}+...+\tilde{q}_0(z)\,, \qquad \tilde{q}_k(z)\neq 0\,,
\end{equation}
where $\tilde{q}_i(z)$ are polynomials and $\theta=\theta_z:=z\partial_z$ are the Euler operators.
Let ${p_i(z):=q_i(z)/q_k(z)}$ for $0<i<k$. The differential equation has an \textit{ordinary point} at $z=z_0$, if the coefficient functions $p_i(z)$ are analytic in a neighbourhood of $z_0$ for all $0<i<k$. 
A point $z_0$ is called a \textit{regular singular point} if $(z-z_0)^{k-i}p_i(z)$ are analytic in a neighbourhood of $z_0$. 
An \textit{irregular singular point} is neither an ordinary nor a regular singular point. We can find all the singular points $z_0\neq\infty$ looking at the zeroes of the discriminant and for the point at infinity $z_0=\infty$, one introduces the variable $t=1/z$ and proceeds with the same analysis around $t=0$.
A differential equation without irregular singular points is called a \textit{Fuchsian} differential equation. 
\\ This equation has $k$ independent solutions 
and we can obtain a basis using \textit{Frobenius} method. 
For every point $z\in\mathbb{C}$, we can construct $k$ linearly independent local solutions. 
Each of them is given in terms of power series which converge up to the nearest singularity. 
These local solutions can be analytically continued to multivalued global solutions over the whole parameter space. 
Unless stated otherwise we will always be looking for the solution around $z_0=0$.\footnote{For solutions away from zero, we can perform the following substitutions: if $z_0\neq\infty$, then $z\rightarrow z'=z-z_0$ and if $z_0=\infty$, then  $z\rightarrow z'=1/z$}
The method starts by solving the indicial equation
\begin{equation}
    \tilde{q}_k(0)\,\alpha^k+\tilde{q}_{k-1}(0)\,\alpha^{k-1}+...+\tilde{q}_0(0)\,\alpha=0\,,
\end{equation}
which we solve for $\alpha$ and its solutions are called the indicials or local exponents at $z_0=0$.
\\For what follows, we are particularly interested in the solutions around a special point where all the indicials are equal\footnote{For simplicity, to show how our method works, we compute the differential equation for our three-loop and four-loop examples explicitly only expanding around the MUM-point. We comment later on how to proceed around a different point.}. 
This point is called \textit{point of maximal unipotent monodromy} (MUM-point) and the basis of solutions can be chosen to be a tower of logarithmic solutions, starting from a power series-type solution $\psi_0$, also called holomorphic solution, a single logarithmic solution $\psi_1$, a double logarithmic solution $\psi_2$ until we reach $\textrm{log}^{k-1}(z)$; 
we call the basis~$\psi_0(z),\psi_1(z),...,\psi_{k-1}(z)$ the \textit{Frobenius basis}.
\\Let us now turn to the multi-parameter Picard-Fuchs operator. 
The first difference is that we have a set of differential operators $\mathcal{D}=\{L_1^{(k_1)},...,L_s^{(k_s)}\}$ and the solutions have to be annihilated simultaneously by all elements in $\mathcal{D}$: 
\begin{equation}
    \textrm{Sol}(\mathcal{D}):=\{f(\mathbf{z})|L_i^{(k_i)}f(\mathbf{z})=0 \textrm{ for all operators } L_i^{(k_i)}\in\mathcal{D}\}\,.
\end{equation}
The set $\mathcal{D}$ generates a left-ideal of differential operators.
If $L_i^{(k_i)}\in\mathcal{D}$ and $f\in\textrm{Sol}(D)$, then~$\hat{L}L_i^{(k_i)} f(\mathbf{z})=0$ for any differential operator $\hat{L}$. 
We can also generalise the method of Frobenius to the multi-variate case.
Around singular points the solutions look like:
\begin{equation}
    \left(\prod_{i+1}^rz_i^{\alpha_i}\right)\sum_{\substack{j_1, \dots, j_r \geq 0 \\ k_1, \dots, k_r \geq 0}}a_{j_1,...,j_r,k_1,...,k_r}\,\textrm{log}^{j_1}(z_1)\cdot\cdot\cdot\,\textrm{log}^{j_r}(z_r)\,z_1^{k_1}\cdot\cdot\cdot z_r^{k_r}\,,
    \label{Frobenius method}
\end{equation}
where $a_{j_1,...,j_r,k_1,...,k_r}$ are polynomials in $\mathbf{z}$ and $\epsilon$. This local basis can be analytically continued to a global solution.
This is, however, a harder problem. 
We are able to write our first example (the unequal masses sunrise integral) using closed solutions.
The three- and four-loop banana integrals with two unequal masses will, however, be $\epsilon$-factorised using local solutions only and we leave a solution in terms of known functions for future work with some recent progress in that direction in ref.~\cite{Duhr:2025ppd}.
\\In the rest of the paper we will consider the following Picard-Fuchs operators:
\begin{align}
\label{genericPF}
L^{(k)} = & L^{(k)}_0 +  L^{(k-1)}_\epsilon \\
=&\sum_{i_1,..,i_r=0}^{| i|=k}q^{(0)}_{i_1,..,i_r}(\mathbf{z})\, \partial_{z_1}^{i_1}...\partial_{z_r}^{i_r}+ \epsilon \sum_{i_1,..,i_r=0}^{| i|=k-1}q^{(1)}_{i_1,..,i_r}(\mathbf{z})\, \partial_{z_1}^{i_1}...\partial_{z_r}^{i_r}\nonumber\\
&+\epsilon^2 \sum_{i_1,..,i_r=0}^{| i|=k-2}q^{(2)}_{i_1,..,i_r}(\mathbf{z})\, \, \partial_{z_1}^{i_1}...\partial_{z_r}^{i_r}+\ldots+\epsilon^{k} q_0^{(k)}(\mathbf{z})\,,\nonumber
\end{align}
with at least one non-vanishing term $q^{(0)}_{i_1,\dots,i_k}$.
Note that the first term corresponds to $L_0^{(k)}$. This generalises the operators considered in some parts of ref.~\cite{Duhr:2025lbz} to several variables.
Also note that, on the maximal cut, this implies
\begin{equation}
     L^{(k)}_0I_1 =- L^{(k-1)}_\epsilon I_1\,,
    \label{genericPFequality}
\end{equation}
i.e., we can trade the $\mathcal{O}(\epsilon^0)$ part of the PF, with the lower order $\epsilon$-dependent PF.

\paragraph{Calabi-Yau differential operators:}\label{CYoperators} 
Since in the following, we will consider Feynman integrals with underlying Calabi-Yau geometry, it is relevant to review a subset of the Picard-Fuchs operators, the so-called \textit{Calabi-Yau differential operators}.
A thorough description of Calabi-Yau operators will be left to the literature~\cite{BognerCY,Bonisch:2021yfw} and we will just review the most relevant properties. 
\\A $(l+1)^{\mathrm{th}}$-order Calabi-Yau operator $L_0^{(l+1)}$   has a MUM-point, which for our purpose we assume to be located at $z=0$. 
At the MUM-point, the Frobenius basis $\psi_i$ for $i=0,1,\hdots,l$, gives a basis of solutions to the operator $L_0^{(l+1)}$. 
It reads
\begin{equation}
    \psi_i = z^r\sum_{j=0}^i\frac{1}{j!}\log^j(z)S_{i-j}(z)  \, ,
\label{eq:frob}
\end{equation}
where $r$ is the local exponent or indicial of our solutions and $S_i(z) = \sum_{j=0}^\infty \sigma_{i,j}z^j$ are local holomorphic series which are normalised by $\sigma_{i,0} = \delta_{i0}$. 
At the MUM-point, the Frobenius basis $\psi_i$ for $i=0,1,\hdots, l$, exhibits the full logarithmic tower up to $\log^l(z)$. From the first two solutions, we can construct\footnote{In the rest of this work, we ignore factors of $2\pi i$ for simplicity.}
\begin{equation}
    t(z) = \frac{\psi_1}{\psi_0} = \log(z) + \hdots\,, \quad q(z) = e^t = z + \hdots \, ,
\label{eq:canvar}
\end{equation}
which generalises the $\tau$ parameter of an elliptic curve and gives a local (at the MUM-point) canonical variable on the moduli space of Calabi-Yau varieties. 
The dots denote local analytic terms in $z$. 
The inverse of this map is also known as the mirror map
\begin{equation}
    z(q) = q + \hdots \, .
\label{eq:mirrormap}
\end{equation}
We can define recursively the $N_j$ operators of ref.~\cite{BognerCY}
 \begin{equation}
     N_0=1\,,\quad N_{j+1}=z\dfrac{\partial}{\partial z} \dfrac{1}{N_j(\psi_j^{(l)})}N_j\,,
 \end{equation}
and the structure series 
\begin{equation}
    \alpha_j=\frac{1}{N_j(\psi_j)}\,.
\end{equation}
This lets us define the $Y$-invariants as
\begin{equation}\label{y-def}
     Y_j=\dfrac{\alpha_1}{\alpha_j}\,,
\end{equation}
which enjoy the symmetry
\begin{equation}
    Y_j=Y_{l-j}\,.
\end{equation}
Note that by definition we have $Y_1= Y_{l-1}=1$. 
The structure series $\alpha_i$ and the $Y$-invariants are local analytic series in $z$ around the MUM-point. 
Finally, we can use the $Y$-invariants to write the Calabi-Yau operator $L_0^{(l-1)}$ in a factorised form
\begin{equation}
    L_0^{(l-1)} = \beta(q)\left[\theta_q^2 \frac{1}{Y_2(q)} \theta_q\frac{1}{Y_3(q)}\theta_q \hdots \theta_q\frac{1}{Y_3(q)}\theta_q \frac{1}{Y_2(q)}\theta_q^2   \right] \frac{1}{\psi_0(q)} \, ,
\label{eq:normalform}
\end{equation}
also known as the local normal form. We have expressed the operator through the variable~$q$ using the mirror map $z(q)$ given in Eq.~\eqref{eq:mirrormap}.
The function $\beta(q)$ is necessary for a proper normalisation of the operator, and we have used the logarithmic derivative (sometimes called Euler operator) $\theta_q = q \partial_q=\partial_{\tau}$.

\subsection{Hodge filtration \& Griffiths transversality}\label{sec:2.4}

Let us review the instruments needed to study the (co)homology of Calabi-Yau manifolds, where we closely follow ref.~\cite{Bonisch:2021yfw}.
Calabi-Yau $l$-folds $M_l$ are complex $l$-dimensional Kähler manifolds. 
They are equipped with a Kähler form $\omega$ of Hodge-type $(1,1)$ that resides in the cohomology group $H^{1,1}(M_l, \mathbb{C})$. 
The extra condition of being Calabi-Yau implies the existence of a non-trivial holomorphic $(l, 0)$-form $\Omega$ spanning $H^{l,0} (M_l, \mathbb{C})$. 
\\
For each point $\mathbf{z}_0$, the fibre $M_l^{\mathbf{z}_0}$ over it enjoys a Hodge decomposition of its middle dimensional cohomology:
\begin{equation}\label{Hodge_dec}
    H^l(M_l,\mathbb{C})=\bigoplus_{p+q=l}H^{p,q}(M_l) \quad\mathrm{with}\quad\overline{H^{p,q}(M_l)}=H^{q,p}(M_l).
\end{equation}
The so-called Hodge numbers $h^{p,q}$ give the dimension of $H^{p,q}(M_l)$ and can be organised into the Hodge diamond:
\begin{align}
\begin{matrix}
    &&&h^{l,l}&&&\\
   &&h^{l,l-1}&&h^{l-1,l}&&\\
   &\iddots&&\vdots&&\ddots&\\
  h^{l,0} &&\dots&&\dots&&h^{0,l}\\
   &\ddots&&\vdots&&\iddots&\\
   &&h^{1,0}&&h^{0,1}&&\\
   &&&h^{0,0}&&&
\end{matrix}.
\end{align}
For instance, when considering a Calabi-Yau one-fold, (which is isomorphic to an elliptic curve) we find the holomorphic $(1, 0)$-form $\Omega=\text{d}x/y$ and its Hodge diamond reads:
\begin{equation}\label{hodge_elliptic}
    \begin{matrix}
        &1&\\
        1&\quad\,\,&1\\
        &1&
    \end{matrix}\,.
\end{equation}
We can now define the \textit{periods} of a Calabi-Yau manifold as pairings between the middle homology and middle cohomology. 
The maximal cuts, at $\epsilon=0$, of Feynman integrals with a single Calabi-Yau geometry are periods and the dimension of the middle cohomology ${|h|=\sum_{k=0}^l h^{l-k,k}}$ corresponds to the number of master integrals on the maximal cut in integer dimension $D=D_0$ for some judicious choice of $D_0$.
In particular we choose an integral topological basis $\Gamma_i$, for $1\leq i\leq |h|$  in the middle homology $H_l(M_l,\mathbb{Z})$, such that the periods are given by
\begin{equation}
    \Pi_{ij}=\int_{\Gamma_i}\hat{\Gamma}^j\,,
    \label{periodsCY}
\end{equation}
with $\hat{\Gamma}^j$ some basis of $H^l(M_l,\mathbb{C})$.
\\
As an example, let us look at an elliptic curve $\mathcal{E}$ in Legendre form:
\begin{equation}
   \mathcal{E}:y^2=x\,(x-1)\,(x-z)\,.
   \label{ellipticcurveE}
\end{equation}
One may choose a symplectic basis such as the usual $\mathcal{A}$-cycle and $\mathcal{B}$-cycle 
\begin{equation}\label{cycles}
    S_a^1,S_b^1\in H_1(\mathcal{E},\mathbb{Z})\,,
\end{equation}
with $S_a^1\cap S_b^1=-S_b^1\cap S_a^1$ and $S_a^1\cap S_a^1=S_b^1\cap S_b^1=0$ in integral homology. 
Then 
\begin{subequations}
\begin{align}
    \Pi_{11}(z)&=\int_{S_a^1}\dfrac{\text{d}x}{y}=2\int_0^1 \dfrac{\text{d}x}{y}\,,\\ 
    \Pi_{12}(z)&=\int_{S_b^1}\dfrac{\text{d}x}{y}=2\int_1^z\dfrac{\text{d}x}{y}\,,
\end{align}
\end{subequations}
can be evaluated in terms of complete elliptic integral of first kind $K(k)$ and $K(1-k)$ for $1\leq z\leq \infty$ through:
\begin{equation}\label{pi_to_K}
    \begin{pmatrix}\Pi_{11}(z)\\\Pi_{12}(z)\end{pmatrix}=4\begin{pmatrix}1&i\\0&-i\end{pmatrix}\cdot\begin{pmatrix}K(z)\\K(1-z)\end{pmatrix},
\end{equation}
with the definition:
\begin{align}\label{ell_1}
    K(k)&=\int_0^1\frac{dt}{\sqrt{(1-t^2)(1-k^2t^2)}}\,.
\end{align}
Introducing $\tau=\Pi_{12}(z)/\Pi_{11}(z)$ and using it as the complex modulus, we can express the periods of the elliptic curve as modular forms of weight one in $\tau$.
Hence, we expect at least two master integrals in the maximal cut of a Feynman integral associated with an elliptic curve. This also matches with the considerations from section~\ref{sec:2.2}.
\\
An important property of Calabi-Yau manifolds is that when varying around a point $z_0$, the holomorphic one-form $\Omega(z_0+\delta)$ gets admixtures of forms of other types. 
We leave a more comprehensive discussion
to the literature, for example ref.~\cite{Voisin_2002} and focus on the consequences that matter to us. 
The main idea we will need is the \textit{Griffiths transversality} property of the derivative:
\begin{equation}\label{griffiths_transversality}
    \partial_{\bf z}^{\bf k} \Omega({\bf z})\in \mathcal{H}^{l,0}\oplus \mathcal{H}^{l-1,1} \oplus \cdot\cdot\cdot \oplus \mathcal{H}^{l-|k|,|k|}\,,
\end{equation}
where $\partial_{\bf z}^{\bf k}:=\partial_{z_1}^{k_1}...\partial_{z_r}^{k_r}$ and $|k|=\sum_{i=1}^{h^{n-1,1}}k_i$ and {\bf z} controls the moduli space of the Calabi-Yau manifold $M_l$. 
Note that just by taking derivatives of $\Omega$ one does not obtain the form $\bar{\Omega}$ we expect from the Hodge decomposition in Eq.~(\ref{Hodge_dec}).
Through \textit{Hypercohomology theory} one can show that the cohomology spanned by the meromorphic differentials obtained by differentiating (as in Eq.~(\ref{griffiths_transversality})) is cohomologically equivalent to the Hodge cohomology~\cite{Griffiths_2, Griffiths_1}, this motivates the introduction of the Hodge bundles $\mathcal{H}^{p,q}$.
\\
Hence, for the Calabi-Yau manifolds that have appeared in the Feynman integrals literature, to span the middle cohomology $\mathcal{H}^l(M_l,\mathbb{C})$ one can just take a basis of derivatives of the holomorphic $l$-form $\Omega$.\footnote{Note that this is not always the case: for Calabi-Yau $l$-folds with $l\geq 4$, taking derivatives of the holomorphic form is not always enough to span the full middle cohomology, i.e., manifolds with Hodge numbers~${(1,\,1,\,h,\,1,\,1)}$ with $h>1$. The examples we consider are at most Calabi-Yau three-folds so we do not encounter this problem.} 
Note that all differentials but $\Omega$ are meromorphic.
This will prove very useful when looking for a good starting basis for Feynman integrals with Calabi-Yau geometry, as the $W_l$ part of the weight filtration coincides with $H^l(\mathrm{CY},\mathbb{Q})$.
This implies that the maximal cut will contain at least $n=\mathrm{dim}(H^l(\mathrm{CY},\mathbb{Q}))$ master integrals, that correspond to the holomorphic period and its independent derivatives for $\epsilon\to 0$.
As an example, let us return to the elliptic curves from Eq.~(\ref{ellipticcurveE}).
The derivative with respect to $z$ of $\Omega=\mathrm{d}x/y$, can be written, up to exact terms, as a linear combination of $\mathrm{d}x/y$ and $x\,\mathrm{d}x/y$, where  the latter is a differential of second kind.
Pairing this differential of second kind with the cycles (\ref{cycles}) will yield the complete elliptic integrals of second kind $E(k^2)$ and $E(1-k^2)$ for $1\leq z \leq \infty$ through:
\begin{equation}\label{pi_to_K_2}
    \begin{pmatrix}
        \Pi_{21}(z)\\\Pi_{22}(z)
    \end{pmatrix}=4\begin{pmatrix}-1&i\\0&-i\end{pmatrix}\cdot \begin{pmatrix}
        E(z)-K(z)\\E(1-z)
    \end{pmatrix}\,,
\end{equation}
with the definition:
\begin{equation}\label{ell_2}
E(k)=\int_0^1\dfrac{\text{d}t(1-k^2 t^2)}{\sqrt{(1-t^2)(1-k^2t^2)}}\,.
\end{equation}
Equation (\ref{griffiths_transversality}) also implies that there are relations between derivatives. 
These relations form the Picard-Fuchs differential ideal $L_0^{(i)}$ that annihilate the periods: $L_0^{(i)}\Pi_j(\mathbf{z})=0$. 
Returning to our elliptic example $(\ref{ellipticcurveE})$, we have $\Omega(z)\in \mathcal{H}^{(1,0)}$, $\partial_z\Omega(z)\in  \mathcal{H}^{(0,1)}\oplus \mathcal{H}^{(1,0)}$, which therefore implies:
\begin{equation}
    \partial_z^2\Omega(z)\in \mathcal{H}^{(1,0)}\oplus \mathcal{H}^{(0,1)}\,,
\end{equation}
which means that $\partial_z^2\Omega(z)$ can be written as a linear combination of $\Omega(z)$ and $\partial_z\Omega(z)$. Hence the same is true for the periods: $\partial_z^2\Pi_{1j}(z)$ can be written as a linear combination of $\Pi_{1j}(z)$ and $\partial_z\Pi_{1j}(z)$. 
Explicitly, we find: 
\begin{align}
    \dfrac{\partial^2}{\partial z^2}\Pi_{11}(z)&=\dfrac{1-2z}{(z-1)z}\dfrac{\partial}{\partial z}\Pi_{11}(z)+\dfrac{1}{4z(1-z)}\Pi_{11}(z)\,,
\end{align}
which can be derived by using the Griﬃths-Dwork reduction~\cite{griffiths1969periods,griffiths1966residue1,griffiths1966residue2,dwork1962zeta,dwork1964zeta2} for the elliptic integrals~$K(k^2)$ and $K(1-k^2)$ and use the translations from Eqs.~(\ref{pi_to_K}) and (\ref{pi_to_K_2}). 
\paragraph{Quadratic relations:} 
Periods (which correspond to maximal cuts) undergo quadratic relations~\cite{Cho_Matsumoto_1995, Duhr:2024rxe}. In particular, for periods of Calabi-Yau manifolds we find:
\begin{equation}
    \mathbf{\Pi}(\mathbf{z})^T\mathbf{\Sigma}\partial_{\bf z}^{\bf k}\mathbf{\Pi}(\mathbf{z})=\int_{M_l}\Omega\,\wedge\,\partial_{\bf z}^{\bf k}\Omega(\mathbf{z})=\begin{cases} 
0 & \text{for } 0 \leq |k| < l, \\
C_{\mathbf{k}}(\mathbf{z}) & \text{for } |k| = l,
\end{cases}
\label{quadrel}
\end{equation}
where $\mathbf{\Sigma}$ is called the {\it intersection matrix} and $C_{\mathbf{k}}(\mathbf{z})$ are rational functions in $\mathbf{z}$ sometimes called the $l$-point {\it Yukawa couplings}. 
This follows from Eq.~(\ref{griffiths_transversality}) as to get a non vanishing integral over $M_l$, an $\omega^{l,l}$-form is needed. 
But since $\Omega\in \mathcal{H}^{l,0}$, we need  $\partial_{\bf z}^{\bf k}\mathbf{\Pi}(\mathbf{z})\in \mathcal{H}^{0,l}\oplus \dots$\,, which is only realised for $|k|=l$.
These relations will be useful in simplifying the differential equation matrix. 
\paragraph{Frobenius algebra:}
A \textit{Frobenius algebra} is a graded vector space $\mathcal{A}=\oplus\mathcal{A}^{(i)}, \, i\geq 0$ with a symmetric non-degenerate bilinear form $\eta$ and a cubic form (also called three-point Yukawa coupling),
\begin{equation}
    C^{(i,j,k)}:\mathcal{A}^{(i)}\otimes\mathcal{A}^{(j)}\otimes\mathcal{A}^{(k)}\rightarrow\mathbb{C}\,,
\end{equation}
where the upper indices are such that 
\begin{equation}
    i+j+k=l\,,
    \label{conditionC}
\end{equation}
with $l$ the dimension of the manifold\footnote{Our examples have manifolds with dimension of at most three ($i=j=k=1$), this is why we drop the upper indices, except in paragraph~\ref{higherdimCY} where they are needed.}.
Among the defining properties of the three-point coupling is the symmetry under any permutation of the indices: $C_{a,b,c}^{(i,j,k)}=C_{\sigma(a,b,c)}^{\sigma(i,j,k)}$.
\\
The Frobenius structure is determined by the Picard-Fuchs ideal combined with Griffiths transversality in Eq.~(\ref{griffiths_transversality}), which means that these are the only needed inputs to compute the three-point coupling.
These are the couplings that will later appear in the generic differential equation (\ref{genericGM}) and  their computation is necessary to write a  basis for master integrals that satisfy an $\epsilon$-factorised differential equation.
Since our examples have an underlying geometry being a $l$-dimensional Calabi-Yau manifold (with $l=1,2,3$), we explicitly compute them for these dimensions.
Note that for $l=1, 2$ the couplings do not appear in the differential equation with respect to the moduli $\tau_a$.
For higher dimensional examples, we refer to ref.~\cite{Ducker:2025wfl}. 
\paragraph{Yukawa couplings for a Calabi-Yau three-fold:}
\label{sec:Yukawacouplings3}
Let us explicitly show how this structure arises for a Calabi-Yau three-fold with hodge numbers:
\begin{equation}
    h^{(3,0)}=h^{(0,3)}=1\,,\quad  h^{(2,1)}=h^{(1,2)}=\hat{h}\,,
\end{equation}
which implies that we can find $2\hat{h}+2$ independent periods
\begin{equation}\label{start_per}
    \underline{\psi}(\mathbf{z})=(\psi_0,\psi_1^{(1)},\dots,\psi_1^{(\hat{h})},\psi_2^{(1)},\dots,\psi_2^{(\hat{h})},\psi_3)\,,
\end{equation}
where the subscript determines the logarithmic power of the period in the Frobenius basis.
These periods have been chosen by requiring the intersection matrix $\Sigma$ to be antidiagonal\footnote{Where here, the identity matrices here should be understood as antidiagonal, e.g. $\mathds{1}_{2\times 2}=\begin{pmatrix}0&1\\1&0\end{pmatrix}\,$.} 
\begin{equation}
    \mathbf{\Sigma}=\begin{pmatrix}
    0&\mathds{1}_{(\hat{h}+1)\times(\hat{h}+1)}\\-\mathds{1}_{(\hat{h}+1)\times(\hat{h}+1)}&0
\end{pmatrix}\,.
\label{sigmaCY3}
\end{equation}
After defining the complex structure moduli as usual:
\begin{equation}
    \tau_a=\dfrac{\psi_1^{(a)}(\mathbf{z})}{\psi_0(\mathbf{z})}\,,\quad a=1,...,\hat{h}\,,
\end{equation}
we can use the Griffiths transversality conditions (\ref{quadrel}) to define the Yukawa three-couplings:
\begin{align}
\label{C_def_1}
    \dfrac{\underline{\psi}(\mathbf{z})}{\psi_0(\mathbf{z})}\cdot\mathbf{\Sigma}\cdot \dfrac{\partial}{\partial \tau_{a}}\dfrac{\partial}{\partial \tau_{b}}\dfrac{\partial}{\partial \tau_{c}}\dfrac{\underline{\psi}(\mathbf{z})}{\psi_0(\mathbf{z})}=C_{a,b,c}(\mathbf{z})\,.
\end{align}
Let us now connect with the notation commonly used in the string theory literature (e.g. section 2.5.5 of ref.~\cite{Clader:2018yyu}) by defining the pre-potential $\mathcal{F}$:
\begin{equation}
    \mathcal{F}=\dfrac{1}{2}\biggl(\dfrac{\psi_3}{\psi_0}+\tau_a \dfrac{\psi_2^{(a)}}{\psi_0}\biggr)\,,
    \label{prepotential}
\end{equation}
which we can connect through Eq.~(\ref{quadrel}) to the periods in Eq.~(\ref{start_per}) by:
\begin{equation}
    \dfrac{\partial}{\partial\tau_a}\mathcal{F}=\dfrac{\psi_2^{(a)}}{\psi_0}\,.
\end{equation}
It also follows that the Yukawa three-couplings can be expressed as
\begin{equation}
    C_{a,b,c}=\dfrac{\partial}{\partial\tau_a}\dfrac{\partial}{\partial\tau_b}\dfrac{\partial}{\partial\tau_c}\mathcal{F}\,.
\end{equation}
With this new notation, let us now write a basis for the cohomology:
\begin{subequations}
\begin{align}
    \hat{\Omega}_0&=\alpha
    _0+\tau^a\,\alpha_a-\dfrac{\partial}{\partial\tau^a}\mathcal{F}\,\beta^a-\biggl(2\mathcal{F}-\tau^a\dfrac{\partial}{\partial\tau^a}\mathcal{F}\biggr)\,\beta^0\,,\\
    \hat{\xi}_a&=\alpha_a-\dfrac{\partial}{\partial \tau^a}\dfrac{\partial}{\partial\tau^b}\mathcal{F}\,\beta^b-\biggl(\dfrac{\partial}{\partial\tau^a}\mathcal{F}-\tau^b\dfrac{\partial}{\partial\tau^a}\dfrac{\partial}{\partial\tau^b}\mathcal{F}\biggr)\,\beta^0\,,\\
    \hat{\xi}^a&=-\beta^a+\tau^a\,\beta^0\,,\\
    \hat{\Omega}^0&=\beta^0\,,
\end{align}
\end{subequations}
where $\alpha_I$ and $\beta^I$, with $I=0,...,\hat{h}$, span a symplectic basis for the cohomology. Notice that this basis is built by starting with the sum of the periods (\ref{start_per}) normalised by the holomorphic period $\psi_0$:
\begin{equation}
    \hat{\Omega}_0=\dfrac{1}{\psi_0}\biggl(\psi_0 \,\alpha_0+\sum_{a=1}^{\hat{h}}\psi_1^{(a)}\,\alpha_a+\sum_{a=1}^{\hat{h}}\psi_2^{(a)}\,\beta^a+\psi_3\,\beta^0\biggr)\,.
\end{equation}
The other elements $\hat{\xi}_a$, $\hat{\xi}^a$ and $\hat{\Omega}^0$ complete $\hat{\Omega}_0$ to a basis in $H^3(M,\mathbb{C})$ and are chosen such that the following differential equation holds:
\begin{equation}\label{frob_ST}
\dfrac{\partial}{\partial \tau^a}\begin{pmatrix}\hat{\Omega}_0\\\hat{\xi}_b\\\hat{\xi}^b\\\hat{\Omega}^0\end{pmatrix}=\begin{pmatrix}0&\delta_{a}^c&0&0\\0&0&C_{a,b,c}&0\\0&0&0&\delta_a^b\\0&0&0&0\end{pmatrix}\cdot \begin{pmatrix}\hat{\Omega}_0\\\hat{\xi}_c\\\hat{\xi}^c\\\hat{\Omega}^0\end{pmatrix}.
\end{equation}
This structure will be of relevance to us in section \ref{genericansatzCY3fold}, when setting up the basis of MIs for an integral with  underlying Calabi-Yau three-fold geometry and in section \ref{sec:banana_4} when computing an explicit example of this type.
 
\subsection{Abelian differentials of first, second and third kind}\label{sec:3.2}
Here we review abelian differentials of first, second and third kind. 
In refs.~\cite{Gorges:2023zgv,Duhr:2024uid,Duhr:2025lbz} it has been pointed out that the master integrals of an elliptic sector can be interpreted as elliptic integrals of the first, second and third kind.
We will use this understanding to interpret the {\it decoupling} master integrals we refer to in section~\ref{sec:4}.
\\Consider a compact Riemann surface $X$ as a complex manifold of dimension one. 
On $X$ we can have zero-forms (scalar functions $f(z,\bar{z})$), one-forms (\textit{differentials}), which locally look like
\begin{equation}
    \omega=\omega_z(z,\bar{z}) \mathrm{d}z+\omega_{\bar{z}}(z,\bar{z}) \mathrm{d}\bar{z}\,,
\end{equation}
and two-forms given by their wedge product. 
For applications to Feynman integrals we are mostly interested in closed one-forms ($\text{d}\omega$=0), as the $\epsilon$-factorised differential equation matrix we want to construct will be a matrix of closed one-forms. 
\\ A differential $\eta$ is called \textit{holomorphic} (\textit{meromorphic}) if, in a local coordinate $z$, it has the form $h(z)\,\mathrm{d}z$ where $h(z)$ is an {\it holomorphic} ({\it meromorphic}) function. 
\\From the Riemann-Roch theorem we know that the dimension of the vector space of holomorphic differentials on $X$ is equal to the genus. 
So, for integrals with an underlying elliptic geometry we have only one holomorphic differential.
\\We can now organise these differentials into three types:
\begin{itemize}
    \item Holomorphic differentials are called Abelian differentials of the \textit{first kind}.
    \item Meromorphic differentials with vanishing residues are called Abelian differentials of the \textit{second kind}.
    \item Meromorphic differentials with non-vanishing residues are called Abelian differentials of the \textit{third kind}.
\end{itemize}
\paragraph{Example: an elliptic curve:}\label{ExEllipticCurve}
Let us come back to our example of the Legendre curve:
\begin{equation}
    \mathcal{E}:\quad y^2=x(x-1)(x-z)\,,
\end{equation}
where we pick the root ordering $1<z<\infty$.
Topologically this corresponds to two Riemann spheres connected by two branch cuts, which we can choose to go from $0$ to $1$ and from $z$ to~$\infty$. 
If we allow differentials of the third kind, with non-vanishing residue at a marked point~$c$, this corresponds to a puncture on each of the Riemann spheres at the marked point~$c$.
 Explicitly one gets the following forms of first, second and third kind:
\begin{align}
   \left\{\begin{array}{ll} \dfrac{\text{d}x}{y}\,\quad &\text{: differential of the first kind},\vspace{2mm}\\
    \dfrac{x\text{d}x}{y}\,\quad &\text{: differential of the second kind},\vspace{2mm}\\ \dfrac{\text{d}x}{y(x-c)}\,\quad &\text{: differential of the third kind,}\end{array}\right.
\end{align}
for $c\neq \{0,1,z,\infty\}$.
With these three differential forms we expect there to be three cycles. 
Two of these cycles correspond to the standard cycles around and across the branch cuts from Eq.~(\ref{cycles}) which are identified with the $\mathcal{A}$- and $\mathcal{B}$-cycles of the torus.
The remaining cycle can be taken to be a contour around the puncture at $x=c$.  
Pairing the differentials of the first and second kind with the cycle $S_a^1$ from Eq.~(\ref{cycles}) will lead to the complete elliptic integrals of the first and second kind $K(k)$ and $E(k)$ according to Eqs.~(\ref{pi_to_K}) and (\ref{pi_to_K_2}), respectively.
If we pair this cycle to differential of the third kind, we will find a complete elliptic integrals of the third kind:
\begin{equation}
    \Pi(a,k)=\int_0^1\dfrac{\text{d}t}{\sqrt{(1-t^2)(1-k^2t^2)}}\dfrac{1}{(1-a t^2)}\,,
\end{equation}
where $a$ encodes the singularity $c$. The marked point $c$ on the elliptic curve $\mathcal{E}$ gets mapped to a point on the torus via {\it Abel's map} $\mathcal{A}$~\cite{Broedel:2017kkb}:
\begin{equation}
   \mathcal{A}(z,c)= 2\int_{0}^c \dfrac{\text{d}x}{y}=4\, F\biggl(\arcsin{\sqrt{\frac{c}{z}}},z\biggr)\,,
   \label{Abelsmap}
\end{equation} 
where the second equality holds for $0\leq\arcsin\sqrt{\frac{c}{z}}\leq \frac{\pi}{2}$ and  we introduce the incomplete integral of first kind:
\begin{align}\label{ell_inc}
    F(\phi,k)&=\int_0^{\sin(\phi)}\frac{\mathrm{d}t}{\sqrt{(1-t^2)(1-k^2t^2)}}\,.
\end{align}
\\Having now defined complete integrals of the first, second and third kind, we can show that they follow a coupled differential equation: 
\begin{equation}\label{decoupling_ex}
    \dfrac{\partial}{\partial k}\begin{pmatrix}K(k)\\E(k)\\\Pi(a,k)\end{pmatrix}=\begin{pmatrix}-\frac{1}{2k}&-\frac{1}{2(k-1)}+\frac{1}{2k}&0\\
    -\frac{1}{2k}&\frac{1}{2k}&0\\
    0&\frac{1}{2(a-1)(k-1)}-\frac{1}{2(a-1)(k-a)}&-\frac{1}{2(k-a)}\end{pmatrix}\cdot\begin{pmatrix}K(k)\\E(k)\\\Pi(a,k)\end{pmatrix},
\end{equation}
where we observe a $2\times 2$ block that couples the elliptic integrals of first and second kind.
The integral of third kind has a homogeneous term and additionally couples back to integrals of the first and second kind.

\paragraph{Extension to higher-dimensional Calabi-Yau manifolds}
It is possible to extend the definition of the differentials of the first and second kind to manifolds beyond Riemann surfaces, however, it goes beyond the scope of this article, so we leave this discussion to the literature~\cite{Griffiths_2}. 
Let us mention that the generalisation of the third kind integrals to higher dimensional geometries is not very well understood and it is topic of on-going research~\cite{alfesneumann2022harmonicweakmaassforms}. 
So, to find an appropriate analogue of integrals of the third kind but in a Calabi-Yau block we will rather use Griffiths transversality from Eq.~(\ref{griffiths_transversality}) and properties of the differential operators, as we will explain later in section~\ref{sec:4}. 

\subsection{\texorpdfstring{$\epsilon$}{epsilon}-factorising L-loop banana integrals of equal mass}
\label{sec:3.1}
We end this review section by summarising the results of ref.~\cite{Pogel:2022vat}. Note that, for this case, it has been shown in ref.~\cite{Duhr:2025lbz} that the methods from ref.~\cite{Pogel:2022vat} and ref.~\cite{Gorges:2023zgv} are equivalent. This allows us to lay the ground for generalising to multi-variate Calabi-Yau Feynman integrals.
The integrals studied in ref.~\cite{Pogel:2022vat} are the so-called banana integrals, which are defined by
\begin{equation}
    I_{\nu_1,\dots,\nu_\text{L},\nu_{\text{L+1}}}(p^2,m^2;\epsilon)=e^{\text{L}\gamma_E \epsilon}\int\prod_{j=1}^\text{L}\left(\dfrac{\text{d}^Dl_j}{i\pi^{D/2}}\frac{1}{(l^2_j+m^2)^{\nu_j}}\right)\frac{1}{((\sum_{i=1}^L l_i-p)^2-m^2)^{\nu_{\mathrm{L}+1}}}\,,
\end{equation}
where $D$ denotes the number of space-time dimensions, $\epsilon$ the dimensional regularisation parameter, $\gamma_E$ the Euler-Mascheroni constant and $\nu=\sum_{j=1}^{\text{L}+1}\nu_j$. We consider these integrals in the natural dimension for two-point functions, which is $D=2-2\epsilon$ and as kinematical variable we choose $x=-\frac{m^2}{p^2}$.
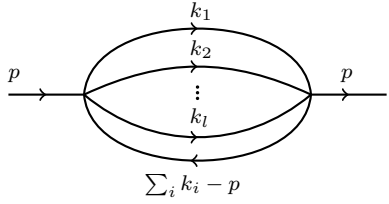
\begin{figure}[!h]
\centering
\begin{tikzpicture}
\coordinate (llinks) at (-2.5,0);
\coordinate (rrechts) at (2.5,0);
\coordinate (links) at (-1.5,0);
\coordinate (rechts) at (1.5,0);
\begin{scope}[very thick,decoration={
    markings,
    mark=at position 0.5 with {\arrow{>}}}
    ] 
\draw [-, thick,postaction={decorate}] (links) to [bend right=40]  (rechts);
\draw [-, thick,postaction={decorate}] (links) to [bend right=-25]  (rechts);

\draw [-, thick,postaction={decorate}] (links) to [bend left=85]  (rechts);
\draw [-, thick,postaction={decorate}] (llinks) to [bend right=0]  (links);
\draw [-, thick,postaction={decorate}] (rechts) to [bend right=0]  (rrechts);
\end{scope}
\begin{scope}[very thick,decoration={
    markings,
    mark=at position 0.5 with {\arrow{<}}}
    ]
\draw [-, thick,postaction={decorate}] (links) to  [bend right=85] (rechts);
\end{scope}
\node (d1) at (0,1.1) [font=\scriptsize, text width=.2 cm]{$k_1$};
\node (d2) at (0,0.6) [font=\scriptsize, text width=.2 cm]{$k_2$};
\draw[fill=black!100] (0,0.025)  circle (0.25pt);
\draw[fill=black!100] (0,0.1)  circle (0.25pt);
\draw[fill=black!100] (0,-0.05)  circle (0.25pt);
\node (d3) at (0,-0.3) [font=\scriptsize, text width=.2 cm]{$k_l$};
\node (d4) at (0.45,-1.2) [font=\scriptsize, text width=2.3 cm]{$\sum_i k_i-p$};
\node (p1) at (-2.0,.25) [font=\scriptsize, text width=1 cm]{$p$};
\node (p2) at (2.4,.25) [font=\scriptsize, text width=1 cm]{$p$};
\end{tikzpicture}
\caption{The $l$-loop banana graph.}
\label{fig:pic_ban}
\end{figure}
As we have reviewed in section \ref{sec:2.4}, a good basis for master integrals in $D=2$ is the derivative basis, that for a $\text{L}-$loop Banana integral takes the form:
\begin{equation}
{\bf{I}_{MC}}=\begin{pmatrix}
    I_{1,1,\dots,1,1}\\
    \partial_xI_{1,1,\dots,1,1}\\
    \partial^2_xI_{1,1,\dots,1,1}\\
    \vdots\\
    \partial^{\text{L}-1}_xI_{1,1,\dots,1,1}
    \end{pmatrix}.
\end{equation}
However, an IBP analysis in $D=2-2\epsilon$ reveals that another master integral is needed, a tadpole $I_{1,1,\dots,1,0}$, that vanishes on the maximal cut.
Defining 
\begin{equation}
    \mathbf{I}=(I_{1,1,\dots,1,0},\,\mathbf{I}_{\mathrm{MC}})^T\,,
    \label{mi_start}
\end{equation} 
we can now set up a differential equation
\begin{equation}\label{deq_start_ban}
\text{d}{\bf{I}}={\bf{A}}(\epsilon,x)\cdot{\bf{I}}\,,
\end{equation}
which is $\epsilon$-factorised after having found the appropriate gauge transformation $\bf U$ such that $\mathbf{M}=\mathbf{U}\cdot\mathbf{I}$ with
\begin{equation}\label{eps_fact_ban}
    \text{d}{\bf{M}}=\epsilon\,\tilde{\bf{A}}(x)\cdot{\bf{M}}.
\end{equation}
For $\text{L}\geq2$, the gauge transformation ${\bf U}$ will introduce periods of a Calabi-Yau $(\text{L}-1)$-fold into~$\tilde{\bf A}$.
 \paragraph{Extracting the geometry from the differential equation:}
 From Eq.~(\ref{deq_start_ban}) we can extract the following inhomogeneous differential equation:
 \begin{equation}\label{deq_ban}
     L^{(\text{L})}I_{1,\dots,1,1}=(-1)^\text{L}\dfrac{(\text{L}+1)!}{x^{\text{L}-1}\prod_{a\in S^{(\text{L})} }(1+a x)}\epsilon^\text{L}I_{1,\dots,1,0}\,,
 \end{equation}
 with $L^{(\text{L})}$ a PF operator of degree L and
 \begin{equation}
 S^{(\text{L})}=\left\{\begin{array}{ll}
 \left\{(2k)^2|k\in\{1,\dots,(\text{L}+1)/2\}\right\},&\text{L}\text{ odd}\,,\\
 \left\{(2k-1)^2|k\in\{1,\dots,(\text{L}+2)/2\}\right\},&\text{L}\text{ even}\,.\end{array}\right.
 \end{equation}
 To find the type of transcendental functions needed to describe the integral, we consider the maximal cut of the integral at $\epsilon=0$, since it solves the homogeneous differential equations.
 The differential equation for the maximal cut corresponds to the homogeneous part of Eq.~(\ref{deq_ban}) at $\epsilon=0$:
 \begin{equation}
     L^{(\text{L})}_0I_{1,\dots,1,1}=0\,,
 \end{equation}
 where we defined $L^{(\text{L})}_0=\lim_{\epsilon\rightarrow 0}L^{(\text{L})}$ as in paragraph~\ref{sec:2.3}. It is possible to show that $L^{(\text{L})}_0$ is a Calabi-Yau differential operator (see paragraph \ref{CYoperators}), hence, the $\text{L}$ solutions $\psi_k$ 
 \begin{equation}\label{periods1}
     L^{(\text{L})}_0\psi_k=0\,,\quad 0\leq k\leq \text{L}-1,
 \end{equation}
 are {\it periods} of a Calabi-Yau $(\text{L}-1)$-fold as in Eq.~(\ref{periodsCY}).

\paragraph{Making an ansatz to $\epsilon$-factorise the differential equation:} 
 
In ref.~\cite{Pogel:2022vat} it was shown that the $\epsilon$-factorising gauge transformation for the $\text{L}$-loop equal mass banana integrals is captured by a set of master integrals ${\bf{M}}={\bf U}\cdot{\bf{I}}$  of the following form:
\begin{align}\label{Bananas_ansatz}
M_0&=\dfrac{\epsilon^\text{L}}{\psi_0}I_{1,\dots,1}\,,\notag\\
    M_k&=\dfrac{1}{Y_{k}}\left(\dfrac{1}{\epsilon}\dfrac{\partial}{\partial \tau} M_{k-1}-\sum_{j=0}^{k-1}F_{(k-1),j}M_j\right)\,,\quad 1\leq k\leq \text{L}-1\,,
\end{align}
where the functions $F_{i,j}$ are independent of $\epsilon$ and can be determined by a set of differential equations (c.f. section 4.1 of ref.~\cite{Pogel:2022vat}).
When all $F_{i,j}$ are set to zero, the diagonal entries of the connection are already in $\epsilon$-form and the only entries which are not yet $\epsilon$-factorised are on the strictly lower triangular part of the connection.
\paragraph{Comparison with the method of refs.~\cite{Gorges:2023zgv,Duhr:2025lbz}:}\label{comparison_section}
As explained in ref.~\cite{Duhr:2025lbz}, the two methods of refs.~\cite{Pogel:2022vat,Gorges:2023zgv,Duhr:2025lbz} in this case are equivalent, let us briefly recapitulate why. Let us look at the ansatz in Eq.~(\ref{Bananas_ansatz}) before the $\epsilon$-rescaling and adding the auxiliary functions $F_{ij}$:
\begin{align}\label{Bananas_ansatz_eps0}
M_0&=\dfrac{1}{\psi_0}I_{1,\dots,1}\,,\notag\\
    M_k&=\dfrac{1}{Y_{k}}\dfrac{\partial}{\partial \tau} M_{k-1}\,,\quad 1\leq k\leq \text{L}-1\,.
\end{align}
The period matrix (or Wronskian) $W_L$ of the maximal cut at $\epsilon=0$ of this system is given by
\begin{equation}
W_L=
    \begin{pmatrix}
      1&\tau&\frac{\psi_2}{\psi_0}&\hdots&\frac{\psi_{L-1}}{\psi_0} \\
      0&1&\partial_{\tau}\left(\frac{\psi_2}{\psi_0}\right)&\hdots&\partial_{\tau}\left(\frac{\psi_{L-1}}{\psi_0}\right) \\
      0&0&\frac{1}{Y_2}\partial^2_{\tau}\left(\frac{\psi_2}{\psi_0}\right)&\hdots&\frac{1}{Y_2}\partial^2_{\tau}\left(\frac{\psi_{L-1}}{\psi_0}\right) \\
      \vdots&\vdots&\vdots&\vdots&\vdots\\
      0&0&\partial_{\tau}\frac{1}{Y_2}\cdot\cdot\cdot\frac{1}{Y_2}\partial^2_{\tau}\left(\frac{\psi_2}{\psi_0}\right)&\hdots&\partial_{\tau}\frac{1}{Y_2}\cdot\cdot\cdot\frac{1}{Y_2}\partial^2_{\tau}\left(\frac{\psi_{L-1}}{\psi_0}\right)
      
    \end{pmatrix}.
\end{equation}
By direct computation it is possible to check that $W_L$ satisfies the following \textit{unipotent} differential equation:
\begin{equation}
    \partial_{\tau}\,W_L(\tau)=N_L(\tau)\,W_L(\tau)\,,
\end{equation}
where $N_L$ is the \textit{nilpotent} matrix:
\begin{equation}
    \begin{pmatrix}
        0 & 1 & 0 &\ldots & & &  \\
        & 0 & Y_2(\tau) & 0 &\ldots & &  \\
        & \ldots& 0 & Y_3(\tau) & 0 &\ldots &  \\
        & & \ldots& 0 & \ddots & 0 &\ldots  \\
        & & & \ldots& 0 & Y_2(\tau) & 0  \\
        & & & & \ldots& 0 & 1  \\
        & & & & & \ldots& 0  \\
    \end{pmatrix} \, .
\end{equation}
Hence we observe that, taking the derivative with respect to $\tau$ corrected by the appropriate $Y_i$ prefactors, effectively corresponds to multiply the differential equation matrix by the inverse of the \textit{semi-simple} part of the Wronskian, which is the first rotation from the method of refs.~\cite{Gorges:2023zgv,Duhr:2025lbz}. 
The second rotation amounts to fixing the $\epsilon$ powers: in refs.~\cite{Gorges:2023zgv,Duhr:2025lbz}, each integral that is acted on by $k$ derivatives gets rescaled by $\frac{1}{\epsilon^k}$.
This corresponds to the same~$\epsilon$-rescaling as in Eq.~(\ref{Bananas_ansatz}).
The final rotation introduces additional functions, that are defined such that the final differential equation is in $\epsilon$-form. 
This corresponds to the functions $F_{i,j}$ in the ansatz from Eq.~(\ref{Bananas_ansatz}), showing that the two methods are equivalent.

\section{Algorithm to construct a canonical basis}\label{sec:4}
Now that we have reviewed the methods from refs.~\cite{Gorges:2023zgv,Duhr:2025lbz,Pogel:2022vat}, commented on the Griffiths transversality conditions in section~\ref{sec:2.4}, and discussed the structure of differentials of the first, second and third kind in section~\ref{sec:3.2}, we are ready to propose our generalisation to several scales.
To this end, we will be focusing on diagonal blocks of the differential equation matrix, where a single geometry is identified.
This proposed basis will be aimed at Feynman integrals with underlying multi-variate Calabi-Yau geometry, for which we will also be presenting some examples in sections \ref{sec:sunrise} and \ref{Computing new integrals}. 
To show that we are in fact doing the same transformations as in refs.~\cite{Gorges:2023zgv,Duhr:2025lbz}, we will use the results presented in section 3 of ref.~\cite{Ducker:2025wfl}, where the authors have explicitly computed the splitting of the period matrix in semi-simple and unipotent part from periods defined on elliptic curves to Calabi-Yau four-folds.

\paragraph{Setting up the problem:}
First of all, we need to identify the geometry of the block we are looking at. 
This can either be done with a Baikov analysis, or finding the periods annihilated by the Picard-Fuchs operator. 
For cases like the banana integrals in two dimensions, the corner integral ($I_{1,\dots,1}$) evaluates to periods of a Calabi-Yau manifold on its maximal cut in $D=2$ dimensions.
This is, however, not always the case and one may have to shift dimensions or take linear combinations of the available master integrals to make the underlying geometry manifest.
\\Once we have found the polynomial $y^2=P_n(z_1,...,z_n)$ defining the geometry, we then call $I_1$ the integral whose maximal cut is a integral of a holomorphic form in a judicious choice of integer dimension $D_0$:\footnote{For higher genus curves we have $g$ holomorphic differential forms. In the Calabi-Yau case there is only one holomorphic differential.}
\begin{equation}
    \mathrm{MaxCut}\,I_1\biggr|_{D=D_0}=\int\frac{\mathrm{d}z_1...\mathrm{d}z_n}{y}\,.
\end{equation}
From the (hyper)cohomology arguments  reviewed in section \ref{sec:2.4} (with the highlighted restrictions), we also know how many derivatives to take to have a basis that spans the middle cohomology in $D=D_0$. 
In summary, the first step amounts to identify the master integrals which can be built from taking derivatives.
For example, in Eq.~(\ref{mi_start}) we identify all except the first master integral (which corresponds to a separate sector), as the first and second kind differentials, obtained by taking derivatives of a single integral.
\\However, when the number of master integrals is greater than the dimension of the cohomology, we need to interpret the extra master integrals differently from the first or second kind integral.
If no separate geometry in the sector is identified, we look for integrals that decouple from the rest of the sector at $\epsilon=0$ (as the integrals of the third kind in Eq.~(\ref{decoupling_ex})). 
We will refer to these as \textit{decoupling} master integrals.
At $\epsilon=0$, they can be treated as a block decoupled from the master integrals built from derivatives. 
In particular, we want their differential equation to read
\begin{equation}
    \dfrac{\partial}{\partial{z_i}}I^{\mathrm{dec}}_{j}=\biggl(q^{(0)}_{(z_i)}\,I_1+\sum_{k=1}^{\text{dim}(\mathbf{z})}q^{(0)}_{k\,(z_i)}\,\dfrac{\partial}{\partial{z_k}}I_1+\epsilon\,\bigl(q^{(1)}_{(z_i)}\,I_1+q^{(1)\,\text{dec}}_{(z_i)}\,I^{\mathrm{dec}}_{j}\bigr)\biggr)\,,
    \label{diffeqdecoupling}
\end{equation}
where $z_i$ are variables from which we can write a Calabi-Yau PF operator for $I_1$ which has a MUM-point at $z_i\rightarrow 0$.\footnote{For elliptics we have only one $z_i$ because there is only one canonical variable $\tau$.} 
Note in particular that there are no derivative terms $\partial_{z_i}^k I_1$ with $k\geq 2$.
 For Feynman integrals with an underlying elliptic curve, this corresponds to a non-vanishing residue around a simple pole that is not one of the roots of the zero locus of the maximal-cut geometry as described in section \ref{sec:3.2}. 
 For instance, Eq.~(\ref{decoupling_ex}) has two master integrals corresponding to the first and second kind differentials which saturate the middle cohomology as evidenced by the Hodge diamond in Eq.~(\ref{hodge_elliptic}), and the last master integral must be a differential of the third kind, since we want it to decouple. 
 However, even in elliptic examples, see, for example, section 4.4 in ref.~\cite{Gorges:2023zgv}, not all the decoupling integrals have been chosen such that their maximal cut is an integral of a  differential form of the third kind when~$\epsilon=0$. 
 Nevertheless, in that case, the differential equation takes the form of Eq.~(\ref{diffeqdecoupling}). 
 Hence, when the integrand analysis is too involved, or finding independent candidates of the third kind is too complicated, we choose master integrals whose differential equation satisfy Eq.~(\ref{diffeqdecoupling}).
 However, this is not the most generic form, for example in section 7.2.2 in ref.~\cite{Duhr:2025lbz}, the decoupling master integral satisfies a more generic differential equation. 
 This is because, unlike integrals of the third kind, this integral has a double pole. 
 In that case, we see that we need more functions than the ones that we use in our examples. 
 Therefore the bases proposed in this paper is valid for decoupling integrals whose differential equation takes the form of Eq.~(\ref{diffeqdecoupling}).
If such a candidate cannot be found, we follow the indications given in ref.~\cite{Gorges:2023zgv,Duhr:2025lbz}.

\paragraph{Some comments on the basis of MIs:}
From the examples we have computed, we have observed that the basis we propose has the following properties:
\begin{itemize}
\item An $\epsilon$-factorisation of the differential equation $\text{d}\mathbf{M}=\epsilon\,\tilde{\mathbf{A}}\cdot \mathbf{M}\,$.
\item A connection $\tilde{\mathbf{A}}$ with poles of at most order one at the MUM-point.
\end{itemize}
Hence, we call it \textit{canonical}.\footnote{So far there is no clear consensus on a definition of canonical differential equation beyond MPLs. In ref.~\cite{Duhr:2025lbz}, a definition was proposed, which includes the independence of the forms in addition to our defining properties. 
Checking the independence of the forms goes beyond the scope of this paper.}
\\As explained in ref.~\cite{Duhr:2025lbz}, such a construction of an 
$\epsilon$-factorised basis is performed \textit{locally}, close to a regular singular point of maximum unipotent monodromy (the MUM-point).
The basis obtained close to one point can then be used globally, upon a suitable redefinition of the periods.
Close to each point, one in general has to redefine the holomorphic solution $\psi_0$ since, the solution that is holomorphic close to a regular singular point, will not be holomorphic when analytically continued to other points.
Moreover, one must make sure that the periods used close to a different point are consistent with the Picard-Fuchs operators.
We leave to the literature~\cite{Duhr:2025lbz,Klemm:2024wtd, Duhr:2024bzt, Driesse:2024feo, Forner:2024ojj, Becchetti:2025rrz} how to proceed when the analysis close to more than one MUM-point is needed.

\paragraph{Starting from a good basis:}
Let us start by considering a sector to $\epsilon$-factorise, for which we assume it to have an underlying geometry of a Calabi-Yau $l$-fold with middle cohomology Hodge numbers $h^{l,0},h^{l-1,1},\dots,h^{0,l}$.
At $\epsilon=0$ it can be shown that one possible solution to the system evaluates to a holomorphic period at the MUM-point $I_1=\psi_0$. Summing up the previous two paragraphs, for such a sector we can assume the following starting basis:
\begin{equation}\label{general_init}
    \mathbf{I}=(\underbrace{I_1,I_2,\dots,I_{n-1},I_{n}}_{\text{first, second kind integrals}},\underbrace{I^\text{dec}_{n+1},\dots,I^\text{dec}_{n+m}}_{\text{decoupling integrals}})\,,
\end{equation}
where for $\mathbf{I_A}=(I_1,\dots,I_{n})$ is the {\it derivative} block and thus $I_2,\dots, I_{n}$ can be expressed as derivatives of $I_1$.
The integrals $\mathbf{I_B}=(I^\text{dec}_{n+1},\dots,I^\text{dec}_{n+m})$ in turn, are the {\it decoupling} master integrals.
The differential equation at $\epsilon=0$ takes the form:
\begin{equation}\label{dif_eq_alg}
    \text{d}\begin{pmatrix}\mathbf{I_A}\\
    \mathbf{I_B}\end{pmatrix}=\begin{pmatrix}\mathbf{A}& 0\\\mathbf{B}& 0\end{pmatrix}\cdot \begin{pmatrix}\mathbf{I_A}\\
    \mathbf{I_B}\end{pmatrix}
\end{equation}
and we additionally require the differential equation for $\mathbf{I}_\mathbf{B}$ to take the form in Eq.~(\ref{diffeqdecoupling}).
Now let us concretely show how this works for $l=1,2,3$ and then propose a generalisation for~${l>3}$.

\subsection{\texorpdfstring{$l=1$}{l=1}, elliptic Feynman integrals}
\label{ellipticgenericansatz}
The Hodge numbers for an elliptic curve are
\begin{equation}
    h^{1,0}=1\,,\quad h^{0,1}=1\,,
\end{equation}
which, by virtue of Eq.~(\ref{griffiths_transversality}), allows for \textit{two} master integrals in a derivative basis. 
Any master integrals in the maximal cut beyond those two are then chosen as decoupling master integrals in the fashion of Eq.~(\ref{dif_eq_alg}).
Hence, the basis for such a Feynman integral reads\footnote{In this section we omit an overall $\epsilon$-rescaling for clarity. In examples it might be needed to preserve uniform transcendental weight.}
\begin{subequations}\label{ellipticansatz}
\begin{align}
    M_0&=\frac{I_1}{\psi_0}\,,\\
    M_j&=I^{\mathrm{dec}}_{j}+F_{j,0}\,M_0\,, \qquad j=1,...,m\\
    M_{m+1}&=\frac{1}{\epsilon}\dfrac{\partial}{\partial {\tau}}M_0+\sum_{j=0}^{m}F_{(m+1),j}\,M_j\,,
\end{align}
\end{subequations}
where $\tau=\frac{\psi_1}{\psi_0}$ and the functions $F_{i,j}$ are fixed by requiring that the terms not proportional to $\epsilon$ vanish. Since the elliptic case, treated as above, effectively behaves as a one variable case (because we need to take only one derivative), Eqs.~(\ref{ellipticansatz}) is equivalent to refs.~\cite{Gorges:2023zgv,Duhr:2025lbz} for the same reasons as in paragraph~\ref{comparison_section}.
\\For the elliptic Feynman integrals we have worked out two bases, that we have called \textit{democratic} and \textit{decoupling} approaches. The basis in Eqs.~(\ref{ellipticansatz}) corresponds to the decoupling approach. We will not write in full generality the basis for the democratic approach, which in spirit is different from refs.~\cite{Gorges:2023zgv,Duhr:2025lbz}, as we observed that it is a good basis only for elliptic integrals. We refer the reader to section~\ref{sec:sunrise_1}, where we show how this works in a specific example. In appendix~\ref{sec:7} we show the proof for the validity of both the approaches in a specific case of two parameters and one decoupling integral\footnote{The generalisation is straightforward. Since for K3 manifold we give the proof in full generality, here we decided to write it for an hands-on example for better clarity.} and in section~\ref{sec:sunrise} we show that for the unequal mass sunrise these two bases are related by a constant rotation.

\paragraph{Punctured surfaces:} 
As discussed in section~\ref{sec:3.2}  for elliptic curves, we can identify differentials of the third kind as coming from punctures on the surface. 
If the dimension of the phase space (which includes all masses and external momenta associated to the sector) is less or equal then the number of moduli of the Calabi-Yau manifold: $\dim \mathbf{z}\leq h^{l-1,1}$, the punctures can be understood as being constant at any one point in the space of moduli.
If instead we find $\dim \mathbf{z}> h^{l-1,1}$, one can interpret the punctures as being free parameters to which the extra parameters of the phase space are mapped to.
Concretely, we encounter this in section \ref{sec:sunrise} when recomputing the unequal sunrise diagram whose underlying geometry is a torus with three punctures.

\subsection{\texorpdfstring{$l=2$}{l=2}, K3 Feynman integrals}
\label{genericK3ansatz}
The hodge numbers for a K3 surface are
\begin{equation}
    h^{2,0}=1\,,\quad h^{1,1}=20\,,\quad h^{0,2}=1\,,
\end{equation}
however, if we have $\text{dim}(\mathbf{z})\leq 20$ variables, we can define $\tilde{h}=\tilde{h}^{1,1}=\text{dim}(\mathbf{z})$, as there are $20-\tilde{h}$ vanishing entries in the period vector, hence $\tilde{h}$ is the number of different moduli $\tau_j=\frac{\psi_1^{(j)}}{\psi_0}$ we can define from the mirror map.
The size of the derivative basis will therefore depend on the number of variables $\tilde{h}$ and we find that there are:
\begin{equation}
    n=h^{2,0}+\tilde{h}+h^{0,2}=2+\text{dim}(\mathbf{z})
\end{equation}
master integrals which can be written in a derivative basis.
Any $m$ master integrals beyond this number is, if possible, once more taken to be decoupled master integrals in the fashion of $\mathbf{I}_\mathbf{B}$ in Eq.~(\ref{dif_eq_alg}).
Hence, the new basis reads: 
\begin{subequations}\label{ansatzK3generalsigma}
\begin{align}
    M_0&=\dfrac{I_1}{\psi_0}\,,\\
    M_{a}&=\dfrac{1}{\epsilon}\dfrac{\partial}{\partial{\tau_{a}}}M_0+F_{a,0}\,M_0\,,\\
    M_{j}&=I^\text{dec}_j+F_{j,0}\,M_0\,,\\
    M_{n+m-1}&=\dfrac{1}{\epsilon}\dfrac{\partial}{\partial{\tau_{\alpha}}}M_{\beta}+\sum_{k=0}^{n+m-2} F_{(n+m-1),k}\,M_k\,, 
\end{align}
\end{subequations}
where 
\begin{equation}
    1\leq a\leq \tilde{h}\,,\qquad \tilde{h}<j\leq m+\tilde{h}\,,
\end{equation}
and $\alpha,\,\beta$ take a single value in the range $\alpha,\,\beta=1,...,\tilde{h}$ with the additional requirement
\begin{equation}\label{condition_1}
    \dfrac{\partial}{\partial\tau_\alpha}M_\beta\neq 0\quad\mathrm{for}\,\,\epsilon=0\,.
\end{equation}
Note that not all the choices are allowed; this depends on the intersection matrix: a general intersection matrix for the periods of a K3 surface with a MUM-point reads
\begin{equation}\label{sigg}
    \begin{pmatrix}
        0&0&1\\
        0& \Sigma &0\\
        1&0&0\\
    \end{pmatrix},
\end{equation}
where $\Sigma$ is a symmetric $\tilde{h}\times\tilde{h}$ matrix with entries $\Sigma_{ij}\in\mathbb{Z}$. Hence, we can use the quadratic relations 
$\underline{\psi}^T\Sigma\underline{\psi}=0\,$, to write the double logarithmic period $\psi_2$ in terms of the others. 
If we now normalise by the holomorphic period, as we do to write the basis in Eq.~(\ref{ansatzK3generalsigma}), we get
\begin{equation}\label{bilinears}
     \dfrac{\psi_2}{\psi_0}=\dfrac{1}{2\psi_0^2}\sum_{i,j=1}^{\tilde{h}}\Sigma_{ij}\,\psi_1^{(i)} \,\psi_1^{(j)}\,=\dfrac{1}{2}\sum_{i,j=1}^{\tilde{h}}\Sigma_{ij}\,\tau^{(i)}\tau^{(j)}\,,
\end{equation}
so the derivative $\partial_{\tau_\alpha}$ in Eq.~(\ref{condition_1}) has to act on $M_{\beta}\sim\partial_{\tau_\beta}\frac{\psi_2}{\psi_0}$, such that $\partial_{\tau_\alpha}\partial_{\tau_\beta}\frac{\psi_2}{\psi_0}\neq 0$ which depends on the intersection matrix as some products of $\tau^{(i)}\tau^{(j)}$ in Eq.~(\ref{bilinears}) can vanish due to the specific coefficients in $\Sigma_{ij}$. 
This leads to a differential equation on the maximal cut (where we take the formal limit $F_{i,j}\rightarrow 0$ in Eq.~(\ref{ansatzK3generalsigma}) and don't consider the terms multiplied by non-positive powers of $\epsilon$ in the strictly lower diagonal part of the matrix) which reads
\begin{equation}
\dfrac{1}{\epsilon}\dfrac{\partial}{\partial \tau_a}
\begin{pmatrix}
M_0\\M_{b}\\M_{n
+m-1}
\end{pmatrix}
=
\begin{pmatrix}
0&\delta_{a,c}&0\\0&0&\tilde{\Sigma}_{a,b}\\0&0&0
\end{pmatrix}\cdot \begin{pmatrix}M_0\\M_{c}\\M_{n+m-1}
\end{pmatrix}, \quad \mathrm{for}\quad  a,\,b,\,c=1,\dots,\,\tilde{h}\,,
\label{ansatzK3eps0}
\end{equation}
where $\tilde{\Sigma}_{a,b}$ depends on $\Sigma_{a,b}$ in Eq.~(\ref{sigg}) and can be written as
\begin{equation}
    \tilde{\Sigma}=\dfrac{\Sigma}{\Sigma_{\alpha,\beta}+\Sigma_{\beta,\alpha}}\,,
\end{equation}
and therefore depends on the choice of $\alpha$ and $\beta$ we made to define $M_{n+m-1}$ in Eq.~(\ref{ansatzK3generalsigma}).
\\Away from the limit $\epsilon\rightarrow 0$, the functions $F_{i,j}$ in Eq.~(\ref{ansatzK3generalsigma}) are needed to put the whole differential equation into $\epsilon$-form.
We refer to appendix~\ref{proofK3} for a constructive derivation of the full form. 
Completing the picture of Eq.~(\ref{ansatzK3eps0}) we find that for the full differential equation in the derivative block will read:
\begin{align}
\dfrac{\partial}{\partial \tau_a}
\begin{pmatrix}
M_0\\M_b\\M_{n+m-1}
\end{pmatrix}
&=\epsilon
\begin{pmatrix}
-F_{a,0}&\delta_{a,c}&0\\ *&*&\tilde{\Sigma}_{a,b}\\ *&*&*
\end{pmatrix}\cdot \begin{pmatrix}M_0\\M_c\\M_{n+m-1}
\end{pmatrix}\,,
\label{ansatzK3eps0withF}
\end{align}
where the entries denoted by $\quad*\quad$ depend on the starting PF operator and the functions~$F_{i,j}$ are fixed by requiring that the terms not proportional to $\epsilon$ vanish.
For more details we again refer to appendix~\ref{proofK3}.
\\The differential equation (\ref{ansatzK3eps0}) coincides with the differential equation for the unipotent part of the period matrix from section 3.1 in ref.~\cite{Ducker:2025wfl}, showing that effectively taking $\tau_a$ derivatives amounts to multiplying the differential equation matrix by the inverse of the semi-simple part of the Wronskian. The $\epsilon$-rescaling is done similarly as in the one variable case, hence, as reviewed in paragraph~\ref{comparison_section}, corresponds to the $\epsilon$-rescaling from refs.~\cite{Gorges:2023zgv,Duhr:2025lbz}. Finally, the auxiliary functions $F_{i,j}$ are required to cancel the terms not proportional to $\epsilon$, as in refs.~\cite{Gorges:2023zgv,Duhr:2025lbz}.

\newpage
\subsection{\texorpdfstring{$l=3$}{l=3}, Calabi-Yau three-fold Feynman integrals}
\label{genericansatzCY3fold}
The Hodge numbers for a generic Calabi-Yau 3-fold are
\begin{equation}
    h^{3,0}=1\,,\quad h^{2,1}=\text{dim}(\mathbf{z})\,,\quad h^{1,2}=\text{dim}(\mathbf{z})\,,\quad h^{0,3}=1\,.
\end{equation}
Hence, the dimension of the block built from Eq.~(\ref{griffiths_transversality})  has size~${(2\,h^{2,1}+2)\times(2\,h^{2,1}+2)}$.
Carrying out the IBP reduction, the maximal cut sector may contain $n+m$ master integrals where we identify the first $n=2\,h^{2,1}+2$ master integrals in the derivative block which span the cohomology of the Calabi-Yau three-fold and $m$ master integrals, if possible, chosen such that they satisfy a decoupling differential equation in the fashion of Eq.~(\ref{dif_eq_alg}). 
We use the following basis to $\epsilon$-factorise the block: 
\begin{subequations}\label{CY3_ansatz}
\begin{align}
    M_0&=\dfrac{I_1}{\psi_0}\,,\\
    M_{a}&=\dfrac{1}{\epsilon}\dfrac{\partial}{\partial{\tau_{a}}}M_0+F_{a,0}\,M_0\,,\\
    M_{h^{2,1}+b}&=\dfrac{1}{\epsilon}\sum_{k,l=1}^{h^{2,1}}\mathcal{Y}_{b,k,l}\,\dfrac{\partial}{\partial\tau_k}M_l +\Biggl(\sum_{k,l=1}^{h^{2,1}}\mathcal{Y}_{b,k,l}\Biggr)\sum_{p=0}^{h^{2,1}}F_{(h^{2,1}+b),p}M_p\,,\label{hard_MI}\\
    M_{j}&=I^\text{dec}_j+F_{j,0}\,M_0\,, \\
    M_{n+m-1}&=\dfrac{1}{\epsilon}\dfrac{\partial}{\partial{\tau_{\alpha}}}M_{h^{2,1}+\beta}+\sum_{k=0}^{n+m-2} F_{(n+m-1),k}\,M_k\,, \qquad 
\end{align}
\end{subequations}
where the indices $a$, $b$ and $j$ cycle through the $n+m$ master integrals with the ranges 
\begin{equation}
    1\leq a\leq h^{2,1}\quad \text{and}\quad 1\leq b\leq h^{2,1}\quad \text{and}\quad n<j\leq n+m\,,
\end{equation}
and the functions $F_{i,j}$ are fixed by requiring that the terms not proportional to $\epsilon$ vanish.
The rational functions $\mathcal{Y}_{i,j,k}$ can be determined through the Yukawa three-couplings defined in Eq.~(\ref{quadrel}). 
In particular they follow from solving for $M_{h^{2,1}+b}$ the set of equations
\begin{align}\label{find_Y}
    \Biggl\{\sum_{b=1}^{h^{2,1}}\biggl(C_{b,c,d}\,M_{h^{2,1}+b}-\sum_{k=0}^{h^{2,1}}F_{(h^{2,1}+b),k}\,M_k\biggr)=\dfrac{1}{\epsilon}\dfrac{\partial}{\partial \tau_c}M_d\,,\quad \begin{matrix}c=\{1,\dots, h^{2,1}\}\\d=\{1,\dots, h^{2,1}\}\end{matrix}\Biggr\}\,.
\end{align}
Note that Eq.~(\ref{find_Y}) only spans $h^{2,1}$ independent equations and that several $\mathcal{Y}_{i,j,k}$ from Eq.~(\ref{hard_MI}) can vanish.
When taking the formal limit $F_{i,j}\rightarrow 0$ and only considering the $n\times n$ derivative block of the maximal cut without the terms multiplied by non-positive powers of $\epsilon$ in the strictly lower triangular part of the matrix, we recover the following differential equation from the basis in Eq.~(\ref{CY3_ansatz}):
\begin{equation}
\dfrac{1}{\epsilon}\dfrac{\partial}{\partial \tau_a}
\begin{pmatrix}
M_0\\M_{b}\\M_{h^{2,1}+c}\\M_{n+m-1}
\end{pmatrix}
=
\begin{pmatrix}
0&\delta_{a,d}&0&0\\0&0&C_{a,b,e}&0\\0&0&0&\delta_{a,c}\\0&0&0&0
\end{pmatrix}\cdot \begin{pmatrix}M_0\\M_{d}\\M_{h^{2,1}+e}\\M_{n+m-1}
\end{pmatrix}\,.
\label{GMCY3fold}
\end{equation}
The structure of this differential equation is once more derived from the Griffiths transversality conditions from Eq.~(\ref{quadrel}).
This corresponds to the structure we reviewed in Eq.~(\ref{frob_ST}).
\\The differential equation (\ref{GMCY3fold}) coincides with the differential equation for the unipotent part of the period matrix from section 3.2 in ref.~\cite{Ducker:2025wfl}, showing that effectively taking $\tau_a$ derivatives, corrected by the appropriate cubic coupling $C_{a,b,c}$, amounts to multiplying the differential equation matrix by the inverse of the semi-simple part of the Wronskian. 
Finally, the $\epsilon$-rescaling and the introduction of the auxiliary functions $F_{i,j}$ is done similarly as in section~\ref{genericK3ansatz}.

\subsection{\texorpdfstring{$l>3$}{l>3}, Calabi-Yau \texorpdfstring{$l$}{l}-fold Feynman integrals}
\label{higherdimCY}
For higher dimensional manifold we generalise the basis following the same procedure, such that, in the appropriate integer dimension, the differential equation reads
\begin{equation}
\dfrac{1}{\epsilon}\dfrac{\partial}{\partial \tau_a}
\begin{pmatrix}
M_0\\M_{i_1}\\M_{i_2}\\\vdots\\M_{i_{l-3}}\\M_{i_{l-2}}\\M_{n+m-1}
\end{pmatrix}
=
\begin{pmatrix}
0&\delta_{a,j_1}&0&0& \hdots & 0 & 0 \\
0&0&C_{a,i_1,j_2}^{(1,1)}&0 &\hdots &0&0\\
0&0&0&C_{a,i_2,j_3}^{(1,2)}&\hdots &0&0\\
\vdots&\vdots&\vdots&\vdots&\vdots&\vdots&\vdots\\
0&0&0&0&\hdots&C_{a,i_{l-3},j_{l-2}}^{(1,l-2)}&0\\
0&0&0&0&\hdots&0&\delta_{a,j_{l-2}}\\
0&0&0&0&\hdots&0&0
\end{pmatrix}\cdot \begin{pmatrix}
M_0\\M_{j_1}\\M_{j_2}\\\vdots\\M_{j_{l-3}}\\M_{j_{l-2}}\\M_{n+m-1}
\end{pmatrix}\,, 
\label{genericGM}
\end{equation}
for $a=1,...,h_{l-1,1}$, that is again the differential equation satisfied by the unipotent part of the period matrix.
The Yukawa three-couplings $C_{i,j,k}^{(1,p)}$ have now different upper indices, because the dimension of the manifold is greater than three as mentioned in Eq.~(\ref{conditionC}). They can be determined as detailed in ref.~\cite{Ducker:2025wfl}, in which the computations are explicitly carried out up to $l=6$. 
This procedure will put the derivative part of the differential equation into $\epsilon$-form up to a strictly lower triangular part which in turn is dealt with the functions $F_{i,j}$. If there are more master integrals in the sector, and if it is possible to choose them such that they satisfy a differential equation in the fashion of Eq.~(\ref{dif_eq_alg}), we proceed as in the previous cases.
\\Note that the auxiliary functions $F_{i,j}$ introduced throughout this section, are defined by a set of differential equations and can, in general, have a complicated analytic structure that can go beyond periods (and their derivatives) of Calabi-Yau manifolds~\cite{paperDM}.
For the purpose of this work, these differential equations will (except for the elliptic case) only be solved as a series expansion around the MUM-point.
We leave for future work~\cite{Duhr:2025xyy}, a more in-depth discussion of these auxiliary functions.
\section{Recomputing the sunrise with unequal masses}\label{sec:sunrise}
The sunrise diagram with unequal masses has first been analytically put into $\epsilon$-form in ref.~\cite{Bogner:2019lfa} with the introduction of elliptic integrals in the gauge transformation.
We will show that the approach presented in section~\ref{sec:4} leads to a canonical basis which only differs by a constant gauge transformation from the results of ref.~\cite{Bogner:2019lfa}.
Let us review the basics and introduce our conventions.
We consider the two-loop unequal mass sunrise family:
\begin{equation}
I_{\nu_1,\nu_2,\dots,\nu_5}(p^2,m_1^2,m_2^2,m_3^2;\epsilon)=e^{2\gamma_\text{E}\epsilon}\int \dfrac{\text{d}^Dl_1}{i\pi^\frac{D}{2}}\dfrac{\text{d}^Dl_2}{i\pi^\frac{D}{2}} \dfrac{1}{D_1^{\nu_1} D_2^{\nu_2}\dots D_5^{\nu_5}}\,,
\end{equation}
with the denominators $D_j$:
\begin{align}
\label{propagatorsSunrise}
    D_1&=l_1^2-m_1^2\,,\quad D_2=l_2^2-m_2^2\,,\quad D_3=(l_1+l_2-p)^2-m_3^2\,,\notag\\
    D_4&=(l_1-p)^2\,,\,\, D_5=(l_2-p)^2\,.
\end{align}
We set the dimensionless kinematic variables $x_i$\footnote{Note that when treating elliptic examples we denote kinematic variables with $\mathbf{x}$ as opposed to $\mathbf{z}$ to avoid a notational conflict since the canonical variables on the torus include the punctures $z_i$.}:
\begin{equation}
    x_1=\dfrac{m_1^2}{p^2}\,,\quad x_2=\dfrac{m_2^2}{p^2}\,,\quad x_3=\dfrac{m_3^2}{p^2}\,.
\end{equation}
\paragraph{Elliptic integrals through the gauge transformation:}
The $\epsilon$-factorisation will be achieved through a gauge transformation as reviewed in section \ref{sec:2.1}.
This gauge transformation will introduce the periods $\psi_0$, $\psi_1$\footnote{In ref.~\cite{Bogner:2019lfa} the periods are denoted $\psi_1$ and $\psi_2$ instead of $\psi_0$ and $\psi_1$.}:
\begin{align}\label{pers_ell}
\psi_0=\dfrac{4K(k)}{c_4}\,,\qquad \psi_1=\dfrac{4iK(1-k)}{c_4}\,,
\end{align}
with
\begin{align}
   k^2&= \frac{16 \sqrt{x_1 x_2
  x_3}}{x_1^2-2 (x_2+x_3+1)
   x_1+x_2^2+(x_3-1)^2-2 x_2
   (x_3+1)+8 \sqrt{x_1 x_2
   x_3}}\,,\notag\\
   c_4&=\sqrt{x_1^2-2 x_1
   (x_2+x_3+1)+8 \sqrt{x_1 x_2
   x_3}+x_2^2-2 x_2
   (x_3+1)+(x_3-1)^2}\,.
\end{align}
The elliptic integral of the first kind has been defined in Eq.~(\ref{ell_1}).
Additionally some incomplete integrals of the first kind will be introduced through Abel's map $\mathcal{A}$ with the definition for $F$ in Eq.~(\ref{ell_inc}):
\begin{align}
\mathcal{A}(u_j)&=\dfrac{4\,F\left(\arcsin\left({\sqrt{u_j}}\right),k\right)}{c_4}\,.
\end{align}
Through these definitions we can now map the kinematic variables $x_1$, $x_2$ and $x_3$ to the moduli space $\mathcal{M}_{1,3}$ of a punctured torus which is defined by the ratio of periods $\tau$ and three punctures $z_1$, $z_2$ and $z_3$ where translation invariance requires $z_1+z_2+z_3=1$:
\begin{align}\label{pullback_fcts}
    \tau=\dfrac{\psi_1}{\psi_0}\,,\quad \quad z_1=\dfrac{\mathcal{A}(u_1)}{\psi_0}\,,\quad \quad z_2=\dfrac{\mathcal{A}(u_2)}{\psi_0}\,,
\end{align}
with
\begin{align}
u_1=\dfrac{(1+\sqrt{x_1})^2-(\sqrt{x_2}-\sqrt{x_3})^2}{(\sqrt{x_2}+\sqrt{x_3})^2-(\sqrt{x_2}-\sqrt{x_3})^2}\,,\qquad u_2=\dfrac{(1+\sqrt{x_2})^2-(\sqrt{x_1}-\sqrt{x_3})^2}{(\sqrt{x_1}+\sqrt{x_3})^2-(1-\sqrt{x_3})^2}\,.
\end{align}
\paragraph{The differential equation:}
Running an IBP algorithm (e.g. ref.~\cite{Lee:2012cn}) we find the following master integrals:
\begin{equation}\label{mis_start_sunrise}
    {\bf I}=\begin{pmatrix}I_{1,1,0,0,0}\\I_{1,0,1,0,0}\\I_{0,1,1,0,0}\\I_{1,1,1,0,0}\\I_{2,1,1,0,0}\\I_{1,2,1,0,0}\\I_{1,1,2,0,0}\end{pmatrix},
\end{equation}
satisfying the differential equation:
\begin{equation}
    \text{d}{\bf I}={\bf A}(x_1,x_2,x_3,\epsilon)\cdot{\bf I}\,,
\end{equation}
where ${\bf A}$ is a matrix where each entry is a rational function in $x_1$, $x_2$, $x_3$ and the dimensional regularisation parameter $\epsilon$.
\\As mentioned in section~\ref{ellipticgenericansatz}, to $\epsilon$-factorise this differential equation we have found two different approaches. 
We start by presenting the {\it democratic approach} and follow it up with the {\it decoupling approach}.

\subsection{The democratic approach}\label{sec:sunrise_1}
We can start from the set of master integrals given by the IBP reduction algorithm where we identify a maximal-cut sector with four entries and three different sub-sectors each of them corresponding to a tadpole:
\begin{equation}
    {\bf{I}}=(\underbrace{I_{1,1,0,0,0},I_{1,0,1,0,0},I_{0,1,1,0,0}}_{\text{tadpoles}},\underbrace{I_{1,1,1,0,0},I_{2,1,1,0,0},I_{1,2,1,0,0},I_{1,1,2,0,0}}_{\text{maximal cut}})^T\,.
\end{equation}
For this choice of basis, the derivatives with respect to the different kinematic parameters are on equal footing. 
We introduce derivatives with respect of to the moduli-space $\mathcal{M}_{1,3}$ to put the $4\times 4$ maximal cut block into $\epsilon$-form.
The basis reads:
\begin{align}\label{ans_sunrise_dem}
M_0^\text{dem}&=\epsilon^2 I_{1,1,0,0,0}\,,\notag\\
M_1^\text{dem}&=\epsilon^2 I_{1,0,1,0,0}\,,\notag\\
M_2^\text{dem}&=\epsilon^2 I_{0,1,1,0,0}\,,\notag\\
    M_3^\text{dem}&=\epsilon^2 \dfrac{I_{1,1,1,0,0}}{\psi_0}\,,\notag\\
    M_4^\text{dem}&=\dfrac{1}{\epsilon}\dfrac{\partial}{\partial z_1} M_3^\text{dem}+F_{4,3}^\text{dem}M_3^\text{dem}\,,\notag\\
     M_5^\text{dem}&=\dfrac{1}{\epsilon}\dfrac{\partial}{\partial z_2} M_3^\text{dem}+F_{5,3}^\text{dem}M_3^\text{dem}\,,\notag\\
      M_6^\text{dem}&=\dfrac{1}{\epsilon}\dfrac{\partial}{\partial \tau} M_3^\text{dem}+F_{6,3}^\text{dem}M_3^\text{dem}\,,
\end{align}
where the derivatives with respect to  $\tau$, $z_1$ and $z_2$ can be determined through the Jacobian~$\mathfrak{J}$:
\begin{equation}\label{jacobian}
\begin{pmatrix}\text{d}x_1\\\text{d}x_2\\\text{d}x_3\end{pmatrix}=\begin{pmatrix}\mathfrak{J}_{1,1}&\mathfrak{J}_{1,2}&\mathfrak{J}_{1,3}\\
\mathfrak{J}_{2,1}&\mathfrak{J}_{2,2}&\mathfrak{J}_{2,3}\\
\mathfrak{J}_{3,1}&\mathfrak{J}_{3,2}&\mathfrak{J}_{3,3}\end{pmatrix}\cdot\begin{pmatrix}\text{d}\tau\\\text{d}z_1\\\text{d}z_2\end{pmatrix}\,.
\end{equation}
Explicit formulas for $F_{5,4}^\text{dem}$, $F_{6,4}^\text{dem}$ and $F_{7,4}^\text{dem}$ as well as the resulting connection matrix $\tilde{\mathcal{A}}$, close to the MUM-point~$\mathbf{x}=0$, can be found in the ancillary file {\bf{\texttt  sunrise.nb}}. We call this \textit{democratic} approach because we take derivatives not only with respect to $\tau$ but also with respect to punctures, placing all the moduli on equal footing. With this basis the differential equation is in canonical form. 
We give all results in terms of elliptic functions and the functions $F_{i,j}^{\mathrm{dem}}$ are determined leveraging the modular properties of the functions as in Refs.~\cite{Giroux:2022wav,Giroux:2024yxu}. We briefly review this method in appendix \ref{newApp}.

\subsection{The decoupling approach}\label{sec:5.2}
Alternatively, it is possible to start from a different basis of MIs which is still rational in $x_1$, $x_2$, $x_3$ and $\epsilon$ (in other words, an IBP algorithm can also find this starting basis):
\begin{equation}
    {\bf I}^\text{dec}=(\underbrace{I_{1,1,0,0,0},I_{1,0,1,0,0},I_{0,1,1,0,0}}_{\text{tadpoles}},\underbrace{I_{1,1,1,0,0},I_{2,1,1,0,0},I_{1,1,1,-1,0},I_{1,1,1,0,-1}}_{\text{maximal cut}})^T.
\end{equation}
We observe that, with this choice of starting master integrals, the differential equation for~$I_{1,1,1,0,0}$ and $I_{2,1,1,0,0}$ at $\epsilon=0$ does not depend on $I_{1,1,1,-1,0}$ and $I_{1,1,1,0,-1}$. 
This corresponds to the structure of differential equations for elliptic integrals of the first, second and third kind reviewed in section \ref{sec:3.2}.
The maximal cut therefore organises itself into a $2\times 2$ minimally coupled elliptic sector and two $1\times 1$ sectors which only couple back to the elliptic sector. This is why we call this alternative choice, the \textit{decoupling} approach.
Following the method outlined in section \ref{sec:4} we split the maximal cut into two parts, first a two-dimensional part which  is expressed through differentials of the first and second kind (at $\epsilon=0$), where we normalise by the holomorphic period and use a derivative, to generate the differential of second kind. 
Secondly a two-dimensional part which is expressed through differentials of the third kind (at $\epsilon=0$) $I_{1,1,1,-1,0}$ and $I_{1,1,1,0,-1}$.
The full basis then reads:
\begin{align}\label{ans_sunrise_dec}
M_0^\text{dec}&=\epsilon^2 I_{1,1,0,0,0}\,,\notag\\
M_1^\text{dec}&=\epsilon^2 I_{1,0,1,0,0}\,,\notag\\
M_2^\text{dec}&=\epsilon^2 I_{0,1,1,0,0}\,,\notag\\
M_3^\text{dec}&=\epsilon^2 \dfrac{I_{1,1,1,0,0}}{\psi_0}\,,\notag\\
M_4^\text{dec}&=\epsilon^2 I_{1,1,1,-1,0}+F_{4,3}^\text{dec} M_3^\text{dec}\,,\notag\\
M_5^\text{dec}&=\epsilon^2 I_{1,1,1,0,-1}+F_{5,3}^\text{dec}M_3^\text{dec}\,,\notag\\
M_6^\text{dec}&=\dfrac{\mathcal{J}_{1,1}}{\epsilon}\dfrac{\partial}{\partial x_1}M_3^\text{dec}+\sum_{j=0}^5 F_{6,j}^\text{dec}M_j^\text{dec}\,,
\end{align}
where\footnote{Note that the choice of variable with respect to which we take a derivative is arbitrary as long as the periods (\ref{pers_ell}) depend on it.} 
\begin{equation}
\mathcal{J}_{1,1}=\left(\dfrac{\partial \tau}{\partial x_1}\right)^{-1}.
\end{equation}
Note that $\mathcal{J}_{1,1}\partial_{x_1}\neq \partial_\tau$ since
\begin{equation}
    \dfrac{\partial}{\partial \tau}=\mathfrak{J}_{1,1}\dfrac{\partial}{\partial x_1}+\mathfrak{J}_{2,1}\dfrac{\partial}{\partial x_2}+\mathfrak{J}_{3,1}\dfrac{\partial}{\partial x_3}\,.
\end{equation}
Instead $\mathcal{J}_{1,1}\partial_{x_1}$ can be interpreted as a $\tau$ derivative when $x_2$ and $x_3$ are held constant.
Similarly we have $\mathcal{J}_{1,1}\neq \mathfrak{J}_{1,1}$ but we can relate $\mathcal{J}_{1,1}$ to the entries of the Jacobian $\mathfrak{J}$ of Eq.~(\ref{jacobian}) through:
\begin{equation}
    \mathcal{J}_{1,1}=\dfrac{\mathfrak{J}_{1,1}\mathfrak{J}_{2,3}\mathfrak{J}_{3,2}+\mathfrak{J}_{1,2}\mathfrak{J}_{2,1}\mathfrak{J}_{3,3}+\mathfrak{J}_{1,3}\mathfrak{J}_{2,2}\mathfrak{J}_{3,1}-\mathfrak{J}_{1,1}\mathfrak{J}_{2,2}\mathfrak{J}_{3,3}-\mathfrak{J}_{1,2}\mathfrak{J}_{2,3}\mathfrak{J}_{3,1}-\mathfrak{J}_{1,3}\mathfrak{J}_{2,1}\mathfrak{J}_{3,2}}{\mathfrak{J}_{1,3}\mathfrak{J}_{3,2}-\mathfrak{J}_{2,2}\mathfrak{J}_{3,3}}.
\end{equation}
The terms $F_{i,j}^\text{dec}$ which we have collected in ancilliary file {\bf{\texttt sunrise.nb}} will nevertheless introduce the entries of the Jacobian $\mathfrak{J}_{i,j}$ of Eq.~(\ref{jacobian}).
We therefore interpret that, the first step of the algorithm (which $\epsilon$-factorises the diagonal terms) solely introduce $\tau$ by holding $x_2$ and~$x_3$ constant.
Then, the second step (which $\epsilon$-factorises the remaining off-diagonal part by introducing the functions $F_{i,j}$) will in turn introduce the punctures $z_1$ and $z_2$. 
With this basis the differential equation is in canonical form, close to the MUM-point at $\mathbf{x}=0$.
The functions $F_{i,j}^{\mathrm{dec}}$ are determined as in section~\ref{sec:5.2}.

\subsection{Comparing the results}
We have shown how to get two canonical forms for the unequal masses sunrise by two methods that combine the ideas reviewed in section~\ref{ellipticgenericansatz}.
It turns out that both  gauge transformations are related by a constant rotation:
\begin{align}
   {\bf M}^\text{dem}&= {\bf U}^\text{dem}\cdot {\bf I}\,,\notag\\
   {\bf M}^\text{dec}&={\bf U}^\text{dec} \cdot {\bf I}\,,
\end{align}
with 
\begin{equation}
    {\bf U}^\text{dem}=\begin{pmatrix}1&0&0&0&0&0&0\\
    0&1&0&0&0&0&0\\
    0&0&1&0&0&0&0\\
    0&0&0&1&0&0&0\\
    -i&-i&-i&0&2i&i&0\\
    -i&-i&-i&0&i&2i&0\\
    0&0&0&0&0&0&1\end{pmatrix}\cdot {\bf U}^\text{dec}\,,
\end{equation}
showing that, effectively, we get only one canonical form.
\\Note that when comparing with the gauge transformation ${\bf U}^\text{BMW}$ of ref.~\cite{Bogner:2019lfa}, we again find a constant rotation which relates it to our results:
\begin{equation}
\setstackgap{L}{1.1\baselineskip}
\fixTABwidth{T}
    {\bf U}^\text{dem}=\parenMatrixstack{1&0&0&0&0&0&0\\0&1&0&0&0&0&0\\
    0&0&1&0&0&0&0\\
    0&0&0&\frac{1}{\pi}&0&0&0\\
    0&0&0&0&\frac{1}{2}&\frac{1}{2}&0\\
     0&0&0&0&\frac{1}{2}&-\frac{1}{2}&0\\
      0&0&0&0&0&0&2i}\cdot {\bf U}^\text{BMW}\,.
\end{equation}
This shows that our method finds, up to a constant rotation, the same differential equation matrix, which
in ref.~\cite{Bogner:2019lfa} is expressed as Kronecker forms recalled in Eq.~(\ref{kronecker}).
\\
We therefore find that in this given case, there is a freedom on the starting set of master integral we can use. 
It turns out that here, both sets of master integrals are easy to determine from the integral reduction.
The main difference between these choices comes from the $\epsilon$-factorisation procedure as the decoupling approach shifts all the functions depending on the marked points into the functions $F_{i,j}^\text{dec}$.
\\
The democratic approach, on the other hand, already incorporates the marked points of Eq. (\ref{pullback_fcts}) the derivatives with respect to the moduli.
This is also evidenced by the ansatz of Eq. (\ref{ans_sunrise_dec}) having 8 functions $F_{i,j}^\text{dec}$ as opposed to the 3 functions $F_{i,j}^\text{dem}$ of Eq. (\ref{ans_sunrise_dem}).
It is therefore a question of preference whether one prefers to make the marked points manifest as in the democratic approach, or whether one keeps this information in the functions $F_{i,j}^\text{dec}$.

\section{Computing new integrals}
\label{Computing new integrals}
In this section we apply the approach presented in section \ref{sec:4} to integrals that, to the best of our knowledge, have not been computed before.
In particular, we will get a canonical differential equation for two three-loop banana integrals with two unequal masses and one four-loop banana integral with two unequal masses.

\subsection{Banana three-loop with two internal unequal masses}\label{sec:banana}
We begin by applying the algorithm from section \ref{sec:4}, in particular from section~\ref{genericK3ansatz}, to get a canonical form for the differential equation of a three-loop integral with an underlying K3 geometry (a Calabi-Yau 2-fold).
We consider the three-loop banana family defined as:
\begin{equation}
    I_{\nu_1,\nu_2,\dots,\nu_9}(p^2,m_{j_1}^2,\dots,m_{j_4}^2;\epsilon)=
    e^{3\gamma_\text{E} \epsilon}
    \int \dfrac{\text{d}^D l_1}{i\pi^{\frac{D}{2}}}\dfrac{\text{d}^D l_2}{i\pi^{\frac{D}{2}}} \dfrac{\text{d}^D l_3}{i\pi^{\frac{D}{2}}} \dfrac{1}{D_1^{\nu_1} D_2^{\nu_2}\dots D_9^{\nu_9}}\,,
\end{equation}
with the denominators $D_j$:
\begin{align}
    D_1&=l_1^2-m_{j_1}^2\,,\quad D_2=l_2^2-m_{j_2}^2\,,\quad D_3=l_3^2-m_{j_3}^2\,,\quad D_4=(l_1+l_2+l_3-p)^2-m_{j_4}^2\,,\notag\\
    D_5&=(l_1-p)^2\,,\,\, D_6=(l_2-p)^2\,,\,\, D_7=(l_3-p)^2\,,\,\,D_8=(l_1-l_2)^2\,,\,\, D_9=(l_1-l_3)^2\,.
\end{align}
In the case with two unequal masses, there are two possible mass configurations, namely the~3-1 configuration where we set:
\begin{equation}\label{3-1 config}
    m_{j_1}=m_{j_3}=m_{j_4}=m_1\,,\quad m_{j_2}=m_2\,,
\end{equation}
and the 2-2 configuration where we set:
\begin{equation}\label{2-2 config}
m_{j_1}=m_{j_2}=m_1\,,\quad m_{j_3}=m_{j_4}=m_2\,.
\end{equation}
In the following, we will work with the dimensionless kinematic parameters:
\begin{equation}
    z_1=\dfrac{m_1^2}{p^2}\,,\quad z_2=\dfrac{m_2^2}{p^2}\,.
\end{equation}
The Hodge diamond for a K3 manifold is 
\begin{align}
\begin{matrix}
    &&&1&&&\\
   &&0&&0&&\\
  1 &&& 20 &&&1\\
   &&0&&0&&\\
   &&&1&&&
\end{matrix}\,,
\end{align}
however, since our manifold depends only on $\text{dim}(\mathbf{z})=2$ moduli ($\mathbf{z}=(z_1,z_2)$), there are~${20-2=18}$ vanishing entries in the period vector. Hence we define
\begin{equation}
    \underline{\psi}(\mathbf{z})=(\psi_0(\mathbf{z}),\psi_{1}^{(1)}(\mathbf{z}),\psi_{1}^{(2)}(\mathbf{z}),\psi_2(\mathbf{z}))^T\,,
\end{equation}
where $\psi_0(\mathbf{z})$ denotes the only period that is holomorphic at the MUM-point and therefore allows for a Frobenius expansion with no logarithms.\footnote{We will from now on leave the dependence on $\mathbf{z}$ implicit.}
The periods $\underline{\psi}_1=(\psi_{1}^{(1)},\psi_{1}^{(2)})^T$ on the other hand are the $\text{dim}(\mathbf{z})({=}2)$ periods that diverge with a single logarithmic power when approaching the MUM-point. 
Consequently, $\psi_2$ is the period which has a double-logarithmic expansion close to the MUM-point. 
From the Griffiths transversality conditions in Eq.~(\ref{quadrel}) we also find that there is a quadratic relation which relates these periods and we are able to eliminate $\psi_2$ in the fashion of Eq.~(\ref{bilinears}). 
We therefore will henceforth only consider $\psi_0$, $\psi_1^{(1)}$ and $\psi_1^{(2)}$ as independent periods.

\subsubsection{The 3-1 configuration}
We begin with the 3-1 configuration, in which we set the masses to the values $m_1$ and $m_2$ according to Eq.~(\ref{3-1 config}).
Running an IBP algorithm and massaging the starting basis into the form of Eq.~(\ref{general_init}) yields:
\begin{equation}
    {\bf I}^\text{3-1}=\begin{pmatrix}I_{1,0,1,1,0,0,0,0,0}\\
    I_{0,1,1,1,0,0,0,0,0}\\
    I_{1,1,1,1,0,0,0,0,0}\\
    I_{2,1,1,1,0,0,0,0,0}\\
    I_{1,2,1,1,0,0,0,0,0}\\
    I_{3,1,1,1,0,0,0,0,0}\\
    I_{1,1,1,1,-1,0,0,0,0}
    \end{pmatrix}\,,
\end{equation}
\\
where the master integral $I_{1,1,1,1,-1,0,0,0,0}$ decouples from the maximal cut in the same fashion as the last two master integrals in section \ref{sec:5.2}.
We therefore are left with a minimally coupled $4\times 4$ sector with the master integrals $I_{1,1,1,1,0,0,0,0,0}$, $I_{2,1,1,1,0,0,0,0,0}$, $I_{1,2,1,1,0,0,0,0,0}$ and~$I_{3,1,1,1,0,0,0,0,0}$. 
By setting $\epsilon$ to zero we can read off two coupled PF operators of the form:
\begin{equation}\label{PFbananas}
\begin{aligned}
0 &= 
\Bigl(
  \frac{\partial^3}{\partial z_1^3}
  + q^{(0)}_{1,1(z_1,z_1,z_1)}\,\frac{\partial^2}{\partial z_1^2}
  + q^{(0)}_{1(z_1,z_1,z_1)}\,\frac{\partial}{\partial z_1}
  + q^{(0)}_{2(z_1,z_1,z_1)}\,\frac{\partial}{\partial z_2}
  + q^{(0)}_{(z_1,z_1,z_1)}
\Bigr)\,I_{1,1,1,1,0,0,0,0,0}
,\\
0 &=
\Bigl(
  \frac{\partial^3}{\partial z_2\partial z_1^2}
  + q^{(0)}_{1,1(z_2,z_1,z_1)}\,\frac{\partial^2}{\partial z_1^2}
  + q^{(0)}_{1(z_2,z_1,z_1)}\,\frac{\partial}{\partial z_1}
  + q^{(0)}_{2(z_2,z_1,z_1)}\,\frac{\partial}{\partial z_2}
  + q^{(0)}_{(z_2,z_1,z_1)}
\Bigr)\,I_{1,1,1,1,0,0,0,0,0}
.
\end{aligned}
\end{equation}
where $q_i^{(0)}$ are rational functions in $z_1$ and $z_2$. 
We can construct three independent solutions in a Frobenius basis close to the MUM-point $(z_1,z_2)=(0,0)$ for $\psi_0$, $\psi_1^{(1)}$ and $\psi_1^{(2)}$ where explicit expressions can be found in {\bf{\texttt banana\_3\_1.nb}}.
This basis satisfies the quadratic relations
\beq
\underline{\psi}^T\cdot\Sigma^\text{3-1}\cdot\underline{\psi}=0\,\,,
\eeq
with the intersection pairing
\beq
\Sigma^\text{3-1} = \begin{pmatrix}
 0 & 0 & 0 & 1 \\
 0 & -6 & -3 & 0 \\
 0 & -3 & 0 & 0 \\
 1 & 0 & 0 & 0 
 \end{pmatrix}.
 \eeq
We can then define two independent moduli:
\begin{equation}
\tau_1=\dfrac{\psi_1^{(1)}}{\psi_0}\,,\qquad\tau_2=\dfrac{\psi_1^{(2)}}{\psi_0}\,,
\end{equation}
and can construct a change of variables through the Jacobian $\mathfrak{J}$:
\begin{equation}
    \begin{pmatrix}
    \text{d}z_1\\\text{d}z_2\end{pmatrix}=\begin{pmatrix}\mathfrak{J}_{1,1}&\mathfrak{J}_{1,2}\\
    \mathfrak{J}_{2,1}&\mathfrak{J}_{2,2}\end{pmatrix}\cdot \begin{pmatrix}\text{d}\tau_1\\\text{d}\tau_2\end{pmatrix}.
\end{equation}
Having defined the variables $\tau_1$ and $\tau_2$ we are now ready to write the basis proposed in section~\ref{sec:4}:
\begin{align}
M_0^\text{3-1}&=\epsilon^3\,I_{1,0,1,1,0,0,0,0,0}\,,\notag\\
M_1^\text{3-1}&=\epsilon^3\,I_{0,1,1,1,0,0,0,0,0}\,,\notag\\
M_2^\text{3-1}&=\epsilon^3\,\dfrac{I_{1,1,1,1,0,0,0,0,0}}{\psi_0}\,,\notag\\
M_3^\text{3-1}&=\dfrac{1}{\epsilon} \dfrac{\partial}{\partial \tau_1}M_2^\text{3-1}+F_{3,2}^\text{3-1}M_2^\text{3-1}\,,\notag\\
M_4^\text{3-1}&=\dfrac{1}{\epsilon} \dfrac{\partial}{\partial \tau_2}M_2^\text{3-1}+F_{4,2}^\text{3-1}M_2^\text{3-1}\,,\notag\\
M_5^\text{3-1}&=\epsilon^3\,I_{1,1,1,1,-1,0,0,0,0}+F_{5,2}^\text{3-1}M_2^\text{3-1}\,,\notag\\
M_6^\text{3-1}&=\dfrac{1}{\epsilon} \dfrac{\partial}{\partial \tau_1}M_3^\text{3-1}+\sum_{j=0}^5F_{6,j}^\text{3-1}M_j^\text{3-1}\,,
\end{align}
where the functions $F_{i,j}^\text{3-1}$ are made up of algebraic functions multiplied by powers of the holomorphic period $\psi_0$ and (iterated) integrals of the same kind of functions. 
This basis brings the differential equation in canonical form, close to the MUM-point. We once again give explicit (expanded in series around the MUM-point) results in the ancillary file {\bf{\texttt banana\_3\_1.nb}}. 
Moreover, we have checked that we get at most simple poles in each of the variables at the MUM-point. 
Note that the expansions of the functions $F_{i,j}^\text{3-1}$ have \textit{integer} coefficients, as we observed in banana integrals with equal masses~\cite{Pogel:2022vat} and other integrals with underlying single-scale Calabi-Yau geometry~\cite{Duhr:2025lbz}.

\subsubsection{The 2-2 configuration}
Let us now consider the 2-2 configuration, in which we set the masses to the values $m_1$ and~$m_2$ according to Eq.~(\ref{2-2 config}).
Running an IBP algorithm and massaging the starting basis into the form of Eq.~(\ref{general_init}) yields:
\begin{equation}
    {\bf I}^\text{2-2}=\begin{pmatrix}I_{1,0,1,1,0,0,0,0,0}\\
    I_{0,1,1,1,0,0,0,0,0}\\
    I_{1,1,1,1,0,0,0,0,0}\\
    I_{2,1,1,1,0,0,0,0,0}\\
    I_{1,2,1,1,0,0,0,0,0}\\
    I_{3,1,1,1,0,0,0,0,0}\\
    I_{1,1,1,1,-1,0,0,0,0}\\
    I_{1,1,1,1,0,0,0,-1,0}
    \end{pmatrix},
\end{equation}
where the master integrals $I_{1,1,1,1,-1,0,0,0,0}$ and $I_{1,1,1,1,0,0,0,-1,0}$ decouple from the maximal cut in the same fashion as the last two master integrals in section \ref{sec:5.2}. Just as in the 3-1 configuration
we are left with a minimally coupled $4\times 4$ sector with the master integrals $I_{1,1,1,1,0,0,0,0,0}$, $I_{2,1,1,1,0,0,0,0,0}$, $I_{1,2,1,1,0,0,0,0,0}$ and $I_{3,1,1,1,0,0,0,0,0}$. 
Two coupled PF operators of the form in Eq.~(\ref{PFbananas}) yield three independent solutions $\psi_0$, $\psi_1^{(1)}$ and $\psi_1^{(2)}$, where we express $\psi_2$ in terms of the other periods by virtue of the quadratic relations in Eq.~(\ref{bilinears}).
Explicit expansions can be found in the ancillary file {\bf{\texttt banana\_2\_2.nb}}.
This basis satisfies the quadratic relations
\beq
\underline{\psi}^T\cdot\Sigma^\text{2-2}\cdot\underline{\psi}=0\,,
\eeq
with the intersection pairing
\beq
\Sigma^\text{2-2} = \begin{pmatrix}
 0 & 0 & 0 & 1 \\
 0 & -2 & -4 & 0 \\
 0 & -4 & -2 & 0 \\
 1 & 0 & 0 & 0 
 \end{pmatrix}.
 \eeq
\\
We can then define two independent moduli:
\begin{equation}
\tau_1=\dfrac{\psi_1^{(1)}}{\psi_0}\,,\quad\tau_2=\dfrac{\psi_1^{(2)}}{\psi_0}\,,
\end{equation}
and we can determine a change of variables through the Jacobian $\mathfrak{J}$:
\begin{equation}
    \begin{pmatrix}
    \text{d}z_1\\\text{d}z_2\end{pmatrix}=\begin{pmatrix}\mathfrak{J}_{1,1}&\mathfrak{J}_{1,2}\\
    \mathfrak{J}_{2,1}&\mathfrak{J}_{2,2}\end{pmatrix}\cdot \begin{pmatrix}\text{d}\tau_1\\\text{d}\tau_2\end{pmatrix}.
\end{equation}
Having defined the variables $\tau_1$ and $\tau_2$ we are now ready to set up the basis as proposed in section~\ref{sec:4}:
\begin{align}
M_0^\text{2-2}&=\epsilon^3 I_{1,0,1,1,0,0,0,0,0}\,,\notag\\
M_1^\text{2-2}&=\epsilon^3 I_{0,1,1,1,0,0,0,0,0}\,,\notag\\
M_2^\text{2-2}&=\epsilon^3 \dfrac{I_{1,1,1,1,0,0,0,0,0}}{\psi_0}\,,\notag\\
M_3^\text{2-2}&=\dfrac{1}{\epsilon} \dfrac{\partial}{\partial \tau_1}M_2^\text{2-2}+F_{3,2}^\text{2-2}M_2^\text{2-2}\,,\notag\\
M_4^\text{2-2}&=\dfrac{1}{\epsilon} \dfrac{\partial}{\partial \tau_2}M_2^\text{2-2}+F_{4,2}^\text{2-2}M_2^\text{2-2}\,,\notag\\
M_5^\text{2-2}&=\epsilon^3 I_{1,1,1,1,-1,0,0,0,0}+F_{5,2}^\text{2-2}M_2^\text{2-2}\,,\notag\\
M_6^\text{2-2}&=\epsilon^3 I_{1,1,1,1,0,0,0,-1,0}+F_{6,2}^\text{2-2}M_2^\text{2-2}\,,\notag\\
M_7^\text{2-2}&=\dfrac{1}{\epsilon} \dfrac{\partial}{\partial \tau_1}M_3^\text{2-2}+\sum_{j=0}^6F_{7,j}^\text{2-2}M_j^\text{2-2}\,,
\end{align}
where, as before, the new functions $F_{i,j}^\text{2-2}$ are a combination of algebraic functions multiplied by powers of the holomorphic period $\psi_0$ and (iterated) integrals of the same kind of functions. 
This basis brings the differential equation in canonical form, close to the MUM-point. We present explicit expressions (expanded in series around the MUM-point) for $F_{i,j}^\text{2-2}$ in the ancillary file {\bf{\texttt banana\_2\_2}}.
Note that, here as well, the expansions of the functions $F_{i,j}^\text{2-2}$ have \textit{integer} coefficients.

\subsection{Banana four-loop with two internal unequal masses}\label{sec:banana_4}
In this section, we apply the algorithm from section~\ref{sec:4}, in particular from section~\ref{genericansatzCY3fold}, to get a canonical form for the differential equation of a four-loop integral with an underlying Calabi-Yau three-fold geometry.
We consider the four-loop banana family:
\begin{equation}
    I_{\nu_1,\nu_2,\dots,\nu_9}(p^2,m_{j_1}^2,\dots,m_{j_4}^2,m_{j_5}^2;\epsilon)=e^{4\gamma_\text{E} \epsilon}\int \dfrac{\text{d}^D l_1}{i\pi^{\frac{D}{2}}}\dfrac{\text{d}^D l_2}{i\pi^{\frac{D}{2}}} \dfrac{\text{d}^D l_3}{i\pi^{\frac{D}{2}}}\dfrac{\text{d}^Dl_4}{i\pi^{\frac{D}{2}}} \dfrac{1}{D_1^{\nu_1} D_2^{\nu_2}\dots D_{14}^{\nu_{14}}}\,,
\end{equation}
with the denominators $D_j$:
\begin{align}
    D_1&=l_1^2-m_{j_1}^2\,,\quad D_2=l_2^2-m_{j_2}^2\,,\quad D_3=l_3^2-m_{j_3}^2\,,\quad
    D_4=l_4^2-m_{j_4}^2\,,\notag\\
    D_5&=(l_1+l_2+l_3+l_4-p)^2-m_{j_5}^2\,,\quad D_6=(l_1-p)^2\,,\quad D_7=(l_2-p)^2\,,\notag\\
    D_8&=(l_3-p)^2\,,\quad D_9=(l_4-p)^2\,,\quad D_{10}=(l_1-l_2)^2\,,\quad D_{11}=(l_1-l_3)^2\,,\notag\\
    D_{12}&=(l_1-l_4)^2\,,\quad D_{13}=(l_2-l_3)^2\,,\quad D_{14}=(l_2-l_4)^2\,.
\end{align}
In the case with two unequal masses, there are two possible mass configurations, namely the 4-1 configuration where we set:
\begin{equation}\label{4-1 config}
    m_{j_1}=m_{j_2}=m_{j_3}=m_{j_4}=m_1\,,\quad m_{j_5}=m_2\,,
\end{equation}
and the 3-2 configuration where we set:
\begin{equation}\label{3-2 config}
m_{j_1}=m_{j_2}=m_{j_3}=m_1\,,\quad m_{j_4}=m_{j_5}=m_2\,.
\end{equation}
We will only analyse the 4-1 configuration, we expect similar results for the 3-2 configuration. In the following, we will work with the dimensionless kinematic parameters:
\begin{equation}
    z_1=\dfrac{m_1^2}{p^2}\,,\quad z_2=\dfrac{m_2^2}{p^2}\,.
\end{equation}
From the differential equation that we generate with \textsc{Kira 2.0}~\cite{Klappert:2020nbg}, we find {\it nine} master integrals among which we find two tadpoles:
\begin{equation}
    I_{1, 1, 1, 1, 0, 0, 0, 0, 0, 0, 0, 0, 0, 0}\,,\quad I_{1, 1, 1, 0, 1, 0, 0, 0, 0, 0, 0, 0, 0, 0}\,,
\end{equation}
and one decoupling master integral:
\begin{equation}
    I_{1, 1, 1, 1, 1, -1, 0, 0, 0, 0, 0, 0, 0, 0}\,.
\end{equation}
The remaining six master integrals form a
 system of coupled differential equations for the master integral $I_{1, 1, 1, 1, 1, 0, 0, 0, 0, 0, 0, 0, 0, 0}$  which have (on the maximal cut) operators of PF type for which we find, at $\epsilon=0$, {\it six} independent solutions:
 \begin{equation}
    \underline{\psi}(\mathbf{z})=\bigl(\psi_0(\mathbf{z}),\psi_{1}^{(1)}(\mathbf{z}),\psi_{1}^{(2)}(\mathbf{z}),\psi_2^{(2)}(\mathbf{z}),\psi_2^{(1)}(\mathbf{z}),\psi_3(\mathbf{z})\bigr)^T\,,
\end{equation}
that satisfy the quadratic equations $\underline{\psi}^T\cdot\Sigma^\text{4-1}\cdot\underline{\psi}=0$, with $\Sigma^\text{4-1}$ antidiagonal as in Eq.~(\ref{sigmaCY3}).\footnote{We will from now on leave the dependence on $\mathbf{z}$ implicit.}
The periods expanded around the MUM-point are available in the ancillary file {\bf{\texttt banana\_4\_1.nb}}.
The notation is motivated as follows: since the point $(z_1,z_2)=(0,0)$ is a MUM-point, we find a Frobenius basis which has one holomorphic period $\psi_0$, two single log periods $\psi_1^{(1)}$, $\psi_1^{(2)}$, two double log periods $\psi_2^{(1)}$, $\psi_2^{(2)}$ and one triple log period $\psi_3$.
This corresponds to the structure of a Calabi-Yau threefold with the following Hodge numbers for its middle cohomology:
\begin{equation}\label{hodge_CY}
    h^{(0,3)}=h^{(3,0)}=1\,,\quad h^{(2,1)}=h^{(1,2)}=2\,.
\end{equation}
Therefore the maximal cut of the master integral of interest $I_{1, 1, 1, 1, 1, 0, 0, 0, 0, 0, 0, 0, 0, 0}$, in $D=2$, takes the form:
\begin{equation}\label{CY_start}
\begin{aligned}
    \mathrm{MaxCut}\,I_{1, 1, 1, 1, 1, 0, 0, 0, 0, 0, 0, 0, 0, 0}=&\biggl\{ \delta_{j,1}\,\psi_0+\delta_{j,2}\,\psi_1^{(1)}+\delta_{j,3}\,\psi_1^{(2)}+\delta_{j,4}\,\psi_2^{(1)}\\
    &\quad+\delta_{j,5}\,\psi_2^{(2)}+\delta_{j,6}\,\psi_3+\sum_{k\geq 1}\epsilon^k f_k^{(j)}\,;\quad j=\{1,2,3,4,5,6\}\biggr\}\,,
\end{aligned}
\end{equation}
where the functions $f_k^{(j)}$ are rational in the kinematic variables $\mathbf{z}$. 
This corresponds to the integral of the first kind differential in the limit $\epsilon=0$ along the independent contours.
\\After defining the complex structure moduli as usual:
\begin{equation}
    \tau_1=\dfrac{\psi_1^{(1)}}{\psi_0}\,,\qquad  \tau_2=\dfrac{\psi_1^{(2)}}{\psi_0}\,,
\end{equation}
we use Eq.~(\ref{C_def_1}) to define the Yukawa three-couplings.
Since the derivatives in Eq.~(\ref{C_def_1}) commute, we find that there are {\it four} possible couplings: $C_{1,1,1}$, $C_{1,1,2}$, $C_{1,2,2}$ and $C_{2,2,2}$.
Additionally, when unpacking Eq.~(\ref{C_def_1}), one finds the following simpler equations for the couplings:
\begin{equation}
    \dfrac{\partial}{\partial \tau_i}\dfrac{\partial}{\partial \tau_j}\dfrac{\psi_2^{(k)}}{\psi_0}=-C_{i,j,k}\,,\quad  \dfrac{\partial}{\partial \tau_i}\dfrac{\partial}{\partial \tau_j}\dfrac{\psi_3}{\psi_0}=\tau_1\,C_{i,j,1}+\tau_2\,C_{i,j,2}\,.
\end{equation}
Through these expressions we can now introduce the {\it pre-potential} $\mathcal{F}$, as in Eq.~(\ref{prepotential}), which is related to the normalised periods as follows:
\begin{equation}
   \dfrac{\psi_2^{(j)}}{\psi_0}=\dfrac{\partial}{\partial \tau_j}\mathcal{F}\,,\quad\quad \dfrac{\psi_3}{\psi_0}=2\mathcal{F}-\tau_1 \dfrac{\partial}{\partial \tau_1}\mathcal{F}-\tau_2  \dfrac{\partial}{\partial \tau_2}\mathcal{F},
\end{equation}
and has the property:
\begin{equation}
    \dfrac{\partial}{\partial \tau_i}\dfrac{\partial}{\partial \tau_j}\dfrac{\partial}{\partial \tau_k}\mathcal{F}=C_{i,j,k}\,.
\end{equation}
With this compact notation, we can now define the master integral $M_2^\text{4-1}$ which normalises Eq.~(\ref{CY_start}) by the holomorphic period $\psi_0$ and express it as:
\begin{align}\label{norm_M3}
    M_2^\text{4-1}&=\dfrac{I_{1, 1, 1, 1, 1, 0, 0, 0, 0, 0, 0, 0, 0, 0}}{\psi_0}\notag\\&
    =\biggl\{ \delta_{j,1}+\delta_{j,2}\tau_1+\delta_{j,3}\tau_2+\delta_{j,4}\dfrac{\partial}{\partial \tau_1}\mathcal{F}+\delta_{j,5}\dfrac{\partial}{\partial \tau_2}\mathcal{F}\notag\\
    &\quad\quad+\delta_{j,6}\biggl(2\mathcal{F}-\tau_1 \dfrac{\partial}{\partial \tau_1}\mathcal{F}-\tau_2  \dfrac{\partial}{\partial \tau_2}\mathcal{F}\biggr)+\sum_{k\geq 1}\epsilon^k \dfrac{f_k^{(j)}}{\psi_0}\,;\quad j=\{1,2,3,4,5,6\}\biggr\}\,.
\end{align}
The other five master integrals can be defined as proposed in section~\ref{genericansatzCY3fold}. 
\\Alternatively, we can notice that if we want to find the differential equation of Eq.~(\ref{GMCY3fold}),
apart from $M_2^\text{4-1}$, $\frac{1}{\epsilon}\partial_{\tau_1}M_2^\text{4-1}$ and $\frac{1}{\epsilon}\partial_{\tau_2}M_2^\text{4-1}$, we need to look for master integrals which are non-vanishing when $\epsilon= 0$, but whose derivative $\partial_{\tau_1}$ and $\partial_{\tau_2}$ either vanish or are proportional to a constant in this limit (after the appropriate $\epsilon$-rescaling).
We find that the following master integrals have this property:
\begin{equation}\label{M_9}
    \dfrac{\partial}{\partial \tau_1}\underbrace{\biggl(\mathcal{Y}_1\dfrac{\partial}{\partial \tau_1}\dfrac{\partial}{\partial \tau_1}-\mathcal{Y}_2\dfrac{\partial}{\partial \tau_2}\dfrac{\partial}{\partial \tau_1}\biggr)M_2^\text{4-1}}_{\text{first candidate}}=\biggl\{ \delta_{j,6}+\mathcal{O}(\epsilon)\,,\quad j=\{1,2,3,4,5,6\}\biggr\}\,,
\end{equation}
and
\begin{equation}
       \dfrac{\partial}{\partial \tau_2}\underbrace{\biggl(\mathcal{Y}_3\dfrac{\partial}{\partial \tau_1}\dfrac{\partial}{\partial \tau_2}-\mathcal{Y}_4\dfrac{\partial}{\partial \tau_2}\dfrac{\partial}{\partial \tau_2}\biggr)M_2^\text{4-1}}_{\text{second candidate}}=\biggl\{ \delta_{j,6}+\mathcal{O}(\epsilon)\,,\quad j=\{1,2,3,4,5,6\}\biggr\}\,,\label{M10}
\end{equation}
where we introduce the following rational functions $\mathcal{Y}_j$ of the Yukawa three-couplings (which can be derived using Eq.~(\ref{find_Y})):
\begin{equation}\label{Ys}
\begin{split}
    \mathcal{Y}_1&=\dfrac{C_{1,2,2}}{C_{1,1,1}C_{1,2,2}{-}C_{1,1,2}^2}\,,\quad \mathcal{Y}_2=\dfrac{C_{1,1,2}}{C_{1,1,1}C_{1,2,2}{-}C_{1,1,2}^2}\,,\notag\\
    \mathcal{Y}_3&=\dfrac{C_{1,2,2}}{C_{1,2,2}^2{-}C_{1,1,2}C_{2,2,2}}\,,\quad \mathcal{Y}_4=\dfrac{C_{1,1,2}}{C_{1,2,2}^2{-}C_{1,1,2}C_{2,2,2}}\,.
\end{split}
\end{equation}
This choice of these two master integrals coincides with what we get by using the basis from section~\ref{genericansatzCY3fold}.
We can choose to define $M_8^\text{4-1}$ to be either of Eqs.~(\ref{M_9}) and (\ref{M10}). 
Choosing $M_8^\text{4-1}=$(\ref{M_9}), we have constructed a canonical basis as in section~\ref{genericansatzCY3fold}.
When considering the maximal cut at $\epsilon=0$ the differential equation takes the form of Eq.~(\ref{GMCY3fold}):
\begin{equation}\label{frob_example}
 \dfrac{\partial}{\partial\tau_j}   \begin{pmatrix}M_2^\text{4-1}\\M_3^\text{4-1}\\M_4^\text{4-1}\\M_5^\text{4-1}\\M_6^\text{4-1}\\M_8^\text{4-1}\end{pmatrix}=\begin{pmatrix}0&\delta_{j,1}&\delta_{j,2}&0&0&0\\
 0&0&0&C_{1,1,j}&C_{1,2,j}&0\\
 0&0&0&C_{1,2,j}&C_{2,2,j}&0\\
 0&0&0&0&0&\delta_{j,1}\\
 0&0&0&0&0&\delta_{j,2}\\
 0&0&0&0&0&0\end{pmatrix}\cdot \begin{pmatrix}M_2^\text{4-1}\\M_3^\text{4-1}\\M_4^\text{4-1}\\M_5^\text{4-1}\\M_6^\text{4-1}\\M_8^\text{4-1}\end{pmatrix}\,.
\end{equation}
where the master integrals come from the following basis: 
\begin{align}
M_0^\text{4-1}&=\epsilon^4\,I_{1, 1, 1, 1, 0, 0, 0, 0, 0, 0, 0, 0, 0, 0}\,,\notag\\
M_1^\text{4-1}&=\epsilon^4\,I_{1, 1, 1, 0, 1, 0, 0, 0, 0, 0, 0, 0, 0, 0}\,,\notag\\
M_2^\text{4-1}&=\epsilon^4\,\dfrac{I_{1, 1, 1, 1, 1, 0, 0, 0, 0, 0, 0, 0, 0, 0}}{\psi_0}\,,\notag\\
M_3^\text{4-1}&=\dfrac{1}{\epsilon} \dfrac{\partial}{\partial \tau_1}M_2^\text{4-1}+F_{3,2}^\text{4-1}M_2^\text{4-1}\,,\notag\\
M_4^\text{4-1}&=\dfrac{1}{\epsilon} \dfrac{\partial}{\partial \tau_2}M_3^\text{4-1}+F_{4,2}^\text{4-1}M_2^\text{4-1}\,,\notag\\
M_5^\text{4-1}&=\dfrac{1}{\epsilon}\biggl(\mathcal{Y}_1\dfrac{\partial}{\partial \tau_1}-\mathcal{Y}_2 \dfrac{\partial}{\partial \tau_2}\biggr)M_3^\text{4-1}+(\mathcal{Y}_1-\mathcal{Y}_2 )\sum_{j=2}^4F_{5,j}^\text{4-1}M_j^\text{4-1}\,,\notag\\
M_6^\text{4-1}&=\dfrac{1}{\epsilon}\biggl(\mathcal{Y}_3\dfrac{\partial}{\partial \tau_1}-\mathcal{Y}_4 \dfrac{\partial}{\partial \tau_2}\biggr)M_4^\text{4-1}+(\mathcal{Y}_3-\mathcal{Y}_4 )\sum_{j=2}^4F_{6,j}^\text{4-1}M_j^\text{4-1}\,,\notag\\
M_7^\text{4-1}&=\epsilon^4\,I_{1, 1, 1, 1, 1, -1, 0, 0, 0, 0, 0, 0, 0, 0}+F_{7,2}^\text{4-1}M_2^\text{4-1}\,,\notag\\
M_8^\text{4-1}&=\dfrac{1}{\epsilon} \dfrac{\partial}{\partial \tau_1}M_5^\text{4-1}+\sum_{j=0}^7F_{8,j}^\text{4-1}M_j^\text{4-1}\,.
\end{align}
The integrals $M_0^\text{4-1}$, $M_1^\text{4-1}$ correspond to tadpoles and $M_7^\text{4-1}$ corresponds to a decoupling master integral.
The derivative structure of Eq.~(\ref{frob_example}) is preserved but we introduce the additional functions $F^\text{4-1}_{i,j}$ to $\epsilon$-factorise the strictly lower triangular part of the matrix in the fashion of section~\ref{genericansatzCY3fold}. With the additional functions, the differential equation is in canonical form, close to the MUM-point.
We once more display all results in the ancillary file {\bf{\texttt banana\_4\_1.nb}}.
Note that all the functions $F_{i,j}$ are holomorphic and have {\it integer} coefficients when expanded at the MUM-point $(z_1,z_2)=(0,0)$.

\section{Conclusion}
\label{conclusion}
In this paper, we have presented a way to construct bases of master integrals which bring the differential equations of multi-scale Feynman integrals with an underlying  Calabi-Yau geometry into canonical form. This is a multi-scale generalisation of ref.~\cite{Pogel:2022vat}, and it corresponds to the first application of the method introduced in refs.~\cite{Gorges:2023zgv,Duhr:2025lbz} to multi-scale Calabi-Yau Feynman integrals.
We have applied this construction to the already well-studied unequal masses sunrise integral and have found the same result as ref.~\cite{Bogner:2019lfa} up to a constant gauge transformation.
\\We have shown that we can apply this method to new integrals, such as three-loop integrals and a four-loop integral, all dependent on two scales.
Remarkably, we find that for the examples we have computed, the basis we propose yields a closed connection with at most simple poles in each of the variables.
 We believe that a basis of this form can be used for any block which can be rewritten as a PF system annihilating the periods of Calabi-Yau manifolds.
 The study of decoupling integrals might need to be appropriately modified when the decoupling integrals have poles of order two or higher~\cite{Duhr:2025lbz}.
 \\One key take-away is that a thorough understanding of the geometry associated to the studied sector is needed to build a sensible basis. 
 We expect that for more complicated integrals, a single sector may be able to mix different geometries and we believe that, choosing a basis of master integrals according to the mixed Hodge structure, may help to disentangle into sub-blocks the differential equation, displaying the properties we observe in banana integrals.
 \\While some recent progress has been made into understanding the differential forms associated to the three-loop banana integrals in canonical form~\cite{Duhr:2025ppd,Duhr:2025tdf}, further investigations need to be made, in particular for higher dimensional Calabi-Yau manifolds. Knowing closed forms for the periods could also help us understand when the coupled differential equations for the additional functions $F_{i,j}$ can be solved.
 \\As banana integrals have a simple structure, they were a good playground to understand Feynman integrals with underlying multi-scale Calabi-Yau geometry, it would now be natural to probe more complicated graphs with such geometries.

\acknowledgments

We thank  Claude Duhr, Mathieu Giroux, Albrecht Klemm, Christoph Nega, Sebastian P\"ogel, Andrzej Pokraka, Franziska Porkert, Oliver Schlotterer, Sven Stawinski, Lorenzo Tancredi, Fabian Wagner and Stefan Weinzierl for fruitful discussions. 
In particular, we are grateful to Janis Dücker and Julian Piribauer for an in depth explanation of the Frobenius algebra and of their results later published in ref.~\cite{Ducker:2025wfl}. We also thank Claude Duhr, Christoph Nega, Oliver Schlotterer, Lorenzo Tancredi and Fabian Wagner for useful comments on the manuscript.
SM wishes to thank Uppsala University for its hospitality. YS wishes to thank the BCTP in Bonn for its hospitality and the Elliptics Summer School 2023 where a lecture by Stefan Weinzierl partly inspired this work. The authors thank the 6th School of Analytic Computing in High-Energy and Gravitational Theoretical Physics where the collaboration started.
\\YS acknowledges support from the Centre for Interdisciplinary Mathematics at Uppsala University and partial support by the European Research Council under ERC- STG-804286 UNISCAMP.
\\This work was co-funded by
the European Union (ERC Consolidator Grant LoCoMotive 101043686 (SM)). Views and opinions expressed are
however those of the author(s) only and do not necessarily reflect those of the European
Union or the European Research Council. Neither the European Union nor the granting
authority can be held responsible for them.

\appendix

\section{Democratic and decoupling approach for elliptic Feynman integrals}\label{sec:7}
Here we argue why the two bases proposed in section \ref{ellipticgenericansatz} work. 
The difference of these approaches lies in the choice of starting master integrals and consequently on the Picard-Fuchs operators which annihilate the maximal cut. 
Hence the properties of PF operators are essential, from which we construct the bases.
We will give an explicit derivation for the sunrise integral with two unequal masses.
\subsection{Democratic approach}
We start by choosing the following master integrals:
\begin{align}
    \mathbf{I}=\begin{pmatrix}I_{1,1,1,0,0}\\\partial_{x_1}I_{1,1,1,0,0}\\\partial_{x_2}I_{1,1,1,0,0}\end{pmatrix}\,,
\end{align}
where the propagators are as defined in Eq.~(\ref{propagatorsSunrise}), with $m_3^2=m_2^2$. $I_{1,1,1,0,0}$ is such that the integrand of its maximal cut for $\epsilon=0$ is a differential of the first kind.
We can compute three Picard-Fuchs operators that annihilate the master integral $I_{1,1,1,0,0}$. 
They have the following form (recall that we define $x_1=\frac{m_1^2}{s}$ and $x_2=\frac{m_2^2}{s}$, which in practice leads to setting $s=1$) 
\begin{align}
    L^{(2)}_{(x_1,x_1)}&=\dfrac{\partial^2}{\partial x_1^2}+q^{(0)}_{1\,(x_1,x_1)}\dfrac{\partial}{\partial x_1}+q^{(0)}_{2\,(x_1,x_1)}\dfrac{\partial}{\partial x_2}+q^{(0)}_{(x_1,x_2)}\notag\\
    &\quad+\epsilon \left(q^{(1)}_{1\,(x_1,x_1)}\dfrac{\partial}{\partial x_1}+q^{(1)}_{2\,(x_1,x_1)}\dfrac{\partial}{\partial x_2}+q^{(1)}_{(x_1,x_1)}\right)+\epsilon^2\, q^{(2)}_{(x_1,x_1)}\,,\nonumber\\
    L^{(2)}_{(x_2,x_2)}&=\dfrac{\partial^2}{\partial x_2^2}+q^{(0)}_{1\,(x_2,x_2)}\dfrac{\partial}{\partial x_1}+q^{(0)}_{2\,(x_2,x_2)}\dfrac{\partial}{\partial x_2}+q^{(0)}_{(x_2,x_2)}\notag\\
    &\quad+\epsilon \left(q^{(1)}_{1\,(x_2,x_2)}\dfrac{\partial}{\partial x_1}+q^{(1)}_{2\,(x_2,x_2)}\dfrac{\partial}{\partial x_2}+q^{(1)}_{(x_2,x_2)}\right)+\epsilon^2\, q^{(2)}_{(x_2,x_2)}\,,\nonumber\\
    L^{(2)}_{(x_1,x_2)}&=\dfrac{\partial^2}{\partial x_1\partial x_2}+q^{(0)}_{1\,(x_1,x_2)}\dfrac{\partial}{\partial x_1}+q^{(0)}_{2\,(x_1,x_2)}\dfrac{\partial}{\partial x_2}+q^{(0)}_{(x_1,x_2)}\notag\\
    &\quad+\epsilon \left(q^{(1)}_{1\,(x_1,x_2)}\dfrac{\partial}{\partial x_1}+q^{(1)}_{2\,(x_1,x_2)}\dfrac{\partial}{\partial x_2}+q^{(1)}_{(x_1,x_2)}\right)+\epsilon^2\, q^{(2)}_{(x_1,x_2)}\,.
    \label{diffops}
\end{align}
We can split all the differential operators as
\begin{equation}
    L_{(x_i,x_j)}^{(2)}=L_{0\,(x_i,x_j)}^{(2)}+L_{\epsilon\,(x_i,x_j)}^{(1)}\,,
\end{equation}
where $L_{0\,(x_i,x_j)}^{(2)}$ is the $\epsilon^0$ part and $L_{\epsilon\,(x_i,x_j)}^{(1)}$ contains all powers of $\epsilon$.
While $L_{0\,(x_i,x_j)}^{(2)}$ is a differential operator of the same order as $L_{(x_i,x_j)}^{(2)}$, the operator $L_{\epsilon\,(x_i,x_j)}^{(1)}$ is of one order less (in this particular case it is of first order).
\\
Let us now focus on $L_{0\,(x_i,x_j)}^{(2)}$. The system of differential equations
\begin{equation}
    \begin{cases}
      L_{0\,(x_1,x_1)}^{(2)}\psi=0\,,\\
      L_{0\,(x_2,x_2)}^{(2)}\psi=0\,,\\
      L_{0\,(x_1,x_2)}^{(2)}\psi=0\,,
    \end{cases}\,
\end{equation}
has three independent solutions that we denote by $\psi_0, \psi_1, \phi$. 
Since we are working with a Calabi-Yau one-fold (an elliptic curve), we are able to find solutions written in a closed form. 
Note that for other geometries this is not always possible, in which case we expand the solution at a convenient point using the method of Frobenius. 
The canonical variables on the moduli space of punctured tori $\mathcal{M}_{1,2}$ are:
\begin{equation}
\label{canonicalVariables}
    \tau=\dfrac{\psi_1}{\psi_0}\,, \qquad z_1=\dfrac{\phi}{\psi_0}\,,
\end{equation}
where the map from $x_1,x_2$ to $\tau$ is the \textit{mirror} map and the map from $x_1,x_2$ to $z_1$ is \textit{Abel}'s map from Eq.~(\ref{Abelsmap}).
It turns out that, with this choice of variables, the differential operator greatly simplify:
\begin{align}
    \tilde{L}_{0\,(\tau,\tau)}^{(2)}&=\dfrac{\partial^2}{\partial\tau^2}\dfrac{1}{\psi_0}\,,\label{PF_ell_1}\\
   \tilde{L}_{0\,(z_1,z_1)}^{(2)}&=\dfrac{\partial^2}{\partial z_1^2}\dfrac{1}{\psi_0}\,,\label{PF_ell_2}\\
    \tilde{L}_{0\,(\tau,z_1)}^{(2)}&=\dfrac{\partial^2}{\partial\tau \partial z_1}\dfrac{1}{\psi_0}\,.\label{PF_ell_3}
\end{align}
These, to us, are the most natural generalisations of the Calabi-Yau operators from Eq.~(\ref{eq:normalform}) for multi-parameter periods.
We can now define our new master integrals in the following way
\begin{align}
    M_0&=\dfrac{I_{1,1,1,0,0}}{\psi_0}\,,\label{ansM_0}\\
    M_1&=\dfrac{1}{\epsilon}\dfrac{\partial}{\partial \tau}M_{0}+F_{1,0}M_0\,,\label{ans_2}\\
    M_2&=\dfrac{1}{\epsilon} \dfrac{\partial}{\partial z_1}M_{0}+F_{2,0}M_0\,,\label{ans_3}
\end{align}
We now want to check that this new basis leads to an $\epsilon$-factorised differential equation:
\begin{equation}
    \dfrac{\partial}{\partial \tau}\begin{pmatrix}M_0\\M_1\\M_2\end{pmatrix}=\epsilon\,\tilde{\mathbf{A}}_\tau\cdot\begin{pmatrix}M_0\\M_1\\M_2\end{pmatrix}\,,\qquad \dfrac{\partial}{\partial z_1}\begin{pmatrix}M_0\\M_1\\M_2\end{pmatrix}=\epsilon\,\tilde{\mathbf{A}}_{z_1}\cdot\begin{pmatrix}M_0\\M_1\\M_2\end{pmatrix}\,.\label{ans_mat}
\end{equation}
The basis in Eqs.~(\ref{ansM_0}), (\ref{ans_2}) and (\ref{ans_3}) straightforwardly implies:
\begin{align}
    \dfrac{\partial}{\partial \tau}M_{0}&=\epsilon(M_1-F_{1,0}M_0)\,,\\
     \dfrac{\partial}{\partial z_1}M_{0}&=\epsilon(M_2-F_{2,0}M_0)\,,
\end{align}
which make the first row of Eq.~(\ref{ans_mat}) $\epsilon$-factorised.
For row two and three we need to go back to handling PF operators, on the maximal cut we have 
\begin{align}
\tilde{L}^{(2)}I_{1,1,1,0,0}=0   \,,
\end{align}
which leads to
\begin{align}
\tilde{L}_{0}^{(2)}I_{1,1,1,0,0}+\tilde{L}_{\epsilon}^{(1)}I_{1,1,1,0,0}=0\,.
\end{align}
Plugging in the operators from Eqs.~(\ref{PF_ell_1}), (\ref{PF_ell_2}) and (\ref{PF_ell_3}) we find
\begin{align}
    \dfrac{\partial^2 }{\partial\tau^2}M_0&=-\tilde{L}_{\epsilon\,(\tau,\tau)}^{(1)}I_{1,1,1,0,0}\,,\\
    \dfrac{\partial^2 }{\partial z_1^2}M_0&=-\tilde{L}_{\epsilon\,(z_1,z_1)}^{(1)}I_{1,1,1,0,0}\,,\\
    \dfrac{\partial^2 }{\partial\tau    \partial z_1}M_0&=-\tilde{L}_{\epsilon\,(\tau,z_1)}^{(1)}I_{1,1,1,0,0}\,.
\end{align}
Therefore any second derivative acting on $M_0$ can be expressed in terms of first and zero derivative terms at higher orders in $\epsilon$. 
Let us now explicitly apply this to the term $\partial_\tau M_1$ and check whether this term is $\epsilon$-factorised.
\begin{align}\label{eps_fact_proof}
    \dfrac{\partial}{\partial\tau}M_1&=\dfrac{\partial}{\partial\tau}\biggl(\dfrac{1}{\epsilon}\dfrac{\partial}{\partial\tau}M_0+F_{1,0}M_0\biggr)\notag\\
    &=\dfrac{1}{\epsilon}\dfrac{\partial^2}{\partial\tau^2}M_0+\biggl(\dfrac{\partial}{\partial\tau}F_{1,0}\biggr)M_0+F_{1,0}\biggl(\dfrac{\partial}{\partial\tau}M_0\biggr)\notag\\
    &=-\dfrac{1}{\epsilon}\tilde{L}_{\epsilon\,(\tau,\tau)}^{(1)}I_{1,1,1,0,0} +\biggl(\dfrac{\partial}{\partial\tau}F_{1,0}\biggr)M_0+\epsilon\, F_{1,0}(M_1-F_{1,0}M_0)\,,
\end{align}
with
\begin{equation}
    \tilde{L}_{\epsilon\,(\tau,\tau)}^{(1)}=\epsilon\,\biggl(\tilde{q}_{1\,(\tau,\tau)}^{(1)}\dfrac{\partial}{\partial\tau}\dfrac{1}{\psi_0}+\tilde{q}_{2\,(\tau,\tau)}^{(1)}\dfrac{\partial}{\partial{z_1}}\dfrac{1}{\psi_0}+\tilde{q}_{(\tau,\tau)}^{(1)}\dfrac{1}{\psi_0}\biggr)+\epsilon^2\,\tilde{q}_{(\tau,\tau)}^{(2)}\dfrac{1}{\psi_0}\,,
\end{equation}
where we can compute the term
\begin{align}\label{L_1I_1}
    \dfrac{1}{\epsilon}\tilde{L}_{\epsilon\,(\tau,\tau)}^{(1)}I_{1,1,1,0,0}=\epsilon\,\biggl(\tilde{q}_{1\,(\tau,\tau)}^{(1)}\bigl(M_1-F_{1,0}M_0\bigr) +\tilde{q}_{2\,(\tau,\tau)}^{(1)}\bigl(M_2-F_{2,0}M_0\bigr)+\tilde{q}_{(\tau,\tau)}^{(2)}M_0\biggr)+\tilde{q}_{(\tau,\tau)}^{(1)}M_0\,.
\end{align}
We can now put together Eqs. (\ref{eps_fact_proof}) and (\ref{L_1I_1}) and find:
\begin{equation}
    \dfrac{\partial}{\partial\tau}M_1=\biggl(\dfrac{\partial}{\partial\tau}F_{1,0}\biggr)M_0-\tilde{q}_{(\tau,\tau)}^{(1)}M_0+\epsilon\,(\dots)
\end{equation} 
and therefore by requiring:
\begin{equation}
    \dfrac{\partial}{\partial\tau}F_{1,0}=\tilde{q}_{(\tau,\tau)}^{(1)}\,,
\end{equation}
we $\epsilon$-factorise this specific term. 
By doing the same exercise for $\partial_{z_1}M_1$, $\partial_{\tau}M_2$ and $\partial_{z_1}M_2$ we can fully fix $F_{1,0}$ and $F_{2,0}$ and find an $\epsilon$-factorised differential equation. The only difference with respect to a one parameter case is that each function $F_{i,j}$ is defined through one differential equation, while here there are more because we have to take different derivatives. For the examples that we have checked the coupled differential equations are solvable. We leave for future work understanding more general conditions.

\subsection{Decoupling approach}
We start by choosing the following master integrals:
\begin{align}\label{anS_0}
    \mathbf{I}=\begin{pmatrix}I_{1,1,1,0,0}\\\partial_{x_1}I_{1,1,1,0,0}\\I_{1,1,1,-1,0}\end{pmatrix}\,,
\end{align}
where the propagators are as defined in Eq.~(\ref{propagatorsSunrise}), with $m_3^2=m_2^2$.
We define the new master integrals as:
\begin{align}
    M_0&=\dfrac{I_{1,1,1,0,0}}{\psi_0}\,,\label{decouplingansatz1}\\
    M_1&=I_{1,1,1,-1,0}+F_{1,0} M_0\,,\label{decouplingansatz2}\\
    M_2&=\dfrac{1}{\epsilon}  \mathcal{J}_{1,1}\dfrac{\partial}{\partial x_1}M_0+F_{2,0} M_0+F_{2,1} M_1\,,
    \label{decouplingansatz}
\end{align}
where $\mathcal{J}_{1,1}=\frac{\partial x_1}{\partial \tau}$ is the first entry of the full Jacobian and $\tau$ is defined in Eq.~(\ref{canonicalVariables}). 
The \textit{decoupling approach} resembles the one variable case in the first step.
The idea is, in a first step to consider $x_2$ as constant, in this way we find that we no longer have the moving puncture on the torus $z_1$ of the democratic approach.
Instead the master integral $M_1$ can be seen as decoupled in the fashion of a differential of third kind satisfying Eq.~(\ref{diffeqdecoupling}).
The choice of master integrals reflects also on the Picard-Fuchs operators: in the \textit{democratic approach} the Picard-Fuchs operators annihilate $\psi_0, \psi_1$ and $\phi$, which can be used to fully characterise the moduli space $\mathcal{M}_{1,2}$ through Eq. (\ref{canonicalVariables}), leading to a $3\times 3$ fully coupled block and generalises to a $(n+1)\times(n+1)$ coupled block for the moduli space $\mathcal{M}_{1,n}$.
In the \textit{decoupling approach} we are considering here, we have a block of size $2\times2$ characterised by only \textit{two} solutions, the two periods $\psi_0$ and $\psi_1$, mimicking the one variable case.
This means that with the choice of starting basis of Eq. (\ref{anS_0}), leads to a set of Picard-Fuchs operators which only annihilate the periods $\psi_0$ and $\psi_1$. \\
Having mentioned the differences with respect to the previous approach, we now need to show how the differential equation $\epsilon$-factorises:
\begin{equation}
    \dfrac{\partial}{\partial x_1}\begin{pmatrix}M_0\\M_1\\M_2\end{pmatrix}=\epsilon\,\tilde{\mathbf{A}}_{x_1}\cdot\begin{pmatrix}M_0\\M_1\\M_2\end{pmatrix}\,,\qquad  \dfrac{\partial}{\partial x_2}\begin{pmatrix}M_0\\M_1\\M_2\end{pmatrix}=\epsilon\,\tilde{\mathbf{A}}_{x_2}\cdot\begin{pmatrix}M_0\\M_1\\M_2\end{pmatrix}\,.
\end{equation}
\paragraph{Derivative w.r.t. $x_1$:}
Let us start by computing $\partial_{x_1}\mathbf{M}$.
We straightforwardly obtain that for $M_0$ we have:
\begin{equation}
    \dfrac{\partial}{\partial x_1}M_0=\frac{\epsilon}{  \mathcal{J}_{1,1}}\,\bigl(M_2-F_{2,0}M_0-F_{2,1}M_1\bigr)
\end{equation}
and we see that $\partial_{x_1}M_0$ is $\epsilon$-factorised. 
For $M_2$ we proceed in the same way as in the democratic approach, noticing that the following differential operator:
\begin{equation}
      \mathcal{J}_{1,1}\dfrac{\partial}{\partial{x_1} }  \mathcal{J}_{1,1}\dfrac{\partial}{\partial{x_1}}\dfrac{1}{\psi_0}\,,
    \label{PF1var}
\end{equation}
is an elliptic Picard-Fuchs operator, as it annihilates the periods $\psi_0$ and $\psi_1$.
One can simply recycle the computation from Eq. (\ref{eps_fact_proof}).
The choice for the decoupling master integrals, $M_1$ comes from the idea introduced in~\cite{Gorges:2023zgv} where the authors lie out that in order to choose an independent candidate we can look at the $\epsilon=0$ limit of maximal cut of the integral.
Since $I_{1,1,1,0,0}$ and $\partial_{x_1}I_{1,1,1,0,0}$ don't have residues (as they are integrals of differentials of the first and second kind respectively), we can identify $I_{1,1,1,-1,0}$ as an integral over a differential of the third kind. 
The maximal cut at $\epsilon= 0$ of $I_{1,1,1,-1,0}$ reveals a non-vanishing residue. 
Moreover, we want to see the elliptic block in the differential equation matrix decouple at $\epsilon= 0$, so we choose $I_{1,1,1,-1,0}$ such that its differential equation reads as in Eq. (\ref{diffeqdecoupling}):
\begin{align}
\label{diffopthirdkind}
    \dfrac{\partial}{\partial{x_1}}I_{1,1,1,-1,0}&=\epsilon \,q^{(1)\,\text{dec}}_{(x_1)}I_{1,1,1,-1,0}+\biggl(q^{(0)}_{(x_1)}+\epsilon \,q^{(1)}_{(x_1)}+q^{(0)}_{1\,(x_1)}\dfrac{\partial}{\partial x_1}\biggr)I_{1,1,1,0,0}\,,\\
    \dfrac{\partial}{\partial{x_2}}I_{1,1,1,-1,0}&=\epsilon \,q^{(1)\,\text{dec}}_{(x_2)}I_{1,1,1,-1,0}+\biggl(q^{(0)}_{(x_2)}+\epsilon \,q^{(1)}_{(x_2)}+q^{(0)}_{1\,(x_2)}\dfrac{\partial}{\partial x_1}\biggr)I_{1,1,1,0,0}\,,\label{diffopthirdkind2}
\end{align}
so that at $\epsilon=0$, the master integral $I_{1,1,1,-1,0}$ doesn't couple to itself. 
So computing the $x_1$ derivative of $M_1$ we get
\begin{align}\label{decoup_proof1}
    \dfrac{\partial}{\partial{x_1}}M_1&=\dfrac{\partial}{\partial{x_1}}I_{1,1,1,-1,0}+\biggl(\dfrac{\partial}{\partial{x_1}}F_{1,0}\biggr) M_0+F_{1,0}\biggl(\dfrac{\partial}{\partial{x_1}}M_0\biggr)\notag\\
&=\biggl(q_{(x_1)}^{(0)}\psi_0 +q_{1\,(x_1)}^{(0)}\dfrac{\partial}{\partial x_1}\psi_0 +\dfrac{\partial}{\partial x_1}F_{1,0}\biggr)M_0\\
&\quad+\epsilon\,\biggl\{\dfrac{1}{  \mathcal{J}_{1,1}}\biggl(q_{1\,(x_1)}^{(0)}\psi_0+F_{1,0}\biggr)\biggl(M_2-F_{2,0}M_0-F_{2,1}M_1\biggr)\notag\\
&\qquad\qquad\qquad\qquad\qquad\qquad\qquad+q^{(1)\,\text{dec}}_{(x_1)}\biggl(M_1-F_{1,0}M_0\biggr)+q_{(x_1)}^{(1)}\psi_0M_0\biggr)\biggr\}\notag\,,
\end{align}
which leads to an $\epsilon$-form if we require:
\begin{equation}
     \dfrac{\partial}{\partial{x_1}}F_{1,0}=- q^{(0)}_{(x_1)} \psi_0-q_{1\,(x_1)}^{(0)} \dfrac{\partial}{\partial{x_1}}\psi_0\,.
\end{equation}
\paragraph{Derivative w.r.t. $x_2$:} 
Let us now look at $\partial _{x_2}\mathbf{M}$. 
Let us start from the differential operator
\begin{equation} \label{newpf}
    \dfrac{\partial}{\partial z_1}\dfrac{1}{\psi_0}\,,
\end{equation}
where we use definition of Eq. (\ref{canonicalVariables}) for $z_1$. 
This operator annihilates the periods $\psi_0$ and $\psi_1$ but not $\phi$ (which was have interpreted as a puncture on the torus as $z_1=\frac{\phi}{\psi_0}$).
Therefore Eq. (\ref{newpf}) is a Picard-Fuchs operator for the $2\times2$ elliptic block: 
\begin{align}
    \biggl(\dfrac{\partial}{\partial {z_1}}\dfrac{1}{\psi_0}+ \epsilon\,q^{(1)}_{(z_1)}\biggr)I_{1,1,1,0,0}=0
\end{align}
which by using the entries of the jacobian $\partial_{z_1}=\mathfrak{J}_{2,1}\partial_{x_1}+\mathfrak{J}_{2,2}\partial_{x_2}$ we find
\begin{equation}
     \biggl( \mathfrak{J}_{2,1}\dfrac{\partial}{\partial{x_1}}\dfrac{1}{\psi_0}+ \mathfrak{J}_{2,2}\dfrac{\partial}{\partial{x_2}}\dfrac{1}{\psi_0}+ \epsilon\,q^{(1)}_{(z_1)}\biggr)I_{1,1,1,0,0}=0\,,
\end{equation}
which we can reorganise into 
\begin{align}
\dfrac{\partial}{\partial{x_2}}M_0&=-\epsilon \,\dfrac{q^{(1)}_{(z_1)}}{ \mathfrak{J}_{2,2}} I_{1,1,1,0,0}-\dfrac{ \mathfrak{J}_{2,1}}{ \mathfrak{J}_{2,2}}\dfrac{\partial}{\partial{x_1}}M_0\notag\\
&=\epsilon \biggl\{-\dfrac{q^{(1)}_{(z_1)}\psi_0}{ \mathfrak{J}_{2,2}} M_0-\dfrac{ \mathfrak{J}_{2,1}}{ \mathfrak{J}_{2,2}  \mathcal{J}_{1,1}}(M_2-F_{2,0}M_0-F_{2,1}M_1)\biggr\}\,,
\end{align}
and is therefore $\epsilon$-factorised.
\\Next, let us check whether $\partial_{x_2}M_1$ is $\epsilon$-factorised.
By recalling Eq. (\ref{diffopthirdkind2}) and carrying out a calculation analogous to Eq. (\ref{decoup_proof1}) we find that the term is $\epsilon$-factorised by requiring:
\begin{equation}
     \dfrac{\partial}{\partial{x_2}}F_{1,0}=- q^{(0)}_{(x_2)} \psi_0-q_{1\,(x_2)}^{(0)} \dfrac{\partial}{\partial{x_1}}\psi_0\,.
\end{equation}
Let us now look at the last master integral $M_2$. 
This master integral is for the same arguments as for $M_0$ annihilated by the following PF operator:
\begin{align}
    \dfrac{\partial}{\partial{z_1}}   \mathcal{J}_{1,1}\dfrac{\partial}{\partial{x_1}}\dfrac{1}{\psi_0}=\mathfrak{J}_{2,1}\dfrac{\partial}{\partial{x_1}}   \mathcal{J}_{1,1}\dfrac{\partial}{\partial{x_1}}\dfrac{1}{\psi_0}+\mathfrak{J}_{2,2}\dfrac{\partial}{\partial{x_2}}  \mathcal{J}_{1,1}\dfrac{\partial}{\partial{x_1}}\dfrac{1}{\psi_0}\,.
\end{align}
We can now follow the same steps as for $\partial_{x_2}M_0$ to find that $\partial_{x_2}M_2$ is $\epsilon$-factorised and we have shown that
\begin{equation}
    \dfrac{\partial}{\partial x_1}\mathbf{M}\mathrm{d}x_1+\dfrac{\partial}{\partial x_2}\mathbf{M}\mathrm{d}x_2=\epsilon\,\tilde{\mathbf{A}}\cdot\mathbf{M}\,.
\end{equation}
\subsection{Extensions to more variables}
Both democratic and decoupling approaches can be used for any Feynman integral with underlying elliptic geometry, as we did not use any specific properties of a specific diagram except for its Picard-Fuchs operators. 
We want to stress that, in the \textit{democratic approach}, it is essential to choose a starting master integral basis such that all periods $\psi_0,\psi_1,\phi$ (which includes $\phi$ which is later identified with a marked point) are annihilated by the set of  differential operators in Eq. (\ref{diffops}).
Thus generalisations to more variables will have to mirror this behaviour: we need to find a set of PF operators which also annihilate the $\phi_n$ marked points. 
In the \textit{decoupling approach}, we need to choose a starting integral basis that can be rewritten into higher-order differential equations with a PF operator of the form of Eq. (\ref{PF1var}), that annihilates only the periods $\psi_0$ and $\psi_1$.
In addition in this approach we will get PF operators of the form of Eq. (\ref{newpf}) (with several of them when we increase the number of variables). 
Note that the integrals that are annihilated by the operators of Eqs. (\ref{diffops}), (\ref{PF1var}) and (\ref{newpf}) are integrals whose maximal cuts for $\epsilon=0$ is an integral of the first kind.

\section{Explicit proof for K3 Feynman integrals}
\label{proofK3}
Here we show how we get an $\epsilon$-form for our basis presented in section~\ref{genericK3ansatz}.
First we notice that Eq. (\ref{ansatzK3eps0}), but also Eqs. (\ref{GMCY3fold}) and (\ref{genericGM}), straightforwardly generalise the single variable differential equation of Calabi-Yau systems to the multi-variable case since in the one variable case it reads~\cite{Duhr:2025lbz}:
\begin{equation}
    \begin{pmatrix}
        0 & 1 & 0 &\ldots & & &  \\
        & 0 & Y_2(\tau) & 0 &\ldots & &  \\
        & \ldots& 0 & Y_3(\tau) & 0 &\ldots &  \\
        & & \ldots& 0 & \ddots & 0 &\ldots  \\
        & & & \ldots& 0 & Y_2(\tau) & 0  \\
        & & & & \ldots& 0 & 1  \\
        & & & & & \ldots& 0  \\
    \end{pmatrix} \, ,
\end{equation}
which for a K3 specialises to 
\begin{equation}
    \begin{pmatrix}
        0&1&0\\
        0&0&1\\
        0&0&0
    \end{pmatrix}\,.
\end{equation}
We will now go through the details of how this choice gives an $\epsilon$-factorised matrix. 
The argument is slightly different from the case $l=1$ in appendix \ref{sec:7}, as we do not find any punctures but can instead define canonical variables through the mirror map.
The derivation we shall provide now for $l=2$ can however easily be uplifted to $l> 2$. 
We begin by considering the starting basis $\mathbf{I}_\mathbf{A}$ which takes the form:
\begin{equation}
    \mathbf{I}_\mathbf{A}=\begin{pmatrix}I_1\\\partial_{z_1}I_1\\
    \vdots\\
    \partial_{z_{\tilde{h}}}I_1\\
    \partial_{z_\alpha}\partial_{z_\beta}I_1\end{pmatrix}\,,
\end{equation}
where $\alpha$ and $\beta$ are a set of fixed integers which depends on the integral we consider, as explained in section~\ref{genericK3ansatz}.
From the connection $\mathbf{A}$ we can now read off a system of PF operators of degrees two $\underline{L}^{(2)}$ and three $\underline{L}^{(3)}$ which decompose into an $\epsilon$ independent part $\underline{L}_0^{(i)}$ and an $\epsilon$ dependent part of lower degree $\underline{L}_\epsilon^{(i-1)}$.
In particular, for a single operator of degree three we have\footnote{Note that we always normalise the PF operators and the last subscript denotes the leading derivative term.}:
\begin{equation}
  L^{(3)}(\mathbf{z})=L_0^{(3)}(\mathbf{z})+L_\epsilon^{(2)} (\mathbf{z}) \,,
\end{equation}
with:
\begin{align}
    L_{0\, (z_a,z_b,z_c)}^{(3)}&=\dfrac{\partial}{\partial z_{a}}\dfrac{\partial}{\partial z_{b}}\dfrac{\partial}{\partial z_{c}}+\sum_{i_1,i_2=1}^{\tilde{h}}q^{(0)}_{i_1,i_2\,(z_a,z_b,z_c)}\dfrac{\partial}{\partial z_{i_1}}\dfrac{\partial}{\partial z_{i_2}} + \sum_{i=1}^{\tilde{h}}q^{(0)}_{i\,(z_a,z_b,z_c)}\dfrac{\partial}{\partial z_{i}}+q^{(0)}_{(z_a,z_b,z_c)}\,,\\
    L_{\epsilon\, (z_a,z_b,z_c)}^{(2)}&=\sum_{i_1,i_2=1}^{\tilde{h}}\epsilon\,q^{(1)}_{i_1,i_2\,(z_a,z_b,z_c)}\dfrac{\partial}{\partial z_{i_1}}\dfrac{\partial}{\partial z_{i_2}} + \sum_{i=1}^{\tilde{h}}\biggl(\epsilon\,q^{(1)}_{i\,(z_a,z_b,z_c)}+\epsilon^2\,q^{(2)}_{i\,(z_a,z_b,z_c)}\biggr)\dfrac{\partial}{\partial z_{i}}\\
    &\quad+\epsilon\,q^{(1)}_{(z_a,z_b,z_c)}+\epsilon^2\,q^{(2)}_{(z_a,z_b,z_c)}+\epsilon^3\,q^{(3)}_{(z_a,z_b,z_c)}
\end{align}
and analogously for degree two.
By changing from the kinematic variables $\mathbf{z}$ to the moduli 
\begin{equation}
    \tau_i=\dfrac{\psi_1^{(i)}}{\psi_0}\,,
\end{equation}
and introducing a normalisation by the holomorphic period $\psi_0$, we trivialise the operators $\tilde{L}_0^{(3)}$ and $\tilde{L}_0^{(2)}$:
\begin{align}
    \tilde{L}_{0\, (\tau_a,\tau_b,\tau_c)}^{(3)}&=\dfrac{\partial}{\partial \tau_{a}}\dfrac{\partial}{\partial \tau_{b}}\dfrac{\partial}{\partial \tau_{c}}\dfrac{1}{\psi_0}\,,\\
    \tilde{L}_{0\,(\tau_a,\tau_b)}^{(2)}&=\dfrac{\partial}{\partial \tau_{a}}\dfrac{\partial}{\partial \tau_{b}}\dfrac{1}{\psi_0}\,.
\end{align}
This implies that together with Eq. (\ref{genericPFequality}) we are now able to trade derivative terms in $\tau_i$ with lower derivative terms at higher orders in $\epsilon$ as:
\begin{equation}\label{PF_trafo_1}
    \dfrac{\partial}{\partial \tau_{a}}\dfrac{\partial}{\partial \tau_{b}}\dfrac{\partial}{\partial \tau_{c}}\dfrac{I_1}{\psi_0}=-\tilde{L}_{\epsilon\,(\tau_a,\tau_b,\tau_c)}^{(2)}I_1\,,
\end{equation}
and
\begin{equation}\label{PF_trafo_2}
    \dfrac{\partial}{\partial \tau_{a}}\dfrac{\partial}{\partial \tau_{b}}\dfrac{I_1}{\psi_0}=-\tilde{L}_{\epsilon\,(\tau_a,\tau_b)}^{(1)}I_1\,.
\end{equation}
With these results let us now go through Eq. (\ref{ansatzK3eps0withF}) and check if everything is $\epsilon$-factorised.
By rewriting the integrals in Eq. ($\ref{ansatzK3generalsigma}$) we get
\begin{equation}
    \dfrac{\partial}{\partial \tau_{a}}M_0=\epsilon\left(M_{a}-F_{a,0}\,M_0\right),
\end{equation}
which immediately implies that the first block in the differential equation is in $\epsilon$-form. 
Let us now analyse the second block  $\partial_{\tau_a}\,M_{b}$, where $a,b=1,...,\tilde{h}$. At $\epsilon=0$, we have two possibilities, either $\partial_{\tau_a}\,M_{b}$=0, which implies $\partial_{\tau_a}\partial_{\tau_b}\frac{I_1}{\psi_0}=0$ at $\epsilon=0$, or $\partial_{\tau_a}\,M_{b}=k\,M_{n+m-1}$, with $k\in\mathbb{Q}$ at $\epsilon=0$. 
The configuration of $a$ and $b$ which gives either of these possibilities will depend on the intersection matrix $\Sigma$ as we saw in the main section~\ref{genericK3ansatz}.
We now consider these two cases separately. 
For the first case, we can begin by plugging in the definition for $M_{b}$ to find:
\begin{align}\label{eps_facto_1}
    \dfrac{\partial}{\partial\tau_a}M_b&=\dfrac{\partial}{\partial\tau_a}\biggl(\dfrac{1}{\epsilon}\dfrac{\partial}{\partial \tau_b} M_0+F_{b,0}M_0\biggr)\notag\\
    &=\dfrac{1}{\epsilon}\dfrac{\partial}{\partial\tau_a}\dfrac{\partial}{\partial\tau_b}M_0+\biggl(\dfrac{\partial}{\partial \tau_a}F_{b,0}\biggr)M_0+F_{b,0}\biggl(\dfrac{\partial}{\partial \tau_a}M_0\biggr)\notag\\
    &=\dfrac{1}{\epsilon}\dfrac{\partial}{\partial\tau_a}\dfrac{\partial}{\partial\tau_b}M_0+\biggl(\dfrac{\partial}{\partial \tau_a}F_{b,0}\biggr)M_0+\epsilon\biggl(F_{b,0}M_a-F_{a,0}F_{b,0}M_0\biggr)\,,
\end{align}
where we can now plug in the relevant operator, as we are considering PF operators of the form of Eq. (\ref{genericPF}) and can, by virtue of Eq. (\ref{genericPFequality}), express the first term in the last line through
\begin{equation}
    \dfrac{\partial}{\partial \tau_a}\dfrac{\partial}{\partial \tau_b}M_0=\tilde{L}_{0\,(\tau_a,\tau_b)}^{(2)}I_1=-\tilde{L}_{\epsilon\,(\tau_a,\tau_b)}^{(1)}I_1\,.
\end{equation}
The PF operator $\tilde{L}_\epsilon$ evaluates to:
\begin{align}
    \tilde{L}_{\epsilon\,(\tau_a,\tau_b)}^{(1)}I_1&=\biggl(\epsilon\,\sum_{i=1}^{\tilde{h}}\tilde{q}^{(1)}_{i\,(\tau_a,\tau_b)}\dfrac{\partial}{\partial \tau_i}+\epsilon \,\tilde{q}^{(1)}_{(\tau_a,\tau_b)} +\epsilon^2\,\tilde{q}^{(2)}_{(\tau_a,\tau_b)}\biggr)\dfrac{I_1}{\psi_0}\notag\\
    &=\epsilon \,\tilde{q}^{(1)}_{(\tau_a,\tau_b)} M_0+\epsilon^2\,\tilde{q}^{(2)}_{(\tau_a,\tau_b)}M_0+\epsilon^2\,\sum_{i=1}^{\tilde{h}}\tilde{q}^{(1)}_{i\,(\tau_a,\tau_b)}(M_i-F_{i,0}M_0)\,,
\end{align}
and it therefore follows that, to cancel the $\mathcal{O}(\epsilon^0)$ terms in Eq. (\ref{eps_facto_1}) we need to require:
\begin{equation}
\tilde{q}^{(1)}_{(\tau_a,\tau_b)}-\dfrac{\partial}{\partial\tau_a}F_{b,0}=0\,,
\end{equation}
which gives a condition for $F_{b,0}$. 
Note that more conditions may be needed to fully fix $F_{b,0}$.
\\Let us now consider the other case $\partial_{\tau_a}\,M_{b}=k\,M_{n+m-1}$ at $\epsilon=0$.
This can be rewritten as 
\begin{equation}
\dfrac{\partial}{\partial{\tau_a}}\dfrac{\partial}{\partial{\tau_b}}M_0=k\,\dfrac{\partial}{\partial{\tau_\alpha}}\dfrac{\partial}{\partial{\tau_\beta}}M_0\,,
\end{equation}
where $\alpha, \beta$ are fixed integers as in the last line in Eq. (\ref{ansatzK3generalsigma}).
Since we are considering only PF operators as in Eq. (\ref{genericPF}), we can then write for $\epsilon\neq0$
\begin{equation}
    \dfrac{\partial}{\partial\tau_a}\dfrac{\partial}{\partial \tau_b}M_0=k\,\dfrac{\partial}{\partial{\tau_\alpha}}\dfrac{\partial}{\partial\tau_{\beta}}M_0+\tilde{L}^{(1)}_{1(\tau_a,\tau_b)}\,,
\end{equation}
hence we get an $\epsilon$-form following the same steps as before:
\begin{equation}
    \dfrac{\partial}{\partial {\tau_{a}}}M_{b}=\epsilon\biggl(M_{n+m-1}-\sum_{k=0}^{\tilde{h}} F_{(n+m-1),k}\,M_k\biggr)\,.
\end{equation}
Next, one can do the same exercise for $\partial_{\tau_a}M_{n+m-1}$ which take the form:
\begin{equation}\label{last_row}
    \dfrac{\partial}{\partial\tau_a}M_{n+m-1}=\dfrac{1}{\epsilon^2}\dfrac{\partial}{\partial\tau_a}\dfrac{\partial}{\partial\tau_1}\dfrac{\partial}{\partial\tau_{\tilde{h}}}M_0+\dfrac{1}{\epsilon}\dfrac{\partial}{\partial\tau_a}\dfrac{\partial}{\partial\tau_1}\biggl(F_{\tilde{h},0}M_0\biggr)+\sum_{k=0}^{n+m-2}\dfrac{\partial}{\partial\tau_a}\biggl(F_{(n+m-1),k}M_k\biggr)
\end{equation}
where after having done the algebra and plugging in the relevant PF operators of higher order in $\epsilon$ as in Eqs. (\ref{PF_trafo_1}) and (\ref{PF_trafo_2}), we find some conditions for the functions $F_{i,j}$ which eliminate the terms proportional to $\epsilon^{-1}$ and $\epsilon^0$ from  Eq. (\ref{last_row}) and the only surviving term will be proportional to $\epsilon$.
\\ Lastly we need to check how the decoupling integrals $M_j$ $\epsilon$-factorise. 
As they are decoupling integrals, their starting differential equation is:
\begin{align}
    \dfrac{\partial}{\partial z_a}I^\text{dec}_j&=\epsilon \,q^{(1)\,\text{dec}}_{(z_a)}\,I^\text{dec}_j+\biggl(q^{(0)}_{(z_a)}+\sum_{b=1}^{\tilde{h}}q^{(0)}_{b\,(z_a)}\dfrac{\partial}{\partial z_b}+\epsilon \,q^{(1)}_{(z_a)}\biggr)I_1\notag\\
&=q^{(0)}_{(z_a)}\psi_0\,M_0+\sum_{b=1}^{\tilde{h}}q^{(0)}_{b\,(z_a)}\biggl(\dfrac{\partial}{\partial z_b}\psi_0\biggr)M_0\notag\\
&\quad+\epsilon\,\biggl( q^{(1)\,\text{dec}}_{(z_a)}M_j-q^{(1)\,\text{dec}}_{(z_a)}F_{j,0}M_0+q^{(1)\,\text{dec}}_{(z_a)}\psi_0M_0+\sum_{b=1}^{\tilde{h}}q^{(0)}_{b\,(z_a)}\psi_0(M_b-F_{b,0}M_0)\biggr)\,,
\end{align}
where in the second line we substituted in $I_1=M_0\psi_0$ and $\partial_{z_b}M_0=\epsilon\,(M_b-F_{b,0}M_0)$.
We now act with a $\tau_a$ derivative on $M_j$ to find:
\newpage
\begin{align}
    \dfrac{\partial}{\partial \tau_a}M_j&=\dfrac{\partial}{\partial \tau_a}I_j^\text{dec}+\biggl(\dfrac{\partial}{\partial \tau_a}F_{j,0}\biggr)M_0+F_{j,0}\biggl(\dfrac{\partial}{\partial \tau_a}M_0\biggr)\notag\\
    &=\sum_{c=1}^{\tilde{h}}\biggl(\dfrac{\partial z_c}{\partial \tau_a} \biggr)\biggl(q^{(0)}_{(z_c)}\psi_0 M_0+\sum_{b=1}^{\tilde{h}}q^{(0)}_{b\,(z_c)}\biggl(\dfrac{\partial \psi_0}{\partial z_b}\biggr)M_0\biggr)+\biggl(\dfrac{\partial}{\partial \tau_a}F_{j,0}\biggr)M_0\notag\\
    &+\epsilon\,\biggl\{\sum_{c=1}^{\tilde{h}}\biggl(\dfrac{\partial z_c}{\partial \tau_a} \biggr)\biggl( q^{(1)\,\text{dec}}_{(z_c)}M_j-q^{(1)\,\text{dec}}_{(z_c)}F_{j,0}M_0+q^{(1)\,\text{dec}}_{(z_b)}\psi_0M_0\biggr)\notag\\
    &\qquad+F_{j,0}\bigl(M_a-F_{a,0}M_0\bigr)+\sum_{c=1}^{\tilde{h}}\biggl(\dfrac{\partial z_c}{\partial \tau_a} \biggr)\sum_{b=1}^{\tilde{h}}q^{(0)}_{b\,(z_c)}\psi_0\bigl(M_b-F_{b,0}M_0\bigr)\biggr\}\,,
\end{align}
which will be $\epsilon$-factorised if we require:
\begin{equation}
    \dfrac{\partial}{\partial \tau_a}F_{j,0}=-\sum_{c=1}^{\tilde{h}}\biggl(\dfrac{\partial z_c}{\partial \tau_a} \biggr)\biggl(q^{(0)}_{(z_c)}\psi_0 +\sum_{b=1}^{\tilde{h}}q^{(0)}_{b\,(z_c)}\biggl(\dfrac{\partial \psi_0}{\partial z_c}\biggr)\biggr)\,.
\end{equation}
Thus, we have shown that the differential equation corresponding to the basis of Eq. (\ref{ansatzK3generalsigma}) is in $\epsilon$-form in the upper triangular block and that the strictly lower triangular part can be brought to $\epsilon$-form by introducing auxiliary functions $F_{i,j}$, which are fixed through differential equations.
We did, however not manage to show that these differential equations always allow for a solution, but they are solvable in all examples we have computed.

\section{Modular bootstrap}\label{newApp}
Here we briefly review a method to solve linear differential equations involving modular forms. 
This method was first introduced in Ref.~\cite{Giroux:2022wav} and applied again in Ref.~\cite{Giroux:2024yxu}.
In Eq.~(\ref{ans_sunrise_dem}) and Eq.~(\ref{ans_sunrise_dec}) we have a set of functions $F_{i,j}$ which $\epsilon$-factorise the differential equation.
They follow a linear differential equation and from the elliptic nature of the sunrise integral we know that they are modular forms.
To determine them, we will use their modular properties instead of solving the differential equation directly.
\\
When $F_{i,j}$ are modular forms of weight one depending on $r$ variables, we find:
\begin{equation}\label{start_mb}
    \text{d}F_{i,j}=R_{i,j}^{(0)} \psi_0+\sum_{n=1}^r R_{i,j}^{(k)}\dfrac{\partial}{\partial x_k}\psi_0\,,
\end{equation}
where $R_{i,j}^{(k)}$ are modular invariant one-forms.
Since $\psi_0$ transforms as a modular form of weight one
\begin{equation}
    \gamma(\psi_0)=(c\tau+d)\psi_0\,,
\end{equation}
with $\gamma(\tau)=\frac{a\tau+b}{c\tau+d}$ for $a d-bc=1$ being the standard modular transformation, we find that Eq.~(\ref{start_mb}) transforms as
\begin{equation}
\text{d}\biggl((c\tau+d) F_{i,j}\biggr)=R_{i,j}^{(0)}(c\tau+d)\psi_0+\sum_{n=1}^r R_{i,j}^{(n)}\dfrac{\partial}{\partial x_n}(c\tau+d)\psi_0\,,
\end{equation}
which leads to 
\begin{equation}
 F_{i,j} \,  \text{d}\tau=\psi_0 \sum_{n=1}^r R_{i,j}^{(n)} \dfrac{\partial\tau}{\partial x_n}\,,
\end{equation}
where the derivatives can be expressed through the Jacobian as $\frac{\partial \tau}{\partial x_n}=(\mathfrak{J}^{-1})_{1,n}\,$.
\\
When the functions $F_{i,j}$ are modular forms of weight two we find a differential equation of the form
\begin{equation}
    \text{d}F_{i,j}=\psi_0\biggl(R_{i,j}^{(0)} \psi_0+\sum_{n=1}^r R_{i,j}^{(k)}\dfrac{\partial}{\partial x_k}\psi_0\biggr)\,,
\end{equation}
which leads to 
\begin{equation}
 F_{i,j} \,  \text{d}\tau=\dfrac{\psi_0^2}{2} \sum_{n=1}^r R_{i,j}^{(n)} \dfrac{\partial\tau}{\partial x_n}\,,
\end{equation}
through analogous arguments.

\bibliographystyle{jhep}
\bibliography{bibfile.bib}

\end{document}